\documentclass{jfm}
\usepackage{graphicx}
\usepackage{epstopdf, epsfig}
\usepackage{subfig,setspace}
\usepackage{color}
\usepackage{pdflscape}
\usepackage{amsmath}

\shorttitle{Effects of Ekman pumping on quasi-geostrophic RB convection}
\shortauthor{M. Plumley et al}

\title{The effects of Ekman pumping on quasi-geostrophic Rayleigh-B\'enard convection}

\author{Meredith Plumley\aff{1}, 
 Keith Julien\aff{1} \corresp{\email{julien@colorado.edu}},
  Philippe Marti\aff{1} \and  Stephan Stellmach\aff{2}}

\affiliation{\aff{1}Department of Applied Mathematics, University of Colorado, Boulder, CO 80309, USA
\aff{2}Institut f\"ur Geophysik, Westf\"{a}lische Wilhelms-Universit\"at M\"{u}nster, D-48149 M\"unster, Germany }

\begin{document}

\maketitle

\begin{abstract}
Numerical simulations of three-dimensional, rapidly rotating Rayleigh-B\'enard convection are performed using an asymptotic quasi-geostrophic model that incorporates the effects of no-slip boundaries through (i) parameterized Ekman pumping boundary conditions, and (ii) a thermal wind boundary layer  that regularizes the enhanced thermal fluctuations induced by pumping. The fidelity of the model, obtained by an asymptotic reduction of the Navier-Stokes equations that implicitly enforces a pointwise geostrophic balance, is explored for the first time by comparisons of simulations against the findings of direct numerical simulations (DNS) and laboratory experiments.
Results from these methods have established Ekman pumping as the  mechanism responsible for significantly enhancing the vertical heat transport.
This asymptotic model demonstrates excellent agreement over a range of thermal forcing for $Pr \approx 1$ when  compared with results from experiments and DNS at maximal values of their attainable rotation rates, as measured by the Ekman number ($E \approx 10^{-7}$); good qualitative agreement is achieved for $Pr > 1$.
Similar to studies with stress-free boundaries, four spatially distinct flow morphologies exists. Despite the presence of frictional drag at the upper and/or  lower boundaries, a strong non-local inverse cascade of barotropic (i.e., depth-independent) kinetic energy persists in the final regime  of geostrophic turbulence and is dominant at large scales.  For mixed no-slip/stress-free and no-slip/no-slip boundaries, Ekman friction is found to attenuate the efficiency of the upscale energy transport and, unlike the case of stress-free boundaries, rapidly saturates the barotropic kinetic energy.  For no-slip/no-slip boundaries, Ekman friction is strong enough to prevent the development  of a coherent  dipole vortex condensate. Instead vortex pairs are found to be intermittent, varying in both time and strength. For all combinations of boundary conditions, a Nastrom-Gage type spectrum of kinetic energy is found where the power law exponent changes from $\approx -3$ to $\approx -5/3$, i.e., from steep to shallow, as the spectral wavenumber increases.
\end{abstract}

\begin{keywords}
B\'enard convection, geostrophic turbulence, quasi-geostrophic flows
\end{keywords}
\section{Introduction}\label{sec:intro}

Rotation and thermal forcing are essential ingredients that  influence the dynamics of  stellar and planetary interiors, planetary atmospheres,
stars, and terrestrial oceans. Examples include solar and stellar envelopes \citep{mM05,gastine2014zonal}, planetary atmospheres and interiors \citep{jmA15,mH2015}, and deep ocean convection on Earth \citep{jM99}. The quintessential paradigm for investigating the fundamentals of rotating, thermally forced flows is rotating 
Rayleigh-B\'enard convection (RRBC) about a vertical axis, i.e., convection
in a layer of Boussinesq fluid confined between flat, horizontal, rigidly rotating upper and lower boundaries that sustain a destabilizing temperature jump $\Delta T>0$. 
Flows are characterized by the nondimensional Rayleigh ($Ra$), convective Rossby ($Ro$) and Ekman ($E$) numbers defined as 
\begin{equation}
\label{eqn:nond}
Ra=\frac{g\alpha \Delta T H^3}{\nu \kappa},\qquad 
Ro=\sqrt{\frac{g\alpha \Delta T}{4\Omega^2 H}},\qquad
E=Ro\sqrt{\frac{Pr}{Ra}}.
\end{equation}
These, respectively, measure the strength of thermal forcing, the strength of system rotation $\Omega$ relative to thermal forcing, and 
the importance of viscous diffusion relative to $\Omega$. Here $g$ denotes gravity, $H$ the layer depth, and $\alpha$ the thermal expansion coefficient.  The Prandtl number, $\Pran$ =  ${\nu}/{\kappa}$, is the ratio of viscous to thermal diffusion rates.
 
Geophysical and astrophysical flows are often turbulent and rotationally constrained. This is characterized by high Rayleigh numbers $Ra\gg 1$ and low Rossby numbers $Ro\ll1$, implying small Ekman numbers $E\ll1$. For example, estimates for these parameters in Earth's outer core are $Ra \approx 10^{20}$, $Ro\approx 10^{-6}$,  and $E \approx10^{-15}$ \citep{Soderlund}. Utilization of isotropic turbulence theory for an order of magnitude estimate predicts that the ratio between the largest and smallest length-scales in a turbulent flow scales as $\sim Ra^{3/8}$ \citep{sP00} or $\mathcal{O}(10^7)$ in the case of the outer core.  The low $Ro$ indicates geostrophy, a primary hydrodynamic balance between the pressure gradient force and the Coriolis force. This balance further increases the complexity  with the presence of  fast inertial waves,  thin Ekman boundary layers adjacent to no-slip surfaces, and disparate timescales characterizing the evolution of mean and fluctuating temperatures. The extreme parameters and their associated properties described above are substantially beyond the spatiotemporal resolution capabilities for DNS and  the thermal and mechanical capabilities of  laboratory experiments.

\begin{figure}
	\centering
		\includegraphics[width=0.47\textwidth]{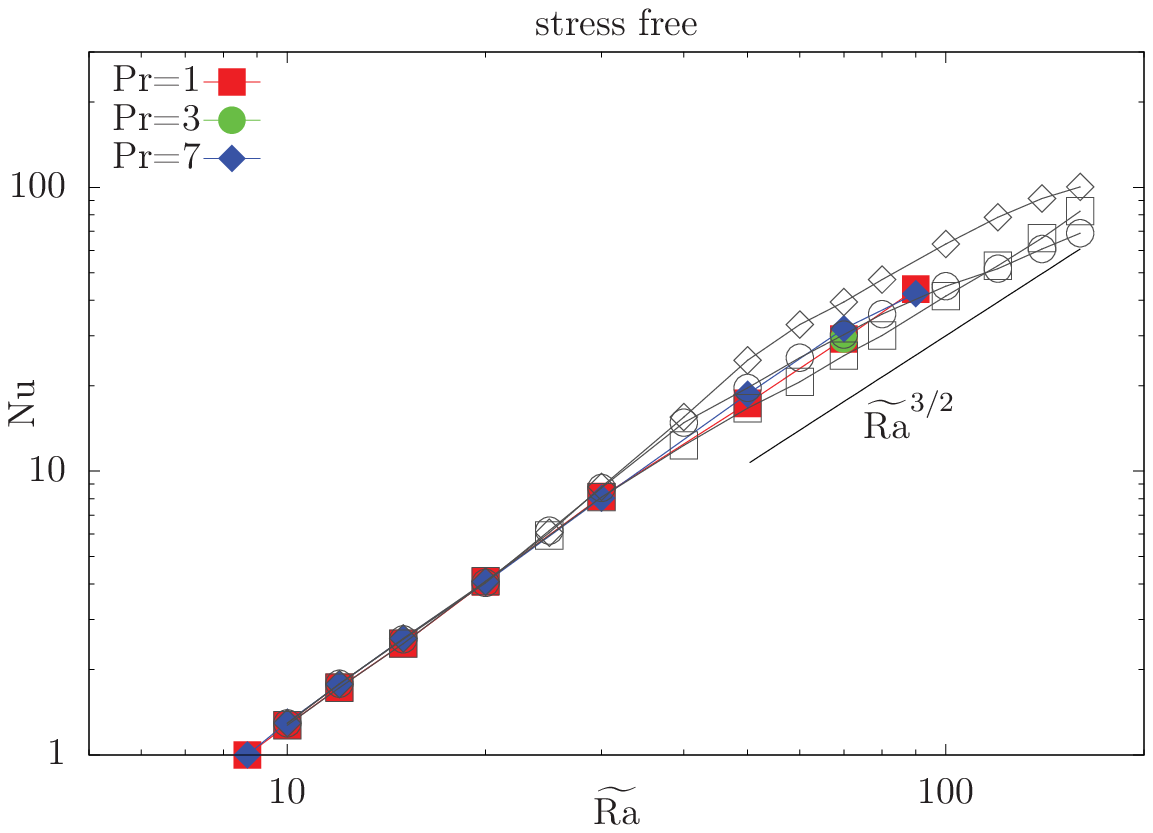} \hspace{.75em}
		\includegraphics[width=0.47\textwidth]{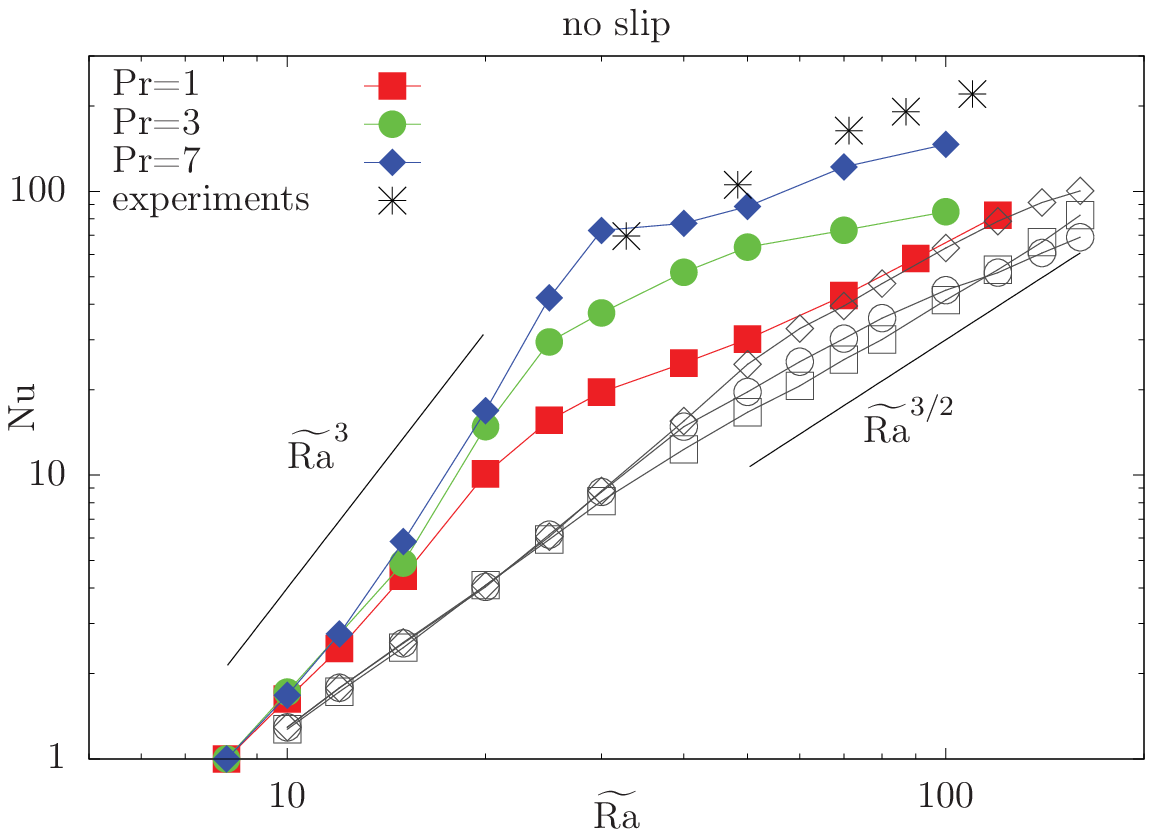}
	\caption{\small{Comparison of DNS for RRBC (filled symbols) at $E=10^{-7}$ and the NH-QGM (open symbols) for
	(a) stress-free and (b) no-slip boundary conditions. The plots show the Nusselt number $Nu$  as a function of reduced Rayleigh number $\widetilde{Ra} = RaE^{4/3}$ for Prandtl numbers $Pr$ of 1 (squares), 3 (circles), and 7 (diamonds). The two methods show good quantitative agreement in the stress-free case (a) but a large discrepancy for no-slip (b). Laboratory data for $5.79 \leq Pr \leq 6.2$ is also included  (asterisks). Plots reproduced from \citet{Stellmach, Cheng}, with an updated  plot (b) including higher $\widetilde{Ra}$ runs.}}
	\label{fig:DNS_SFvsNHQGE}
	\vspace{-3ex}
\end{figure}

 While the extreme parameter values present a challenge for DNS \citep{bF14,cG14,Stellmach,Horn}, they allow for the use of perturbation methods that rely on the small size of $Ro$ as an asymptotic expansion parameter. 
The reduced Non-Hydrostatic Quasi-Geostrophic Model (NH-QGM) \citep{kJ98a, kJ06,Sprague} was developed as a simplification, or reduction, of the Boussinesq equations at low $Ro$.
The model implicitly enforces geostrophy as a pointwise constraint and  is asymptotically exact for RRBC in the presence of stress-free boundary conditions as $Ro\sim E^{1/3}\rightarrow 0$. Of particular interest is the dependence of the heat transport, as measured by the nondimensional Nusselt number $Nu = {qH}/{\rho_0 c_p \kappa \Delta T}$, on the external control parameters $(Ra,E,Ro)$ and the molecular fluid property $Pr$; here $q$ is the heat flux and $c_p$ is the volumetric heat capacity. 
Importantly, $Nu$ characterizes the fluid state and transitions between fluid states and is commonly reported due to the relative ease with which it can be measured.  To validate results obtained from the NH-QGM, comparisons were made with DNS of the Boussinesq equations at  $E=10^{-7}$ \citep{Stellmach}. Figure \ref{fig:DNS_SFvsNHQGE}a illustrates the high level of agreement in the $Nu$-$Ra$ relation for stress-free boundaries with the asymptotic scaling of $\sim 3/2$.

For the case of no-slip boundaries (see figure \ref{fig:DNS_SFvsNHQGE}b) substantial disagreement is found  in comparison with the NH-QGM results for the $Nu$-$Ra$ relation. DNS and laboratory data in water ($Pr\sim7$) show markedly increased values of $Nu$ for all values of $Ra$ with a steep scaling exponent of $\gtrsim 3$ \citep{Cheng,Stellmach}. While there is qualitative agreement between DNS and experiments, differences in the Nusselt number results can be attributed to differing Prandtl numbers for water and different geometries (DNS are performed in a horizontally periodic box with $Pr = 7$, whereas experiments are performed in cylindrical containers with $5.79 \leq Pr \leq 6.2$) 
More specifically, it is shown that the divergence in $Nu$ between the NH-QGM, where Ekman boundary layers are presumed passive, and DNS data grows as the Ekman number $E$ is lowered in the latter.
A recent study by \cite{Cheng} (see their figure 6) indicates that the  heat transport exponent $\alpha$ in the $Nu$-$Ra$ relation $Nu \propto Ra^{\alpha}$  increases monotonically from $1$ to $3.6$  as $E$ decreases to $\sim 10^{-7}$. By replacing the no-slip boundaries with parameterized Ekman pumping boundary conditions and establishing quantitative agreement, \cite{Stellmach} clearly demonstrate that the root cause of this divergence is the presence of Ekman boundary layers. This was an important finding given that the boundary layer width and the associated vertical pumping (or suction) velocities, which respectively scale as  $E^{1/2}H$ and $w_E = E^{1/2} \boldsymbol{\widehat{z}}\cdot\nabla\times\boldsymbol{u}/\sqrt{2}$, have diminishing values as $E\rightarrow 0$ (The latter formulae for $w_E$ represents an exact parameterization of the laminar Ekman boundary layer \citep{Greenspan}). Indeed, the asymptotic NH-QGM assumes that  Ekman boundary layers become increasingly insignificant as $E\rightarrow 0$ and are filtered from the dynamics, thus implicitly assuming impenetrable stress-free boundaries \citep{kJ98a,Sprague}. 

  An asymptotic analysis of the boundary layer regions of RRBC using the Boussinesq equations was recently reported in \cite{Julien15}. Specifically, it was established that the effects of  Ekman pumping become significant within the rotationally constrained regime when $E^{-1/9} \lesssim Ra E^{4/3}  \lesssim E^{-1/3}$, i.e., for sufficiently strong thermal forcing. Consequently, the NH-QGM was updated to include the effect of no-slip boundaries and the resultant Ekman pumping through the use of the parameterized pumping boundary conditions and the introduction of a thermal wind layer.  The thermal wind layer was shown to be capable of supporting the enhanced thermal fluctuations necessary for the amplification of heat transport. Both effects are explicitly controlled by the magnitude of the Ekman number $E$. Thus, the model, called the Composite NH-QGM (CNH-QGM), now includes the Ekman number $E$ as an explicit control parameter. The capabilities of the composite model were explored in \cite{Julien15} by locating and tracking exact single mode (i.e., single horizontal wavenumber) solutions as a function of reduced Rayleigh number. Single-mode solutions represent an unstable solution branch of the CNH-QGM, despite this, their presence strongly influence the path of the realized solution in phase space and therefore capture and characterize the qualitative behavior of the enhanced heat transport in the fully nonlinear regime. 

In this paper, we extend the work of  \cite{Julien15} and explore simulations of the asymptotically reduced CNH-QGM that, similar to DNS, capture all dynamically active spatial scales and thus the realized nonlinear solution. The explicit appearance of the Ekman number $E$ permits a direct comparison with DNS investigations of the Boussinesq equation  for RRBC and provides the capability to assess both the fidelity of the reduced model and the degree to which the DNS has entered the asymptotic regime. 
 In Section 2, the CNH-QGM  derived in  \citet{Julien15} is documented and the multi-layer structure unique to the presence of no-slip  boundaries identified.  In Section 3 the numerical method used to solve the CNH-QGM is discussed and  results obtained from a suite of simulations presented. Concluding remarks are given in Section 4.


\section{Composite Model}\label{sec:model}

The full derivation and discussion of the CNH-QGM is presented in \citet{Julien15}; here we simply provide an overview of their asymptotic model. The leading order velocity field outside the Ekman layer, $\boldsymbol{u}=(\boldsymbol{u}_\perp,W)$, is in geostrophic balance where $ \boldsymbol{\widehat{z}} \times \boldsymbol{u}_\perp = -\nabla_\perp p$. 
It follows that the horizontal velocity field solved by  $\boldsymbol{u}_\perp =(-\partial_{y},\partial_{x},0)\Psi$
is  non-divergent  where the pressure   $p=\Psi$ becomes  the geostrophic streamfunction. On defining $\epsilon=E^{1/3}$ $=Ro$, the reduced CNH-QGM system governing the evolution of the fluid is given by
\begin{equation}\label{eq:xi}
D_t^{\bot} \zeta - \partial_Z W = \nabla^2_{\bot} \zeta \, ,
\end{equation}
\begin{equation}\label{eq:W}
D_t^{\bot} W + \partial_Z \Psi = \frac{\widetilde{Ra}}{Pr} \theta + \nabla^2_{\bot} W \, ,
\end{equation}
\begin{equation}\label{eq:Theta}
D_t^{\bot} \theta + W\partial_Z \overline{\Theta} + \epsilon \left ( \nabla_{\bot}\cdot\left ( \boldsymbol{u}^{ag}_{1\perp}\theta \right )
+ \partial_Z(W \theta - \overline{W\theta})\right ) = 
\frac{1}{Pr} (\nabla^2_{\bot} + \epsilon^2 \partial_{ZZ}) \theta \, ,
\end{equation}
\begin{equation}\label{eq:mTheta}
\partial_\tau \overline{\Theta}  +  \partial_Z\left ( \overline{W\theta}\right ) = \frac{1}{Pr} \partial_{ZZ}  \overline{\Theta} \, ,
\end{equation}
capturing, respectively, the evolution of vertical vorticity $\zeta=-\nabla^2_\perp\Psi$, vertical velocity $W$,  and temperature $\Theta = \overline{\Theta} + \epsilon  \theta$ at reduced Rayleigh number $\widetilde{Ra}=Ra \epsilon^4$ for a given Prandtl number $Pr$. The temperature  is decomposed into a mean (horizontally-averaged) component $\overline{\Theta}$ evolving on a slow timescale $\tau = \epsilon^2 t$ and a small fluctuating component $\theta$ characteristic of RRBC. The overbar indicates the horizontal and fast time averaging operator given by 
\begin{equation} \label{eq:avg}
\overline{f}(Z,T) =\lim_{\tau,A \rightarrow \infty} \frac{1}{\tau A} \int_{A} f(\textbf{x},Z,t,T) d\textbf{x}dt \, ,
\end{equation}
where $A$ is the area of the box. 
Here $D_t^{\bot}=\partial_t  +  \boldsymbol{u}_{\perp} \cdot \nabla_\perp$ denotes the material derivative with horizontal geostrophic advection  
and $\boldsymbol{u}^{ag}_{1\perp}$ in  (\ref{eq:Theta}) denotes the ageostrophic velocity field determined through the three-dimensional incompressibility condition 
\begin{equation}
\nabla_\perp \cdot \boldsymbol{u}^{ag}_{1\perp} + \partial_Z W =0 \, .
\end{equation}
 The system is solved for the RRBC configuration with  fixed temperature boundaries
\begin{equation}\label{eq:bdT}
\overline{\Theta}=1,  \ \ \theta = 0 \textrm{ at} \  Z = 0 \, ,\qquad 
\overline{\Theta}=0,  \ \ \theta = 0 \textrm{ at} \  Z = 1 \, .
\end{equation}
Velocity boundary conditions are either parameterized Ekman pumping for no-slip (NS)  
\begin{equation}\label{eq:bdw}
\mbox{NS-NS:} \quad W =  \epsilon^{1/2} \, \frac{\zeta}{\sqrt{2}}  \textrm{ at}\  Z = 0 \, , \qquad 
W =  -\epsilon^{1/2} \, \frac{\zeta}{\sqrt{2}}  \textrm{ at}\  Z = 1 \, ,
\end{equation}
or implicit impenetrable stress-free boundaries ( SF)
\begin{equation}\label{eq:bdw_SF}
\mbox{SF-SF:} \quad W =  0  \textrm{ at}\  Z = 0,1
\end{equation}
with $\epsilon = 0$ or mixed (NS-SF) combinations thereof (e.g., NS at $Z=0$ and SF at $Z = 1$).
Equations (\ref{eq:xi})-(\ref{eq:bdT}) with one of the above choices of velocity boundary conditions form a closed reduced system that filters fast inertial waves occurring on  $\mathcal{O}(\Omega^{-1})$ 
timescales and $\mathcal{O}(L)$ spatial scales. The imposition of strong spatial anisotropy by nondimensionalizing vertical and horizontal scales with $H$ and  $L=\epsilon H$ is a key step in the filtering that gives  rise to the reduced dynamics \citep{Julien07}. Velocity and time are  nondimensionalized using $\nu/L$ and $L^2/\nu$ and temperature is nondimensionalized by $\Delta T$.

The evolution of the barotropic vertical vorticity $\langle \zeta \rangle$ is a key quantity in the understanding of the inverse energy cascade
recently observed in reduced simulations of RRBC \citep{Rubio} as well as DNS \citep{bF14,cG14,Stellmach}. Depth averaging the vertical vorticity equation (\ref{eq:xi}) according to $\langle\  \rangle =\int_0^1 \ dZ$, with the vortical variables decomposed as $f = \langle f \rangle + f'$, gives
\begin{equation}\label{eq:bxi}
\partial_t \langle \zeta \rangle=-   J[ \langle \Psi \rangle, \langle \zeta \rangle] -  \langle J[  \Psi^\prime ,  \zeta^\prime ]  \rangle
- \frac{\epsilon^{1/2}}{\sqrt{2}} \left ( \zeta^\prime (0) + \zeta^\prime (1) \right )
- \frac{2 \epsilon^{1/2}}{\sqrt{2}}  \langle \zeta \rangle 
 + \nabla^2_\perp \langle \zeta \rangle \, ,
\end{equation}
indicating that the growth of the barotropic vertical vorticity $\langle \zeta \rangle$ is dependent on the net balance between right-hand-side terms respectively for the barotropic self-interaction, baroclinic  (i.e. convective) forcing, barotropic and baroclinic Ekman friction, and barotropic viscous dissipation. Here $J[f,g]=\partial_x f \partial_y g - \partial_y f \partial_x g$ denotes the Jacobian for the horizontal advection of $g$, when $f$ is either $\langle \Psi \rangle $ or $\Psi' $.

We spectrally analyze equation (\ref{eq:bxi}) where barotropic and baroclinic fields are  decomposed according to the discrete Fourier transform
\begin{equation}
\langle f \rangle = \sum_{\boldsymbol{k}}  \langle \hat{f}_{\boldsymbol{k}}\rangle e^{  i \boldsymbol{k}\cdot\boldsymbol{x}_\perp} \, , \, f'  = \sum_{\boldsymbol{k}}  \hat{f}'_{\boldsymbol{k}}(Z) e^{  i \boldsymbol{k}\cdot\boldsymbol{x}_\perp} 
\end{equation} 
with $\boldsymbol{k} = (k_x, k_y)$.
Specifically, we  consider the evolution of the  spectral barotropic kinetic energy  $E_{bt}(\boldsymbol{k}) = \frac{1}{2}|\boldsymbol{k}|^2| \langle \hat{\Psi}_{\boldsymbol{k}} \rangle|^2$  at each wave number $\boldsymbol{k}$
 \begin{equation}
  \label{eq:bxk}
 \partial_{t} E_{bt}(\boldsymbol{k}) = T_{\boldsymbol{k}}+F_{\boldsymbol{k}} + Ef^{bc}_{\boldsymbol{k}} + Ef^{bt}_{\boldsymbol{k}}+ D_{\boldsymbol{k}}
 \end{equation}
 obtained  by multiplication of the Fourier decomposition of   (\ref{eq:bxi})  by $-\langle \hat{\Psi}_{\boldsymbol{k}} \rangle e^{i \boldsymbol{k}\cdot\boldsymbol{x}_\perp} $, followed by integration over physical space.
Here  $T_{\boldsymbol{k}} =  \sum_{\boldsymbol{p,q}} T_{{\boldsymbol{k p q}}} $ and $F_{\boldsymbol{k}} =  \sum_{\boldsymbol{p,q}} F_{{\boldsymbol{k p q}}} $ describe how power is converted to barotropic Fourier amplitude $\langle \hat{\Psi}_{\boldsymbol{k}}  \rangle$ through the barotropic and baroclinic triadic interactions involving  modes $\boldsymbol{p}$ and $\boldsymbol{q}$ such that $\boldsymbol{k}+\boldsymbol{q}+\boldsymbol{p}=\boldsymbol{0}$
 \citep{Rubio, bF14}, where
 

 \begin{landscape} {
\begin{table}
	\centering
		\begin{tabular}{c c c c c c c c c c}
		\ \ &\ \ & Box Dimensions & CNH-QGM Resolution & CNH-QGM & DNS Resolution &  CNH-QGM $\Delta t$   & DNS $\Delta t$  & CNH-QGM & DNS \\
			$\widetilde{Ra} $ 
			\ \ &  $Pr$   &  $E^{1/3}H$ x $E^{1/3}H$ x $H$ &  N$_x$ x N$_y$ x N$_z$  &  $\Delta$x/L$_{kol}$  & N$_x$ x N$_y$ x N$_z$ & $ Pr^{-1} E^{2/3} \displaystyle{\frac{H^2}{\kappa}}$  & $ \displaystyle{\frac{H^2}{\kappa}}$ &   $\left\langle Nu\right\rangle_t$  &  $\left\langle Nu\right\rangle_t$  \\
			\hline
			15 & 1/2 & 10$L_c$ x 10$L_c$ x 1& 192 x 192 x 192 & 0.868 &  384 x 384 x 385 & 2x$10^{-3}$ & 1.66x$10^{-8}$ &  4.445 & 4.526 \\
			15 & 1 & 10$L_c$ x 10$L_c$ x 1& 192 x 192 x 192 & 0.633 &  288 x 288 x 385 & 2x$10^{-3}$ & 1.64x$10^{-8}$ &  4.398 & 4.414 \\
			15 & 3 & 10$L_c$ x 10$L_c$ x 1& 192 x 192 x 192 & 0.811 &  384 x 384 x 385 & 2x$10^{-3}$ & 1.86x$10^{-8}$ &  5.020 & 4.897 \\ 
			15 & 7 & 10$L_c$ x 10$L_c$ x 1*& 192 x 192 x 192 & 0.547 & 288 x 288 x 385 & 2x$10^{-3}$ & 1.71x$10^{-8}$ & 6.162 & 5.825\\ 
			20 & 1/2 & 10$L_c$ x 10$L_c$ x 1& 288 x 288 x 192  & 1.04 & 384 x 384 x 385 & 1x$10^{-3}$ & 9.78x$10^{-9}$ &  8.623 &  8.491 \\
			20 & 1 & 10$L_c$ x 10$L_c$ x 1& 192 x 192 x 192 & 0.903 & 384 x 384 x 385 & 3x$10^{-3}$ & 6.04x$10^{-9}$  &  10.18 & 10.11 \\
			20 & 3 & 10$L_c$ x 10$L_c$ x 1& 192 x 192 x 192 & 1.22 & 384 x 384 x 385 & 3x$10^{-3}$ & 5.54x$10^{-9}$ &  22.91 & 14.83 \\
			20 & 7 & 10$L_c$ x 10$L_c$ x 1*& 192 x 192 x 192 & 0.826 &  384 x 384 x 385 & 2x$10^{-3}$ & 4.82x$10^{-9}$ & 28.92 & 16.87\\
			30 & 1/2 & 10$L_c$ x 10$L_c$ x 1& 384 x 384 x 192 & 0.729 &  384 x 384 x 385 & 5x$10^{-4}$ & 1.07x$10^{-8}$ &  14.60 & 15.00 \\
			30 & 1 & 10$L_c$ x 10$L_c$ x 1& 384 x 384 x 192 & 0.601 &  288 x 288 x 385 & 1x$10^{-4}$ & 5.01x$10^{-9}$ &  20.44 & 19.61 \\
			30 & 3 & 10$L_c$ x 10$L_c$ x 1& 288 x 288 x 192 & 1.09 &  384 x 384 x 385 & 1x$10^{-3}$ & 2.14x$10^{-9}$ &  50.56 & 37.29\\
			30 & 7 & 10$L_c$ x 10$L_c$ x 1*& 288 x 288 x 216 & 0.850 & 288 x 288 x 385 & 2x$10^{-3}$ & 2.36x$10^{-9}$  &  112.2 & 72.86 \\
			50 & 1/2 & 10$L_c$ x 10$L_c$ x 1& 384 x 384 x 288 & 1.00 &  384 x 384 x 385 & 3x$10^{-4}$ & 5.74x$10^{-9}$ &  29.82 &  30.73 \\
			50 & 1 & 10$L_c$ x 10$L_c$ x 1& 384 x 384 x 288  & 0.739 & 384 x 384 x 385 & 7x$10^{-4}$ & 2.87x$10^{-9}$ &  31.02 & 30.15\\
			50 & 3 & 10$L_c$ x 10$L_c$ x 1& 288 x 288 x 288 & 1.39 & 384 x 384 x 385 & 1x$10^{-3}$ & 1.51x$10^{-9}$ &  74.70 & 62.87 \\
			50 & 7 & 10$L_c$ x 10$L_c$ x 1*& 462 x 462 x 384  & 0.436 &  384 x 384 x 385 & 1x$10^{-3}$ & 1.10x$10^{-9}$ &  157.2 & 89.17\\
			70 & 1/2 & 10$L_c$ x 10$L_c$ x 1& 520 x 520 x 384 & 0.898 &  768 x 768 x 769 &  1x$10^{-4}$& 1.78x$10^{-9}$ &  46.16  & 48.37 \\
			70 & 1 & 10$L_c$ x 10$L_c$ x 1& 384 x 384 x 384  & 0.687 &  384 x 384 x 385 & 1x$10^{-4}$ & 2.12x$10^{-9}$ &  42.96 & 42.63 \\
			70 & 3 & 10$L_c$ x 10$L_c$ x 1& 288 x 288 x 192  & 1.61 &  384 x 384 x 385 & 1x$10^{-3}$ & 1.40x$10^{-9}$ &  84.21 & 72.89 \\
			70 & 7 & 10$L_c$ x 10$L_c$ x 1*& 384 x 384 x 384  & 0.923 &  384 x 384 x 385 & 1x$10^{-3}$ & 9.04x$10^{-10}$ & 179.4 & 122.0\\
			90 & 1/2 & 10$L_c$ x 10$L_c$ x 1& 520 x 520 x 384  & 1.15 &  768 x 768 x 1025 &  1x$10^{-4}$ & 1.12x$10^{-9}$ & 68.34  & 67.36 \\
			90 & 1 & 10$L_c$ x 10$L_c$ x 1& 384 x 384 x 384  & 1.02 & 576 x 576 x 513 & 1x$10^{-4}$ & 1.11x$10^{-9}$ &  59.22 & 58.04 \\
			100 & 3 & 10$L_c$ x 10$L_c$ x 1& 288 x 288 x 192  & 1.76 &  384 x 384 x 385 & 1x$10^{-3}$ & 1.28x$10^{-9}$ &  95.71 & 84.81\\
			100 & 7 & 10$L_c$ x 10$L_c$ x 1*& 384 x 384 x 384  & 1.03 &  384 x 384 x 513 & 5x$10^{-4}$ & 7.47x$10^{-10}$ &  182.7 & 146.4
		\end{tabular}
\caption{Model parameters used in NS-NS simulations for the CNH-QGM and DNS at $E=10^{-7}$. Horizontal box dimensions are a function of the critical length scale for stress-free convection, $L_c = 4.8154$ for $Pr \geq 1$ and $L_c = 6.0292$ for $Pr < 1$ \citep{Chandrar}. Box dimensions marked by an asterisk were simulated with box size 20$L_c$ x 20$L_c$ x $1$ for CNH-QGM.  Simulation resolutions were verified \textit{a posteriori}. 
The vertical resolutions ensure at least 8 points in both the $\mathcal{O}(E^{1/3}H)$ thermal wind layer for the CNH-QGM and the $\mathcal{O}(E^{1/2}H)$ Ekman layer for DNS. 
For the horizontal direction, the ratio of the Kolmogorov length scale to grid resolution was bounded below 2.
Time averaging for the $Nu$ results used intervals that ranged from 3.6x$10^{-2}$ to 8.5x$10^{-4}$ for DNS and 2x$10^{-1}$ to 1x$10^{-3}$ for CNH-QGM. 
}
	\label{tab:TABLEOfValues}
	\vspace{-1ex}
\end{table} }
\end{landscape}
  \begin{eqnarray}
   \label{eq:bxkT}
 T_{\boldsymbol{k p q}}  &=& b_{\boldsymbol{p q}} Re \left [ \langle \hat{\Psi}_{\boldsymbol{k}} \rangle \langle \hat{\Psi}_{\boldsymbol{p}}\rangle \langle \hat{\Psi}_{\boldsymbol{q}} \rangle  \right ] \delta_{\boldsymbol{k}+\boldsymbol{p} + \boldsymbol{q},\boldsymbol{0}}\\
  \label{eq:bxkF}
 F_{\boldsymbol{k p q}}  &=& b_{\boldsymbol{p q}} Re \left [ \langle \hat{\Psi}_{\boldsymbol{k}} \rangle \langle \hat{\Psi}^\prime_{\boldsymbol{p}} \hat{\Psi}^\prime_{\boldsymbol{q}} \rangle  \right ] \delta_{\boldsymbol{k}+\boldsymbol{p} + \boldsymbol{q},\boldsymbol{0}}\\
 b_{\boldsymbol{p q}}  &=& b_{\boldsymbol{q p}}  =\frac{1}{2} \left ( p^2-q^2 \right )  \left (p_xq_y-p_yq_x  \right )
 \end{eqnarray}
\noindent and $Re[\cdots]$ denotes the real part and $\delta_{\boldsymbol{k}+\boldsymbol{p} + \boldsymbol{q},\boldsymbol{0}}$ is the Kronecker delta function. 
 The latter terms  in (\ref{eq:bxk})  are defined as 
 \begin{eqnarray}
 \label{eq:edissp}
 D_{\boldsymbol{k}} &=& -|\boldsymbol{k}|^2 E_{bt}(\boldsymbol{k}) 
 \\
 Ef^{bc}_{\boldsymbol{k}} &=&-\frac{1}{\sqrt{2}}\varepsilon^{1/2}  \left (  E_{bc}(\boldsymbol{k},Z=0) + E_{bc}(\boldsymbol{k},Z=1)  \right ) \\
  \label{eq:barotropicE}
 Ef^{bt}_{\boldsymbol{k}} &=&-\frac{2}{\sqrt{2}}\epsilon^{1/2}   E_{bt}(\boldsymbol{k}) 
 \end{eqnarray}
\noindent  where
 \begin{eqnarray}
 E_{bc}(\boldsymbol{k},Z)= \langle \hat{\Psi}_{\boldsymbol{k}} \rangle \hat{\Psi}_{\boldsymbol{k}}^{'*}(Z) + c.c.
  \end{eqnarray}
Here the asterisk denotes the conjugate. We find that baroclinic Ekman friction is negligible compared to barotropic Ekman friction, i.e., $\vert Ef^{bc}_{\boldsymbol{k}}\vert \ll \vert  Ef^{bt}_{\boldsymbol{k}} \vert$.

From (\ref{eq:bxk}), one can consider the evolution of the 1D energy spectra, defined as
 \begin{eqnarray} \label{eq:1Denergy}
 E_{bt}(k) = \int_0^{2\pi} E_{bt}(\boldsymbol{k}) k d\phi_{\boldsymbol{k}}
  \end{eqnarray}
where, in the polar representation, $k = |\boldsymbol{k}|$ and $\tan{\phi_{\boldsymbol{k}}} = {k_y}/{k_x}$. 
Also of particular interest are the transfer maps
 \begin{eqnarray} \label{eq:tkp}
T_{kp} = \int k d \phi_{\boldsymbol{k}} \int p d \phi_{\boldsymbol{p}}  \sum_{\boldsymbol{q}} T_{{\boldsymbol{k p q}}} \\
 F_{kp} = \int k d \phi_{\boldsymbol{k}} \int p d \phi_{\boldsymbol{p}}  \sum_{\boldsymbol{q}} F_{{\boldsymbol{k p q}}}
  \end{eqnarray}
detailing the transfer of energy from all wavenumbers of magnitude $p$ to magnitude $k$. 

\section{Results}\label{sec:results}
In this section, we present our findings obtained from the suite of numerical simulations documented in Table~\ref{tab:TABLEOfValues}. These span a range of reduced Rayleigh numbers ${Ra}E^{4/3} \le 100$ and Prandtl numbers $1/2 \le Pr \le 7$.
All simulations of the CNH-QGM and the DNS employ a spectral Fourier discretization in the periodic horizontal directions,  a sparse Chebyshev-tau discretization in the vertical direction \citep{Watson}, and a mixed semi-implicit explicit (IMEX) timestepping scheme. The DNS code utilizes the second-order, IMEX Adams Bashforth/backward-differentiation scheme (AB2/BDF2)  \citep{sS08} and the CNH-QGM employs the third-order IMEX Runge-Kutta acheme (RK3) \citep{SMR91}. The aspect ratio for all simulations is selected to permit a sufficiently large sampling of linearly unstable convection modes to ensure convergence in the averaged statistics and to permit adequate room to capture the inverse energy cascade. 
The grid resolutions in all simulations are selected to adequately resolve dynamics on the most restrictive scales.  In the 
vertical direction this is controlled by the $\mathcal{O}(E^{1/3}H)$ thermal wind layer for the CNH-QGM and the  $\mathcal{O}(E^{1/2}H)$ Ekman layer in the presence of no-slip boundaries for the DNS. Specifically, the resolutions selected ensure a minimum of $8$ grid points in the $\mathcal{O}(E^{1/3}H)$ thermal wind layer and a minimum of  $8$ grid points in the Ekman boundary layer, which is calculated using the distance $z_E = (3/4) \pi \sqrt{2E}$ to the first maxima in the vertical spiral profile \citep{Stellmach}.  The resolution in the horizontal direction is controlled by the viscous dissipation, which is estimated by the Kolmogorov length scale 
\begin{equation} \label{eq:kol}
L_{kol}=(\nu^3/\varepsilon)^{1/4}.
\end{equation}
The energy dissipation rate $\varepsilon$ is estimated as the minimum of the horizontally averaged square of the viscous strain rate. The horizontal resolution ensures that the ratio of the smallest resolved length scale to the  Kolmogorov length scale is below $2$ (see column 5, Table~\ref{tab:TABLEOfValues}).  This criteria was adequate for convergence in the Nusselt number. Moreover, under this criteria, simulations with sufficiently large box sizes (i.e., greater than 10$L_c$ x 10$L_c$ x 1) resulted in the same $Nu$ values as those reported in the table.

The time resolutions utilized in the simulations of the CNH-QGM and the DNS are displayed in Table~\ref{tab:TABLEOfValues} (columns 8 and 9). Each is limited by the CFL constraint imposed by the explicit treatment of the nonlinear advection terms.  In the case of DNS, 
geostrophy and the separation of timescales between the mean and fluctuating temperature fields are not implicitly imposed, which, therefore, increases the degree of stiffness of any  IMEX time integration method. As is evident from the comparison, the asymptotic approach permits a temporal rescaling that automatically factors an  $\mathcal{O}(E^{-2/3} Pr)$ magnitude from the timestep (see column 7 of Table~\ref{tab:TABLEOfValues}). Hence, there is no explicit
Ekman number dependence on the time step and the increase in allowable non-dimensional time steps demonstrates a numerical advantage over DNS.  

Due to spatiotemporal constraints, DNS with $E=10^{-7}$ still remains the lower bound for investigations of rotating convection \citep{Stellmach}. Similar bounds are also present in laboratory experiments because of thermal and mechanical constraints  \citep{Ecke14,Cheng}. The CNH-QGM is capable of explorations at much lower $E$ as a result of the asymptotic design. Specifically, we note that $E=0$ represents the case of purely stress-free boundaries considered elsewhere \citep{Sprague,Julien12, kJ12b,Rubio}. In the following three subsections, the fidelity of the asymptotic model is explored via  qualitative and quantitative comparisons with DNS imposing no-slip boundaries at $E=10^{-7}$. In subsection~\ref{sec:LowEk}, results are reported from simulations with the Ekman number varying below $E=10^{-7}$.



\begin{figure}
	\centering
	 \subfloat[Cellular]{\includegraphics[width=0.23\textwidth]{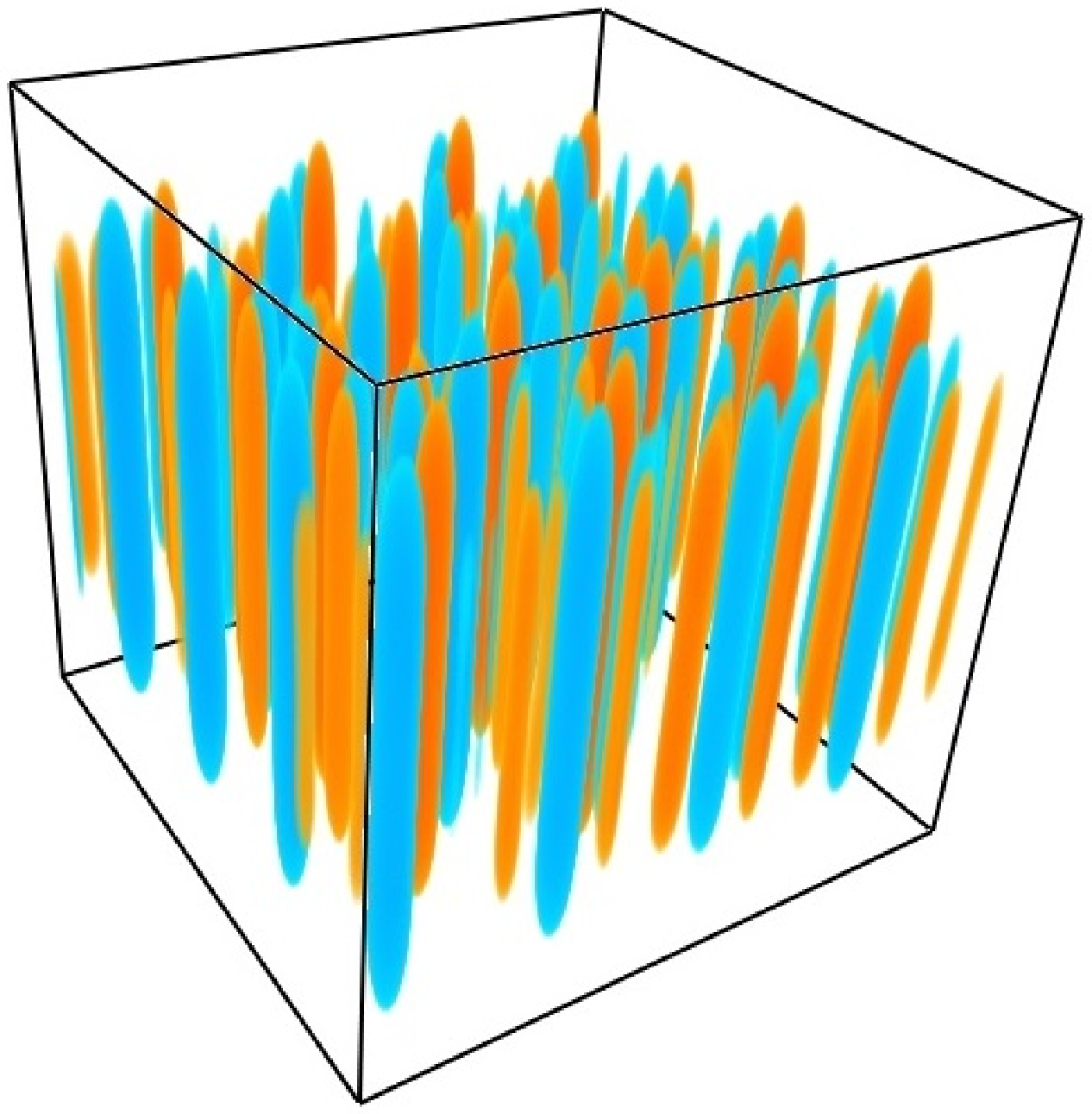}}
	 \subfloat[CTCs]{\includegraphics[width=0.24\textwidth]{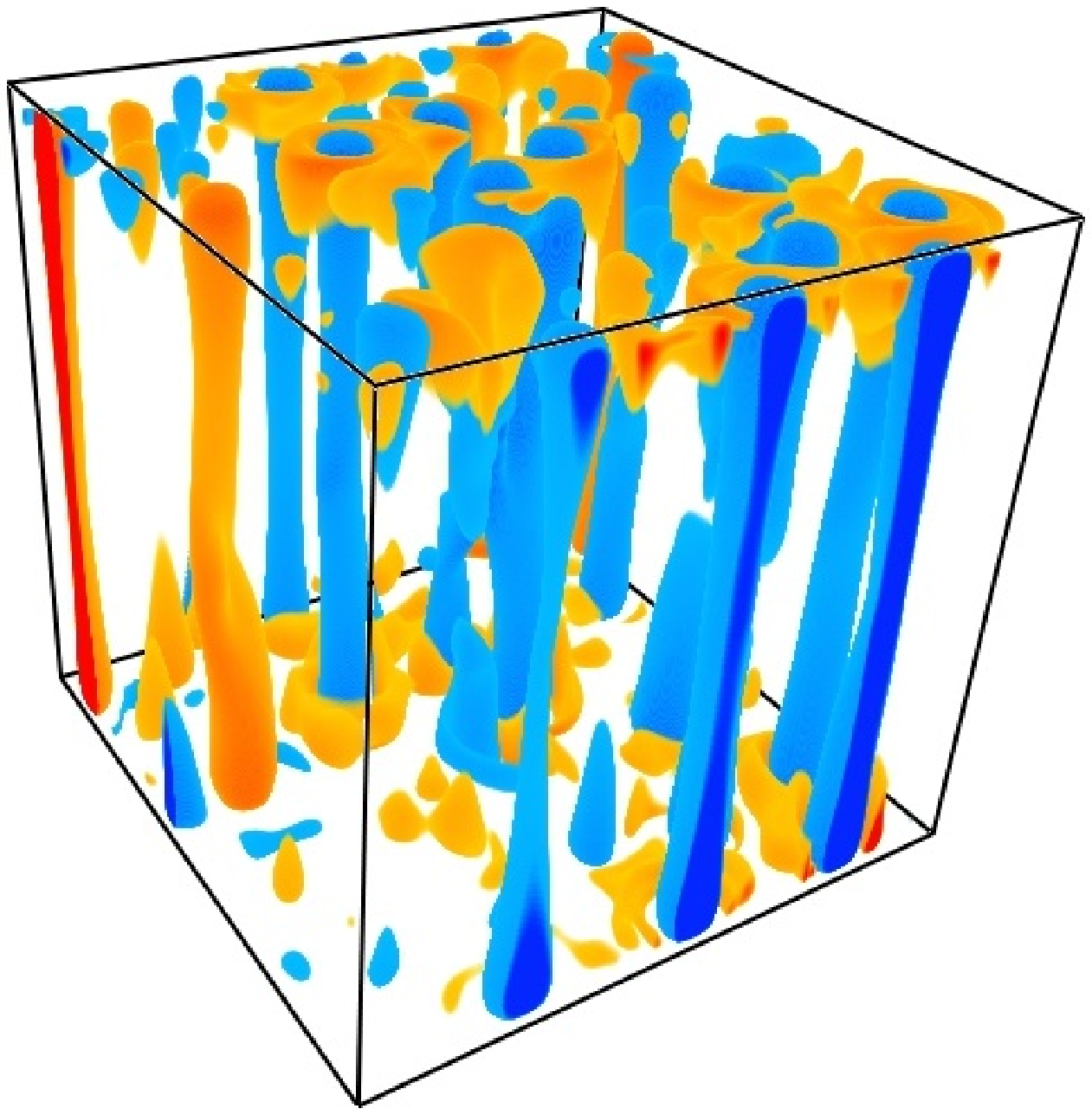}}
         \subfloat[Plumes]{\includegraphics[width=0.23\textwidth]{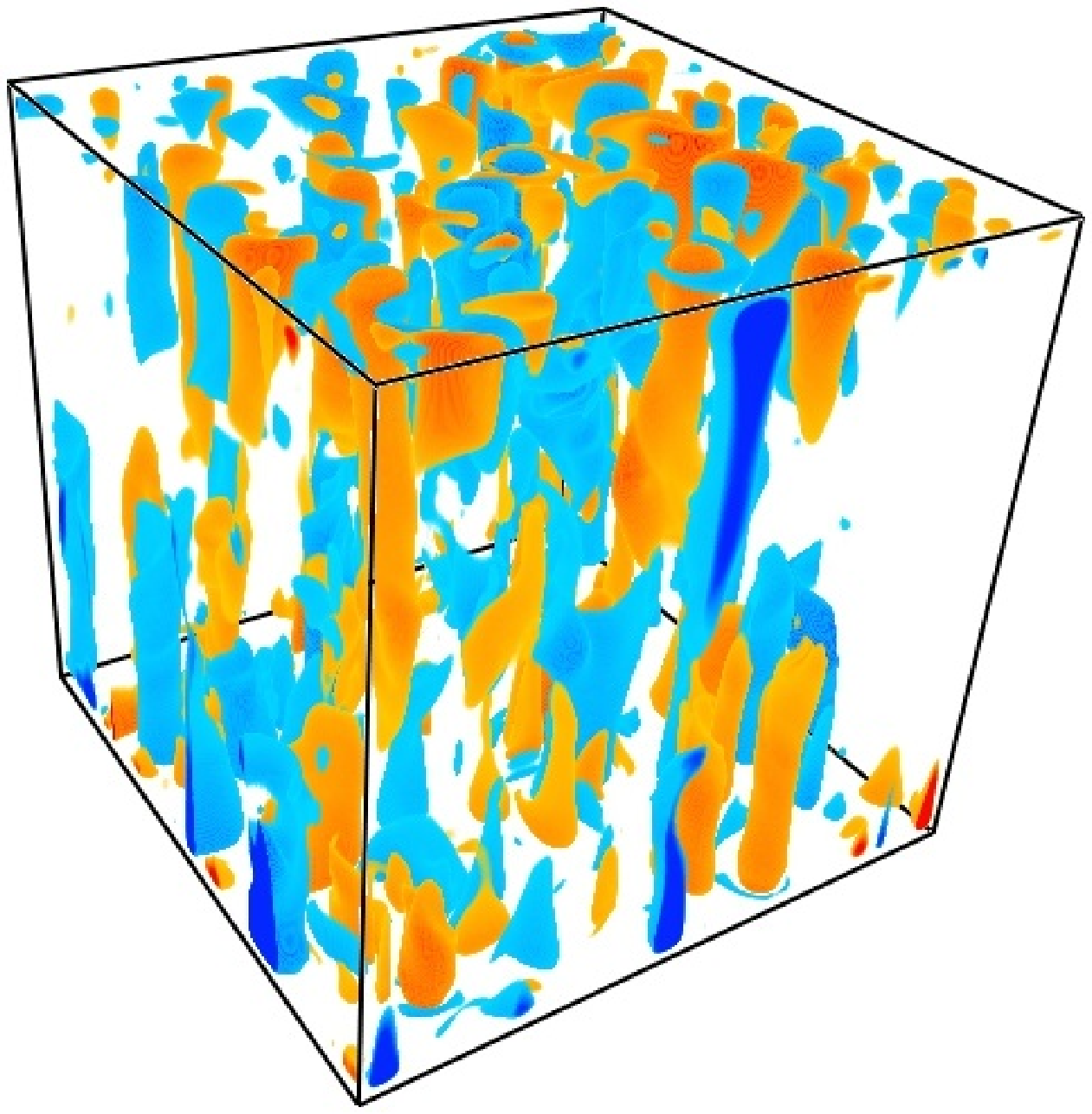}}
         \subfloat[GT]{\includegraphics[width=0.23\textwidth]{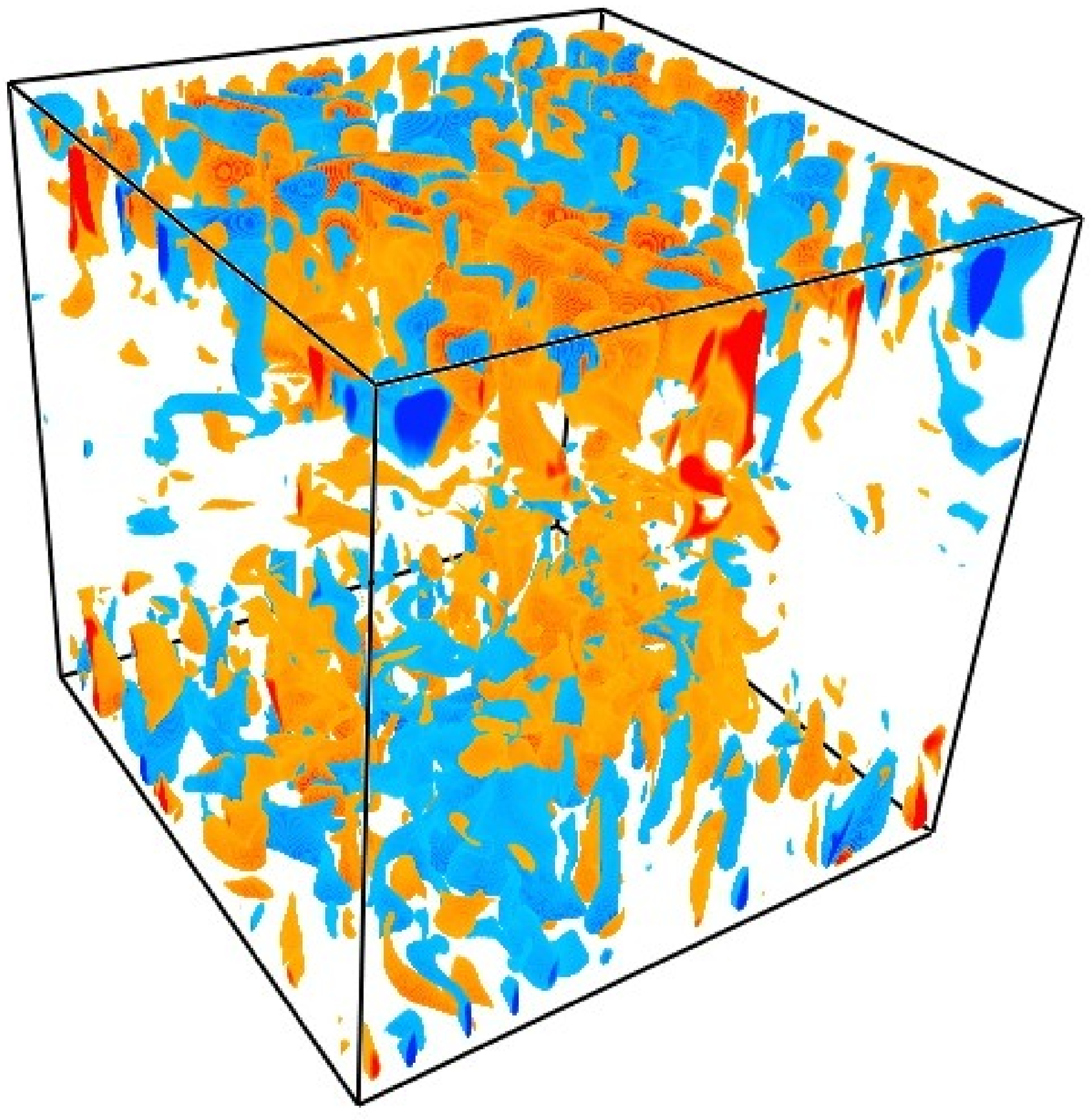}}
	\caption{\small{Perspective  
	view of the fluctuating temperature field $\theta$ illustrating the four geostrophically balanced flow morphologies obtained from simulations of the CNH-QGE with NS-NS boundaries at $E=10^{-7}$ (or $\epsilon=E^{1/3}=0.0046$).	
The plots show (a) cellular convection at $\widetilde{Ra} = 10, \Pran=7$; 
(b) convective Taylor columns at $\widetilde{Ra}=30, \Pran=7$;
(c) plumes at $\widetilde{Ra}=70, \Pran=7$; and 
(d) geostrophic turbulence  at $\widetilde{Ra}=90, \Pran=1$. These results should be compared with figure 2 in \citet{Stellmach}.}}
	\label{fig:Flows}
\vspace{-1ex}
\end{figure}
\subsection{Flow Morphologies}
The four  convective flow morphologies and their domains of existence as a function of $\widetilde{Ra}$ and $Pr$ has been well documented using the NH-QGM for RRBC in the presence of top and bottom stress-free (SF-SF) boundaries (c.f. figure 2 in \cite{Julien12}).  Figure \ref{fig:Flows} illustrates that simulations of the CNH-QGM using the parameterized no-slip (NS-NS) conditions yield the same four geostrophic regimes.  Ordered by increasing $ \widetilde{Ra}$ these include a cellular regime, a regime of shielded convective Taylor columns (CTCs), a plume regime and ultimately the geostrophic turbulence (GT) regime.  The DNS with spatially resolved Ekman boundary layers at similar values of $\widetilde{Ra}, \Pran$ and $E$ yield comparable visualization results (c.f. figure 2 in \cite{Stellmach}), thereby establishing a first marker for the fidelity of the CNH-QGM.

\begin{figure}
	\centering
		\includegraphics[width=0.5\textwidth]{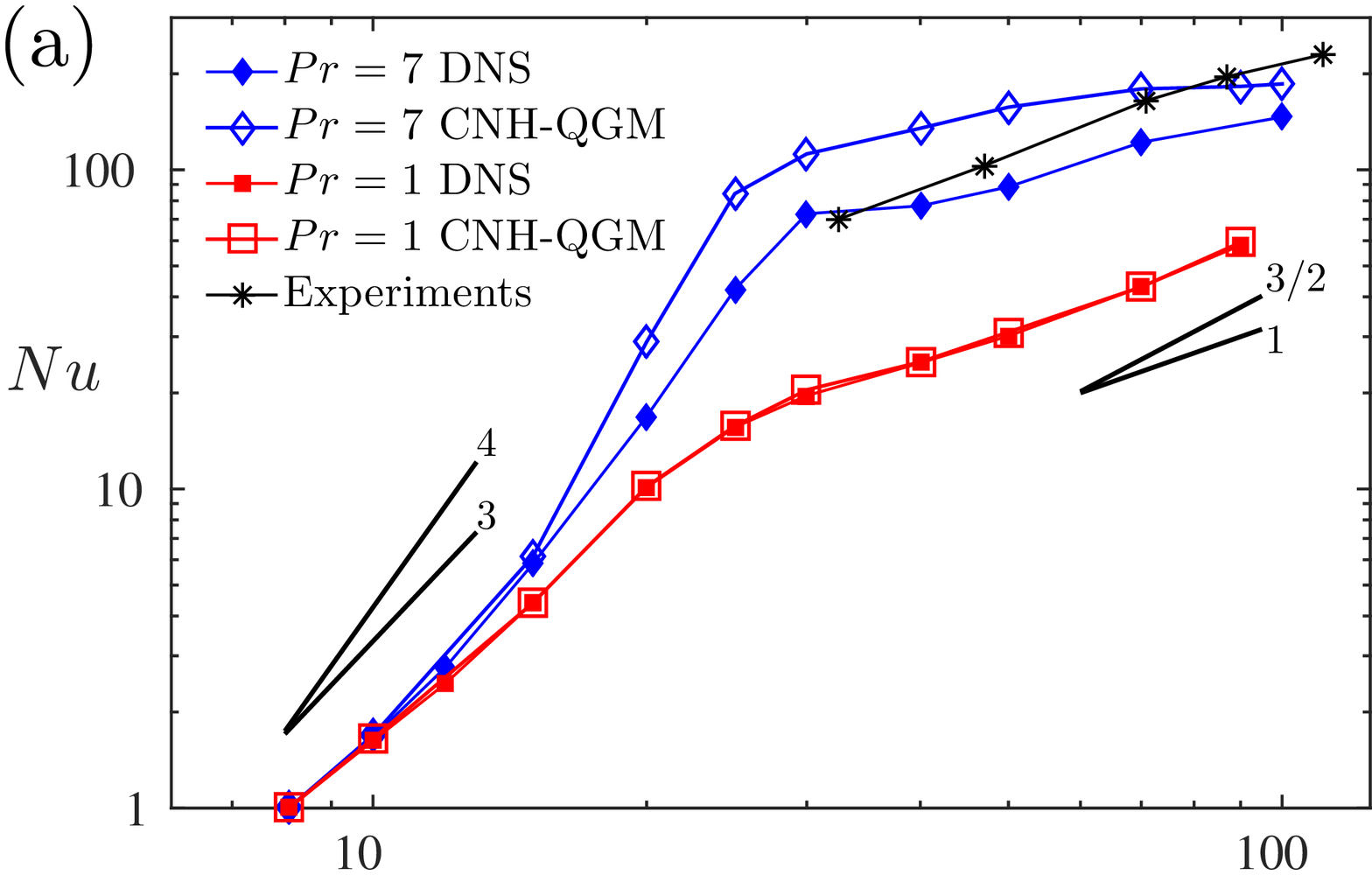} \hspace{-1em}
		\includegraphics[width=0.5\textwidth]{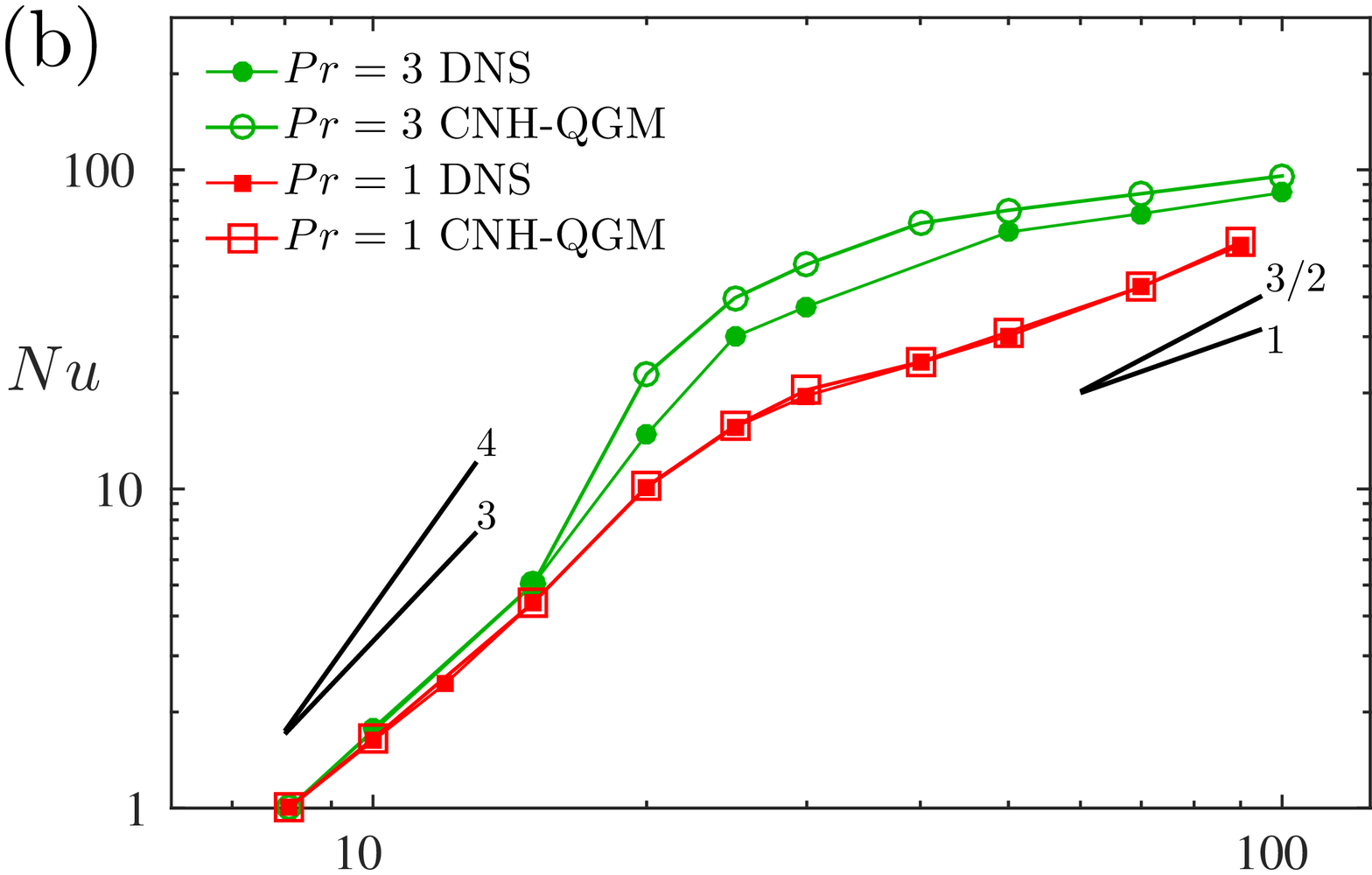} \\
		\includegraphics[width=0.5\textwidth]{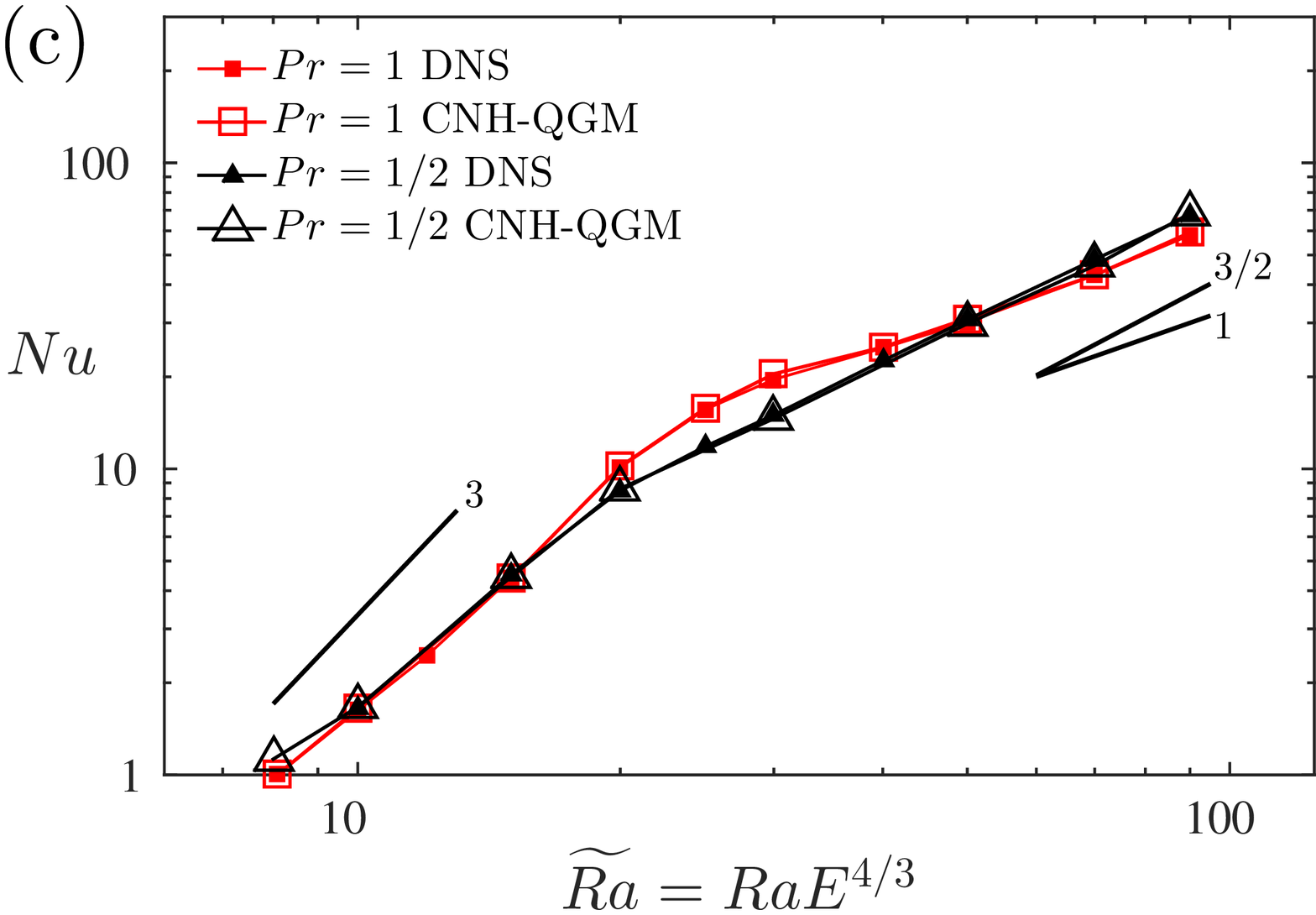} \hspace{-1em}
		\includegraphics[width=0.5\textwidth]{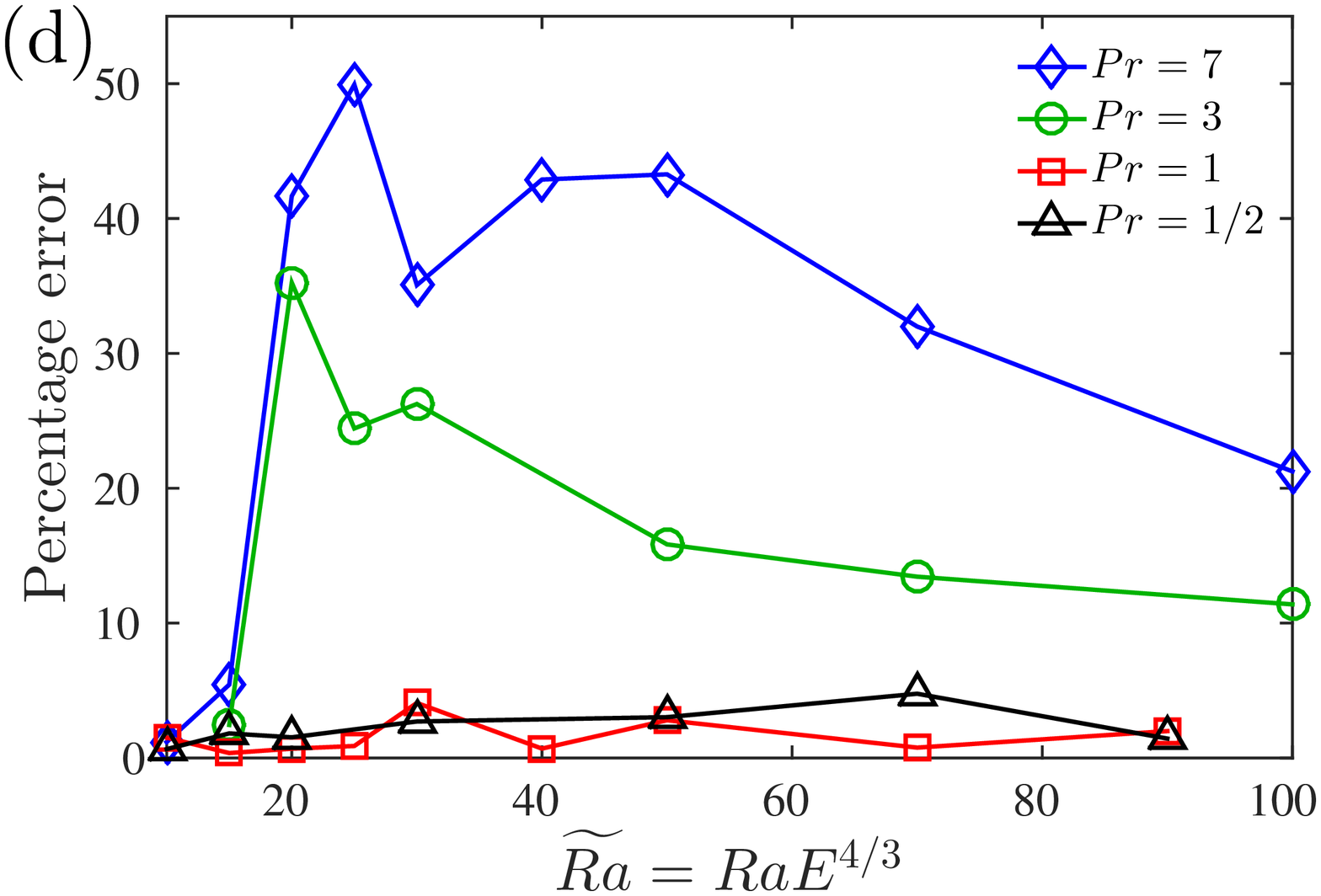}
	\caption{\small{Comparison of heat transport $Nu$  as a function of $\widetilde{Ra}$ between DNS (closed markers) and the CNH-QGM (open markers) with NS-NS  boundaries at $E=10^{-7}$.  
	The plots show the Nusselt number as a function of  (a) $\Pran =  7$ (diamonds), (b) $\Pran =  3$ (circles), (c) $\Pran = 1/2$ (triangles), and (a,b,c) $\Pran =  1$  (squares).  Data from experiments by \citet{Cheng} for $Pr \approx 7$ are marked by asterisks in (a). Note the increased level of heat transport compared with the stress-free results presented in figure \ref{fig:DNS_SFvsNHQGE}. Reference slopes are included as a guide. Plot (d) shows the relative percent errors of the Nusselt number (calculated as 100\% $\times$ ($|Nu_{\text{CNH}} - Nu_{\text{DNS}}|$/$Nu_{\text{CNH}}$)) for all $Pr$ values as a function of $\widetilde{Ra}$.}}
	\label{fig:Comparison}
\end{figure}

\subsection{Heat Transport}
A comparison of the $Nu$-$\widetilde{Ra}$ heat transport relations at $E=10^{-7}$ for DNS and CNH-QGM   is shown in figure \ref{fig:Comparison} for $Pr=7$ (plot a), $Pr=3$ (plot b) and $Pr=1/2$ (plot c). 
Comparisons with $Pr=1$ are included in each plot for reference. 
Confirming the findings of \citet{Stellmach} and \cite{Cheng}  (see  figure~\ref{fig:DNS_SFvsNHQGE}), Ekman pumping results in a significantly enhanced $Nu$ for all $\widetilde{Ra}$ compared to simulations with the SF-SF boundaries. Moreover, as with the stress-free simulations \citep{Julien12,Nieves}, transitions between regimes are observable in  figure \ref{fig:Comparison} by abrupt changes in the \textit{local} heat transport exponent $\beta=d \ln Nu/  \ln\widetilde{Ra}$ (reference slopes are provided to guide the eye). 

For $Pr > 1$, the heat transport law  (figure~\ref{fig:Comparison}a,b)  begins with   $\beta\approx 3$ for cellular convection, followed by a significant steepening of the exponent for the CTC regime, and finally by an abrupt slope reduction to $\beta\approx 1$. This result appears to be consistent with the findings of \cite{Julien15} using single-mode theory (see their figure 12).
To first order, it can be seen that the topology of the curves for the CNH-QGM agree qualitatively with those obtained for DNS for $Pr>1$. Transitional values from cells to CTCs and CTCs  to plumes occur at $\widetilde{Ra}\approx 15$ and $\widetilde{Ra}\approx 30$ respectively. By contrast the much delayed transitional values of $\widetilde{Ra}\approx 22$ and $\widetilde{Ra}\approx 55$ are obtained for the SF-SF case \citep{Nieves}. Explanation for this difference stems from the fact that CTCs form as consequence of a thermal boundary layer instability \citep{Julien12}, and the presence of Ekman pumping enables the threshold value for instability to be achieved at lower $\widetilde{Ra}$. A further inspection of figure \ref{fig:Comparison}a,b reveals that quantitative agreement between  CNH-QGM and DNS is only achieved in the cellular regime for $Pr>1$.  For larger $\widetilde{Ra}$ the asymptotic model yields larger values for $Nu$ in the CTC and plume regimes before showing signs of convergence again as the GT regime is approached (figure~\ref{fig:Comparison}d).
 Estimates for exploring the geostrophic regime indicate a more computationally expensive undertaking that, while possible, is not the focus of the present work. For instance, in the stress-free case, estimated values of $\widetilde{Ra}=220$ for $Pr=3$  and  $\widetilde{Ra}=290$ for $Pr=7$ are required to enter the geostrophic regime, simulations of which would demand resolutions beginning at 500 x 500 x 500.  \citep[page 4, column 2 of][]{kJ12b}. 

Comparisons with DNS reveal that coherent structures, such as  CTCs and plumes, have an increased vertical spatial coherence in the asymptotic model. These structures are stable conduits that result in an increased heat transport efficiency. 
We speculate that the difference at this lower bound threshold value of $E=10^{-7}$, which is only just entering the geostrophically balanced regime,  is due to subdominant ageostrophic dynamics present in the DNS but absent from the asymptotic model.   Subdominant ageostrophic effects include nonlinear vertical and horizontal advection of momenta and heat. These terms destabilize the degree of vertical coherence of CTCs and permit an earlier transition to the geostrophic turbulence regime. Such effects are more prominent in the boundary layer region where velocity and thermal amplitudes are greatest.  These findings are consistent with recent reports in the literature that 
indicate  $E=10^{-7}$  to be a threshold value 
for geostrophically balanced flows \citep{Ecke14,Cheng}.
Advancement will require reducing $E$ in both laboratory experiments and DNS.

\begin{figure}
	\centering
		\includegraphics[width=0.33\textwidth]{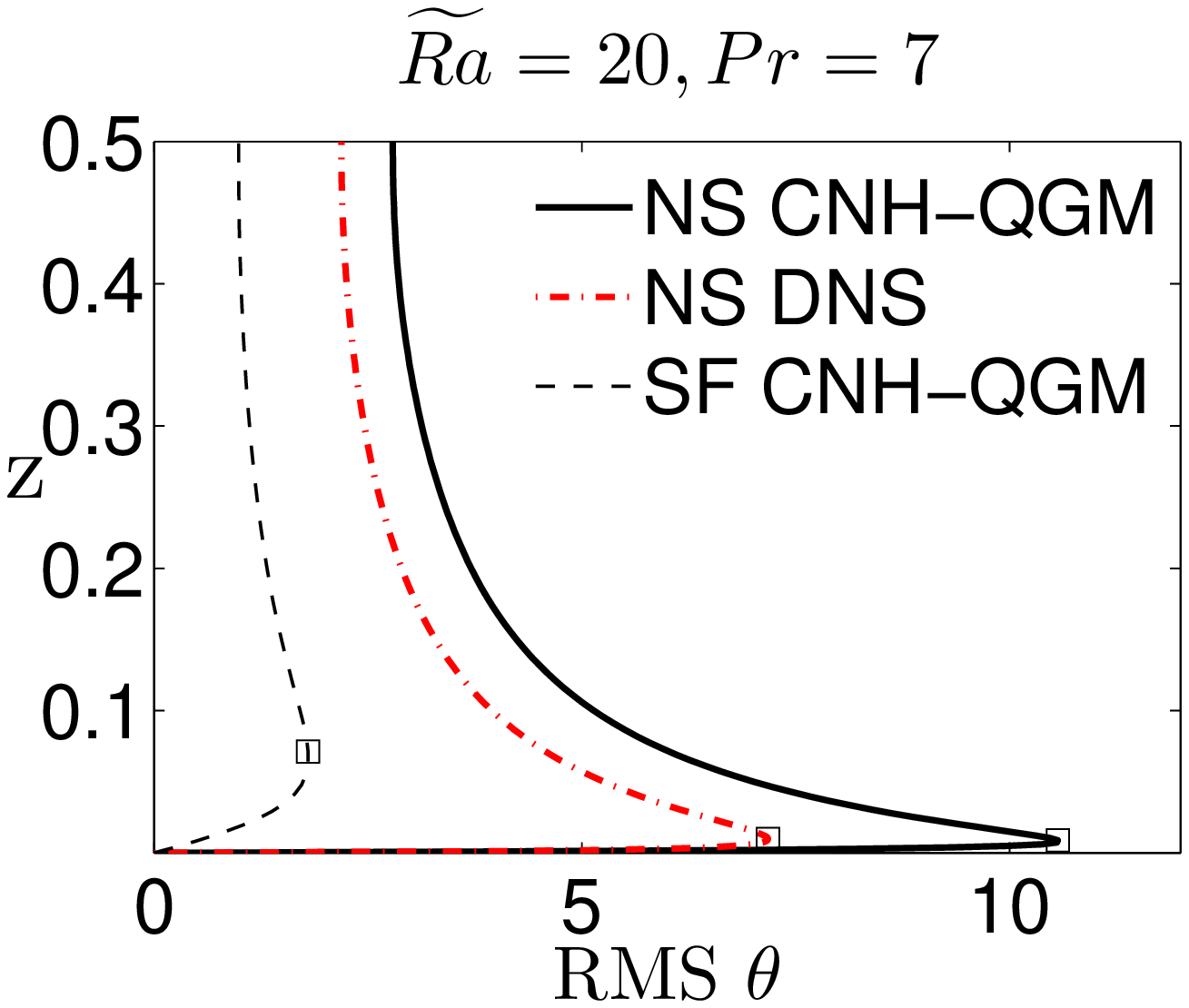} \hspace{-.5em}
		\includegraphics[width=0.33\textwidth]{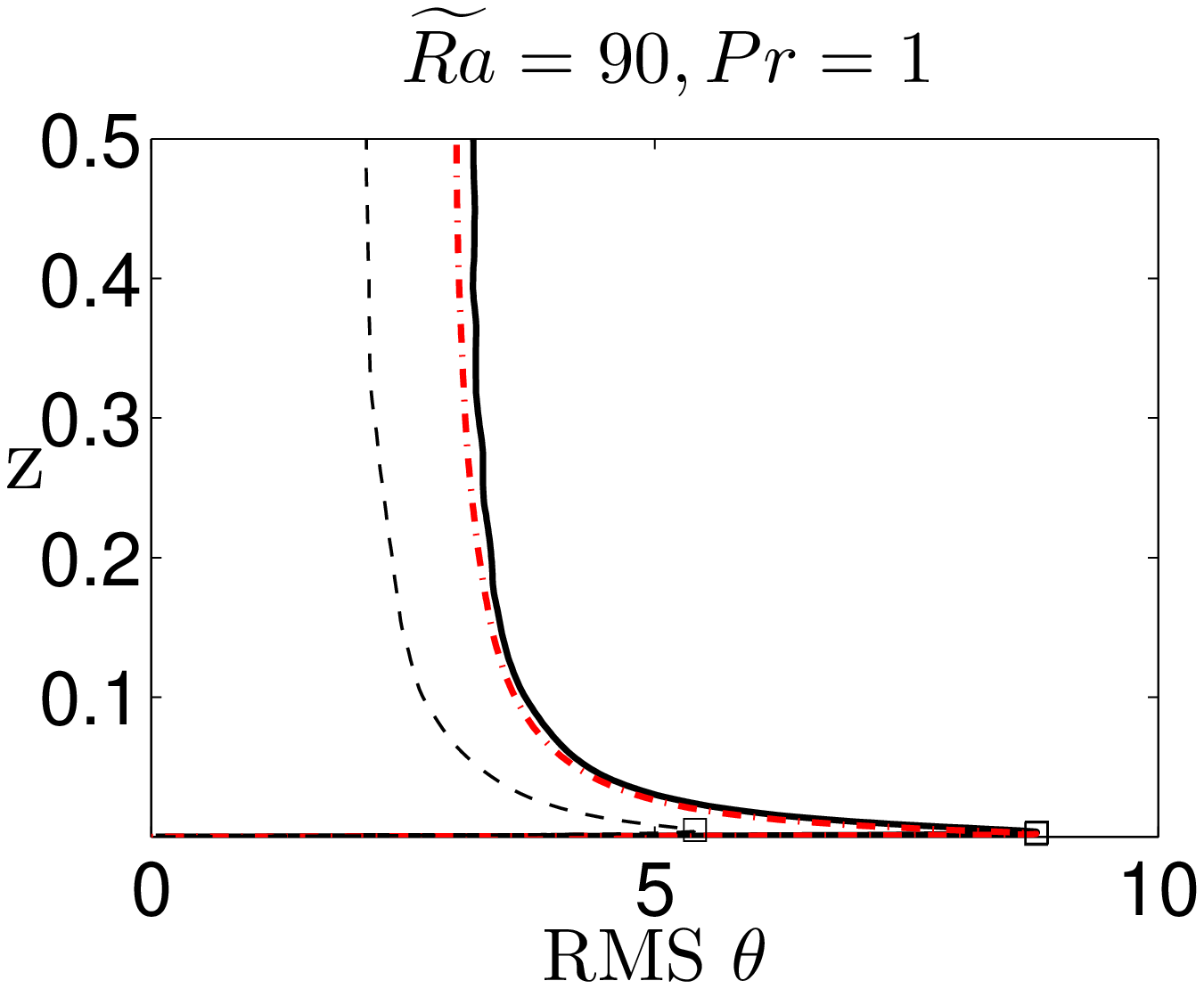}
		\hspace{-.75em}
		\includegraphics[width=0.33\textwidth]{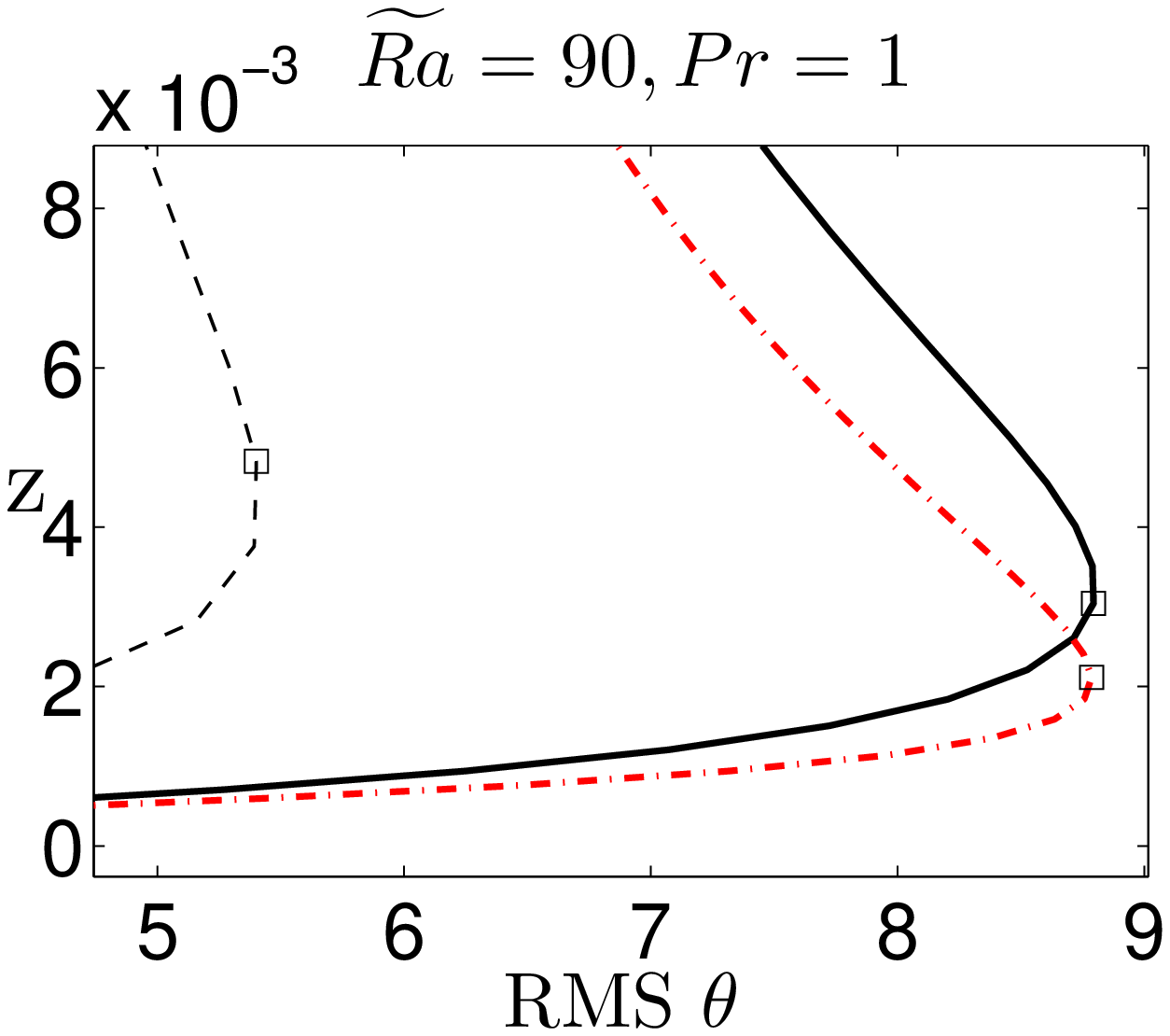}
%
\caption{\small{RMS fluctuating temperature $\theta$ profiles at $E=10^{-7}$ for (a)  CTCs at  $\widetilde{Ra}=20$, $\Pran=7$,  (b) GT at 
$\widetilde{Ra}=90$, $\Pran=1$ and (c) enlargement of the GT boundary layer. Results of CNH-QGM SF-SF (gray, dashed), CNH-QGM NS-NS (black, solid), and DNS NS-NS (red, dot-dash) cases are displayed. The squares denote the locations of the max RMS. }} \label{fig:rms_theta}
\end{figure}

Figure~\ref{fig:rms_theta}a illustrates vertical RMS profiles for the fluctuating temperature fields in the CTC regime at $\widetilde{Ra}=20$, $\Pran=7$. The enhanced thermal fluctuations within the thermal wind layer are evident, with the CNH-QGM overestimating the magnitude of the DNS data, but good topological agreement is found. As a notable reminder, consistent with this observation, both the cellular and GT regimes are dominated by interior dynamics; no thermal boundary exists in the cellular regime, while, the bulk in the GT regime has been shown to be the thermal bottleneck for heat transport and thus insensitive to the properties of the more thermally efficient boundary layers \citep{kJ12b,Julien12}. Hence, the observation of increased quantitative agreement in the cellular and GT regimes implies good agreement of bulk dynamics.


It is well-known that for $Pr\lesssim 0.7$ linear onset is oscillatory. Fully nonlinear oscillatory regimes known to be prevalent at $Pr\ll1$ \citep[e.g.,][]{King13} are not the focus of the present study, and we consider only the range $1/2 \le Pr \le 1$.  For $Pr\le 1$, we find that the cellular convection regime rapidly gives way to geostrophic turbulence. This regime is characterized by the dissipation free scaling exponent $\beta\approx 3/2$ \citep{kJ12b}. Excellent quantitative agreement is achieved between the CNH-QGM and DNS for heat transport (figure~\ref{fig:Comparison}c) and flow morphology (figure~\ref{fig:rms_theta}b,c). Specifically, we observe  Nusselt numbers of similar magnitudes in the geostrophic regime. The heat transport results thus establish a second marker for the fidelity of the CNH-QGM.

%

\subsection{Non-local inverse cascade}
\begin{figure}
	\centering
		\includegraphics[width=0.24\textwidth]{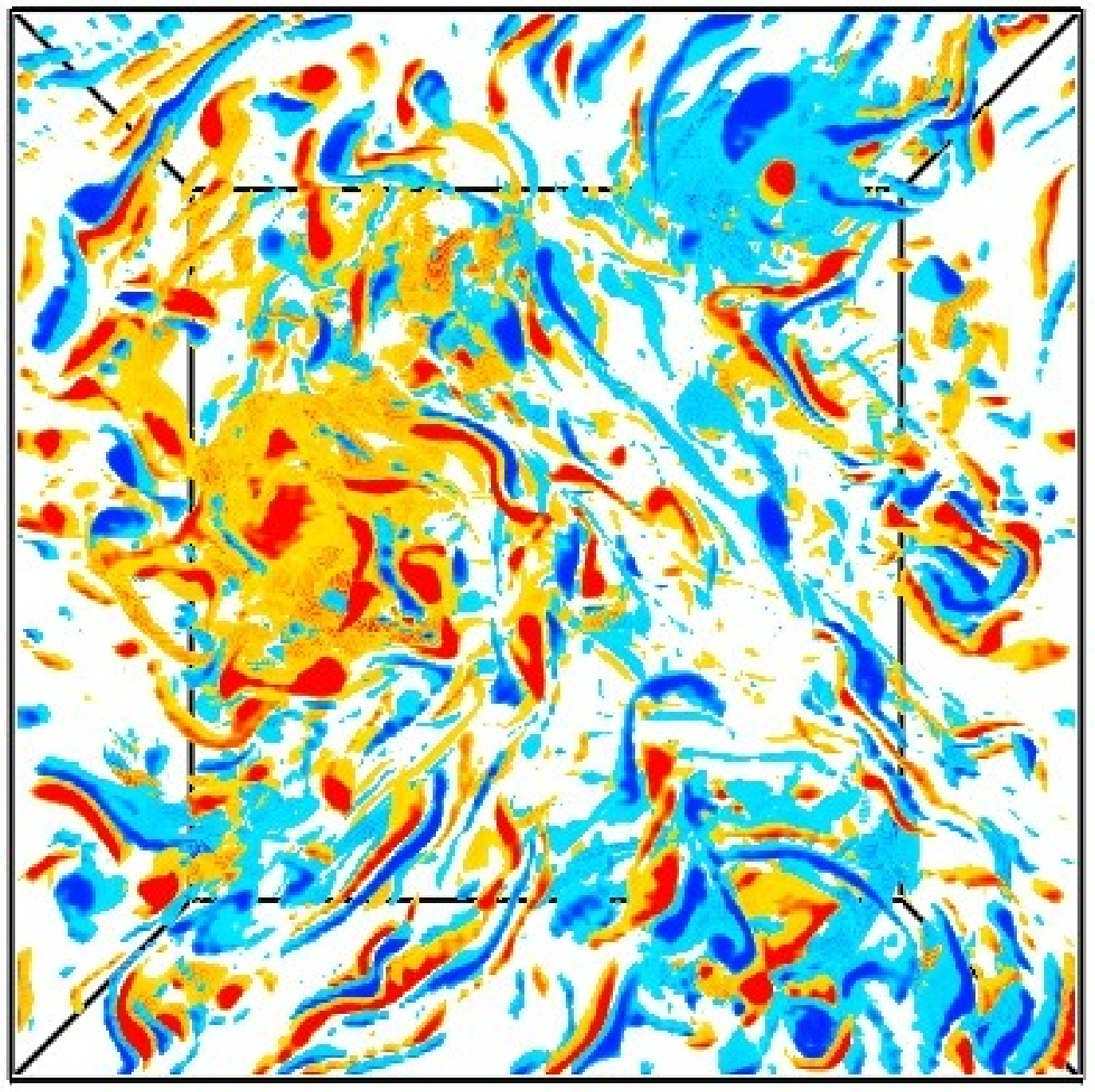} \hspace{3.2em} 
		\includegraphics[width=0.24\textwidth]{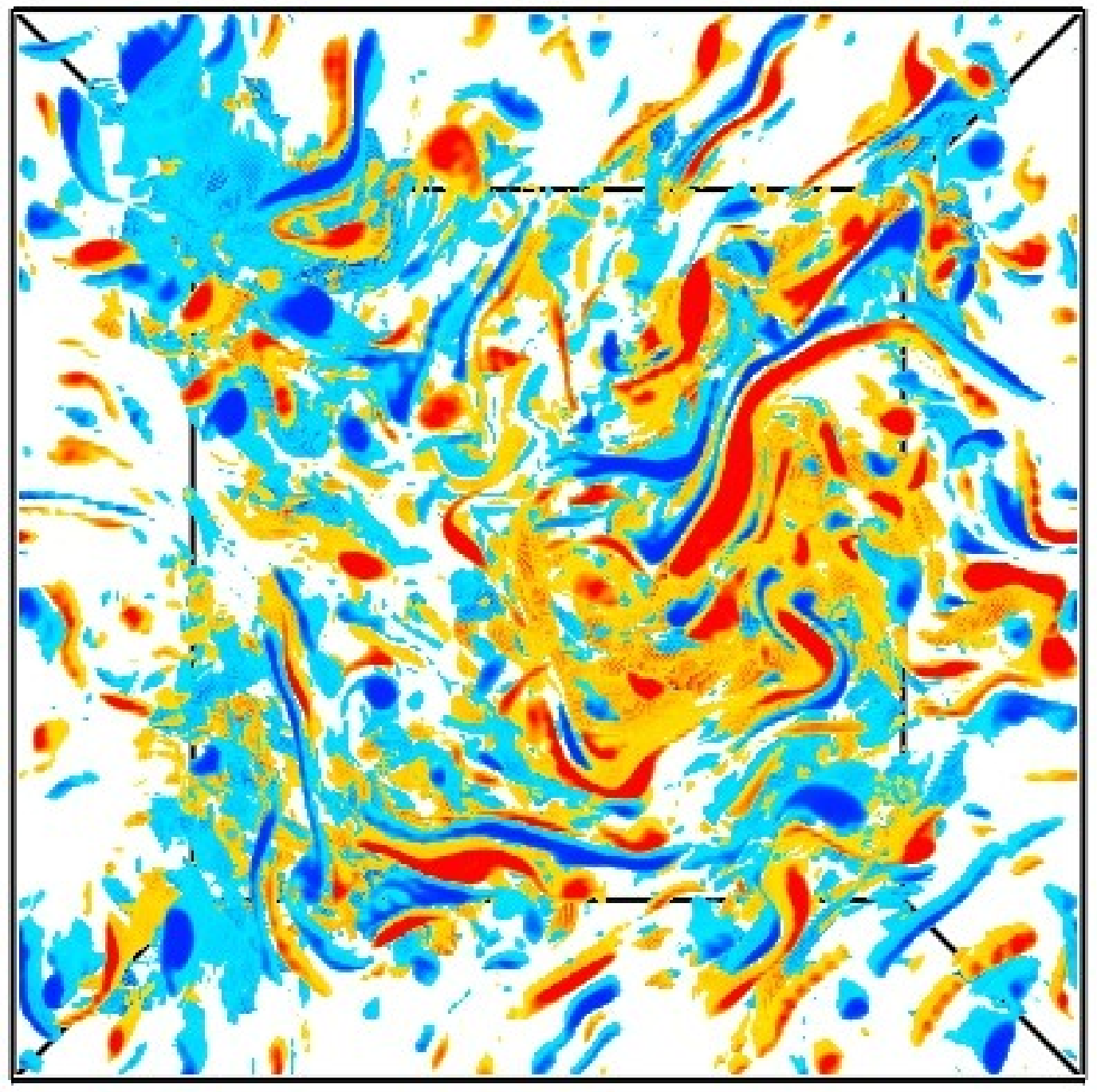} \hspace{3.2em} 
		\includegraphics[width=0.24\textwidth]{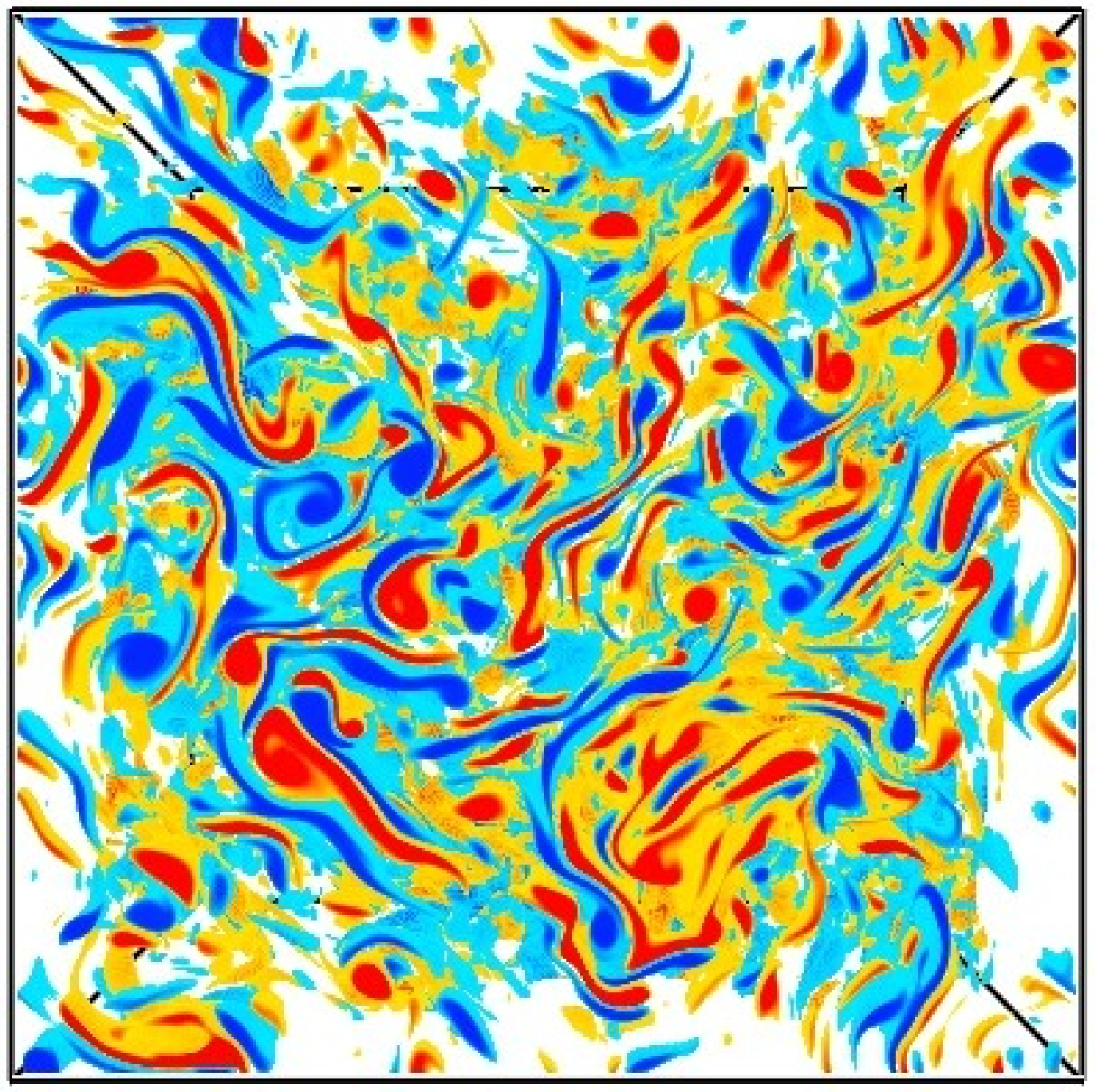} 
		\\
		\subfloat[SF-SF]{\includegraphics[width=0.24\textwidth]{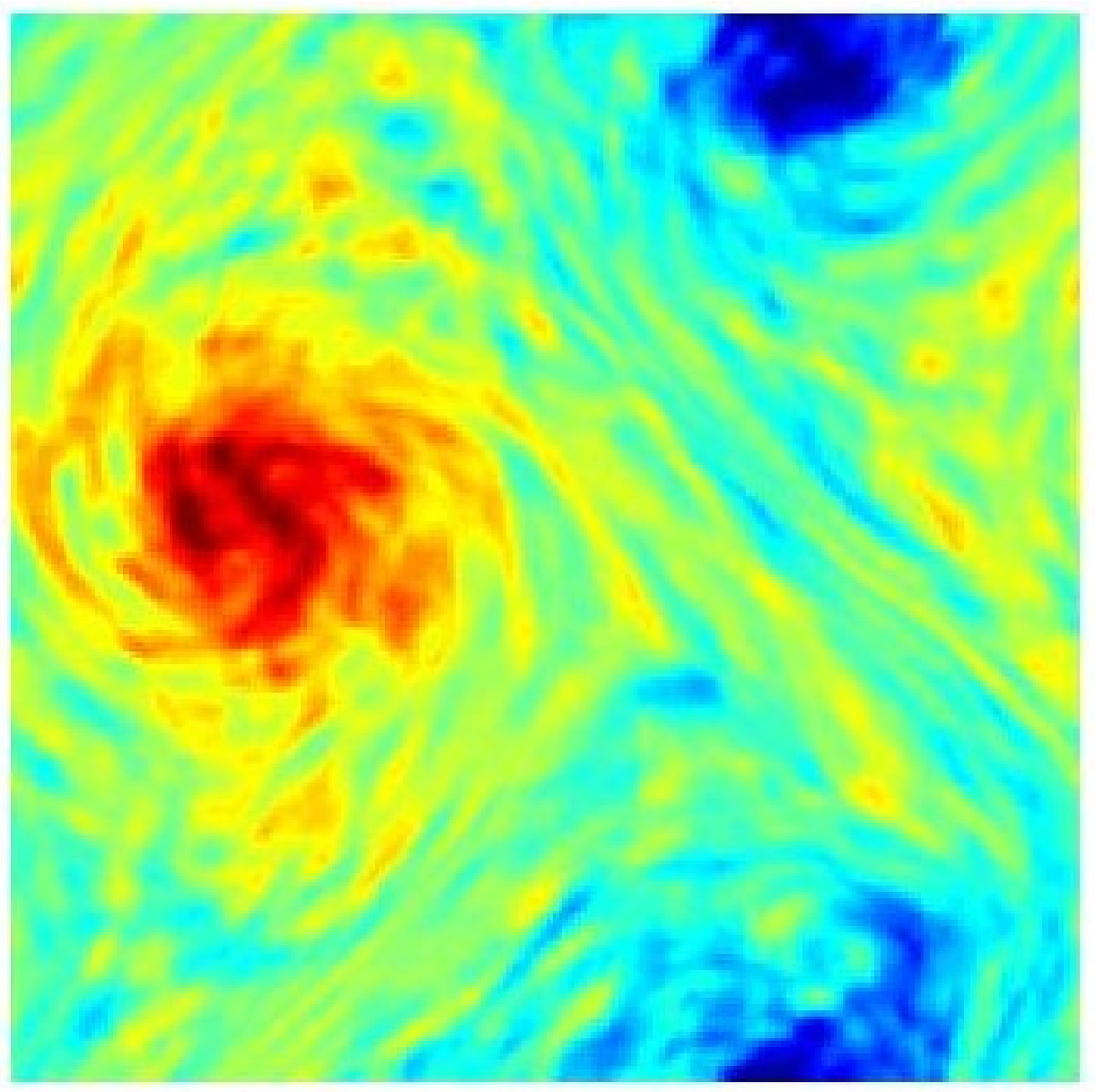}} \hspace{3.2em} 
		\subfloat[NS-SF]{\includegraphics[width=0.24\textwidth]{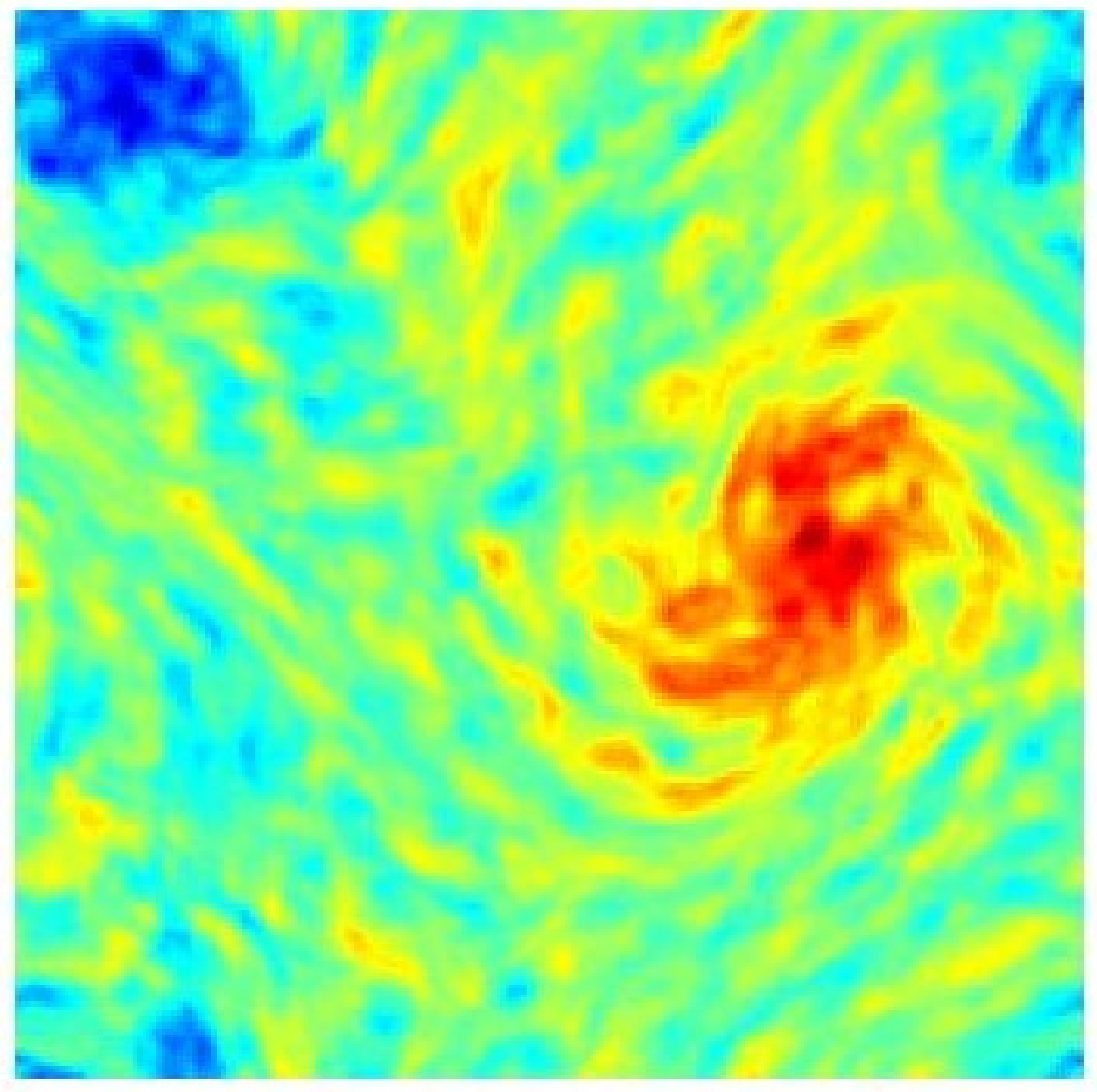}} \hspace{3.2em} 
		\subfloat[NS-NS]{\includegraphics[width=0.24\textwidth]{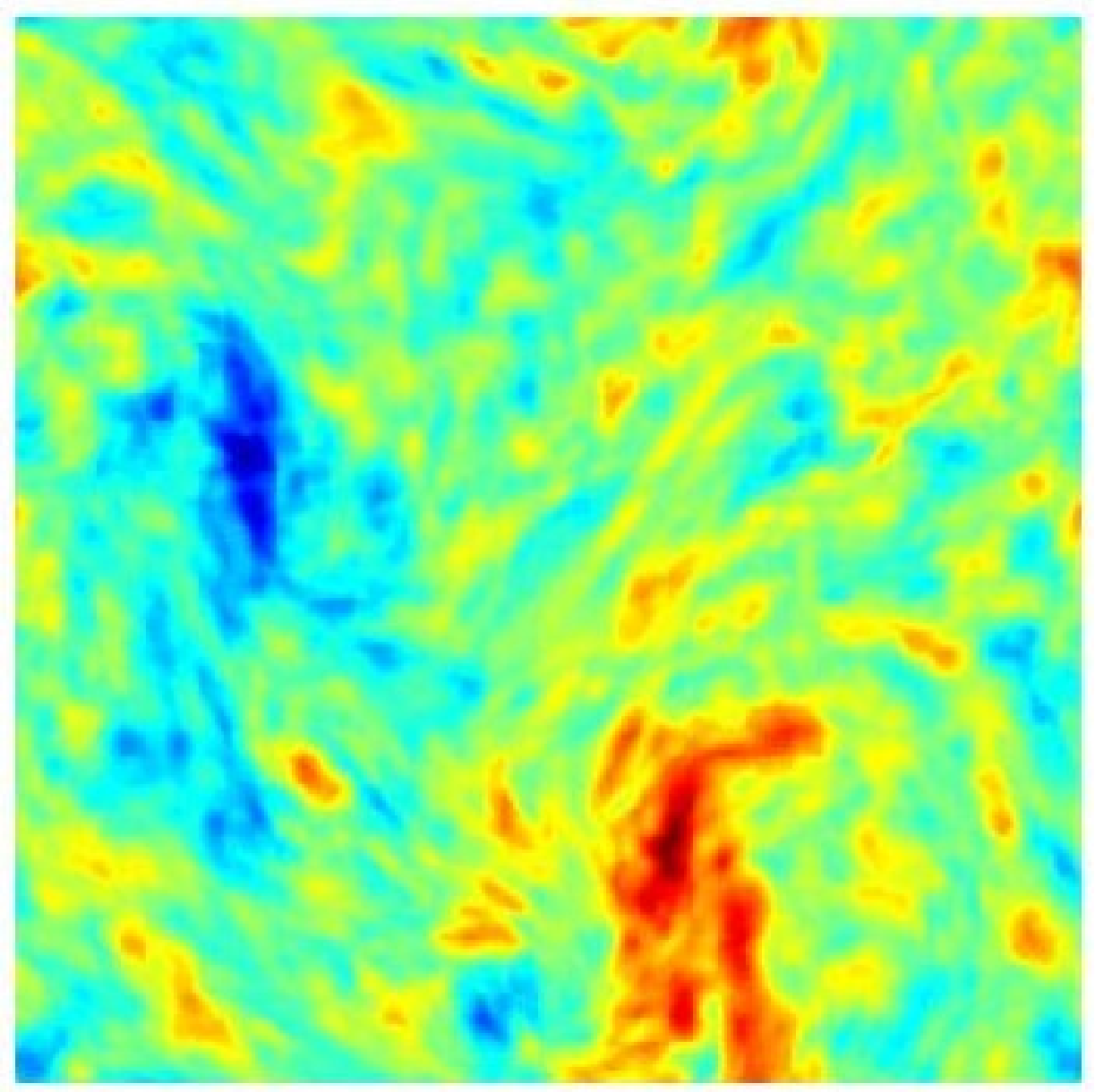}} 
	\caption{\small{Top view visualizations of the total vertical vorticity $\zeta$ in the GT regime at $\widetilde{Ra}=90$, $\Pran=1$
	for CNH-QGM (a) SF-SF $(E=0)$, (b) mixed NS-SF  $(E=10^{-7})$, and (c) NS-NS $(E=10^{-7})$ boundaries. The lowest row displays the barotropic (depth-averaged) vorticity $\langle \zeta \rangle$ illustrating the persistence of large scale cyclonic and anti-cyclonic vortices in the stress-free and mixed cases.  Large scale structures form in the presence of no-slip boundaries, but they are transient and less coherent. Color in the images is scaled between 60 (red) and -60 (blue).}}
	\label{fig:barotropic}
\end{figure}

\begin{figure}
	\centering
	\includegraphics[width=0.7\textwidth]{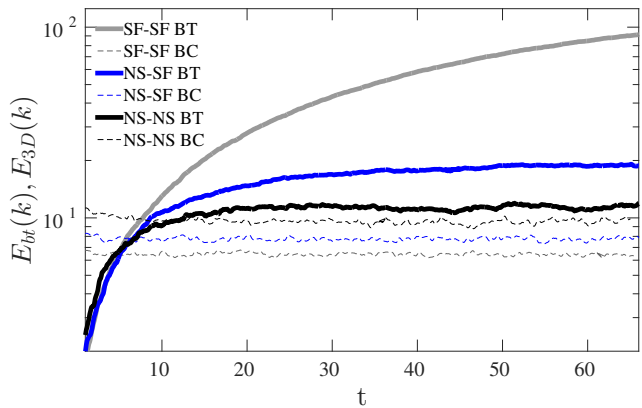}
	\vspace{-2ex}
	\caption{\small{Growth and saturation of the barotropic kinetic energy $E_{bt}(k)$ using the CNH-QGM in the presence of Ekman friction for mixed NS-SF (solid blue) and  NS-NS (solid black) boundaries at $E = 10^{-7}$ for $\widetilde{Ra} = 90$ and $Pr = 1$. The barotropic energy for the SF-SF, $E=0$ case ( solid gray) saturates on a much longer time scale \citep{Rubio}. The total baroclinic kinetic energy $E_{3D}(k)$ (dashed lines) saturates in all cases.}}
	\label{fig:bargrowth}
	\vspace{-2ex}
\end{figure}

\begin{figure} 
	\centering
		\includegraphics[width=0.23\textwidth]{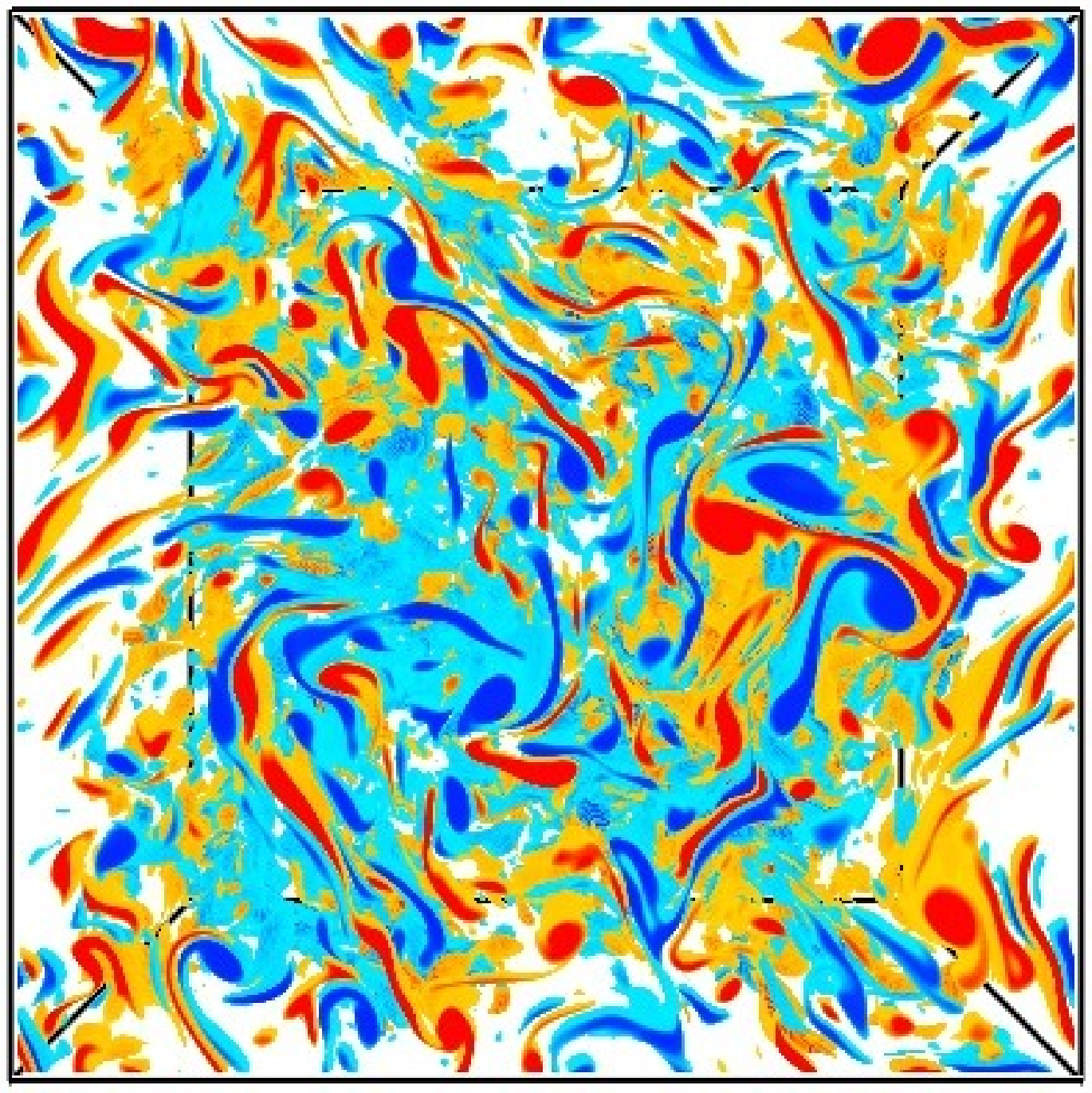} \hspace{.5em} 
		\includegraphics[width=0.23\textwidth]{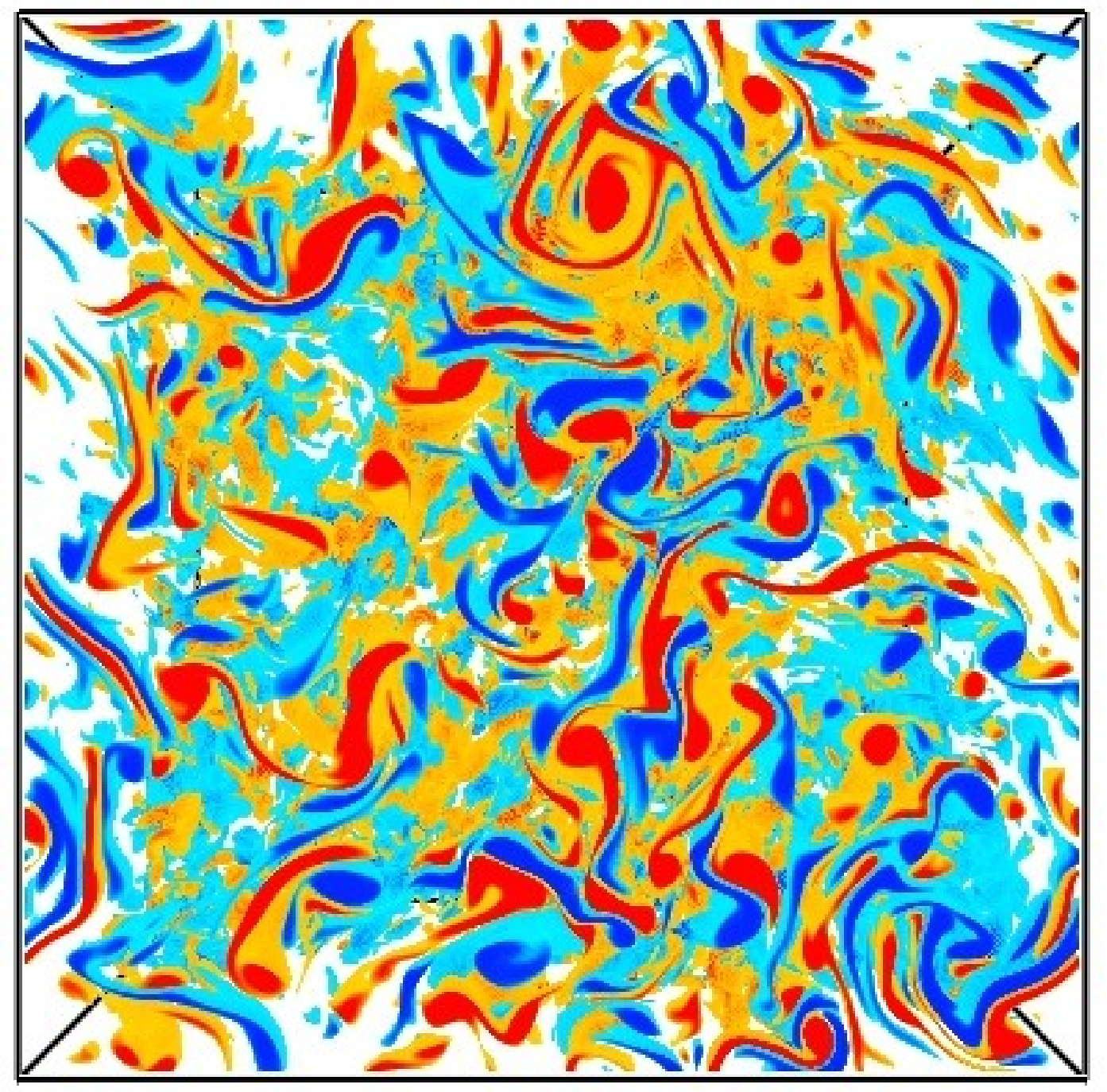} \hspace{.5em} 
		\includegraphics[width=0.23\textwidth]{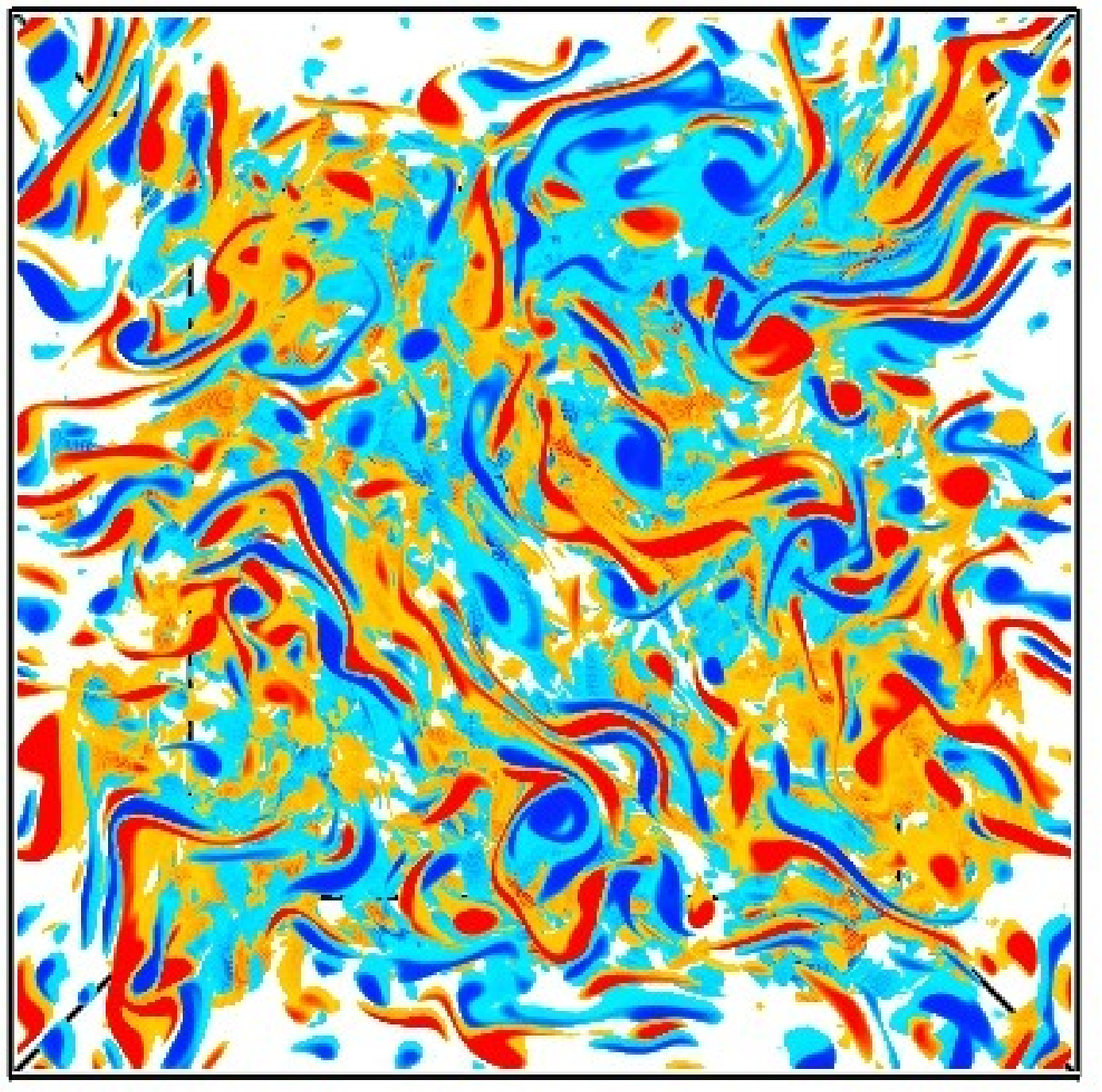} \hspace{.5em} 
		\includegraphics[width=0.23\textwidth]{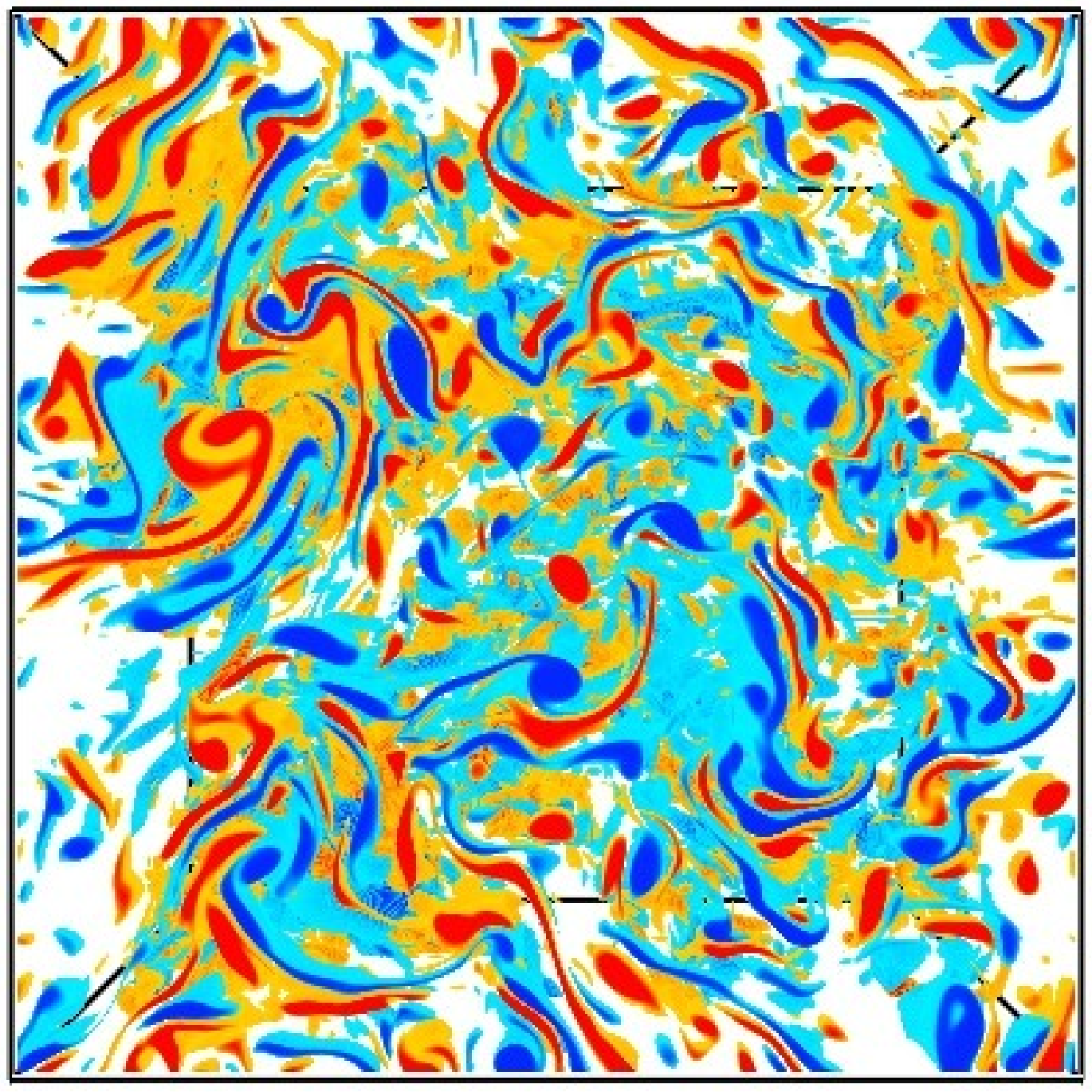} 
				\\
				\subfloat[NS-NS at t $\approx$ 46]{\includegraphics[width=0.23\textwidth]{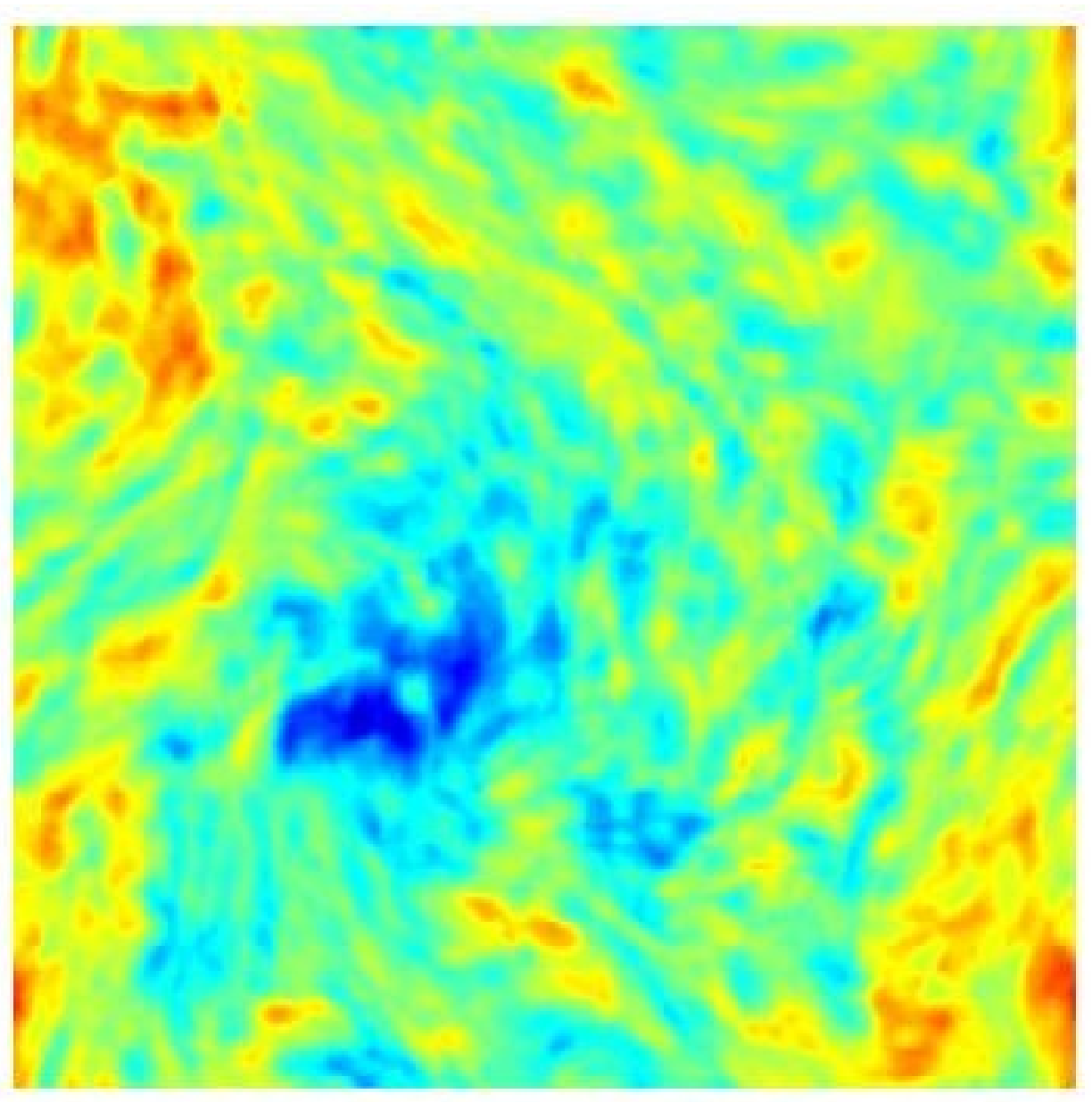}} \hspace{.75em} 
				\subfloat[t $\approx$ 50]{\includegraphics[width=0.23\textwidth]{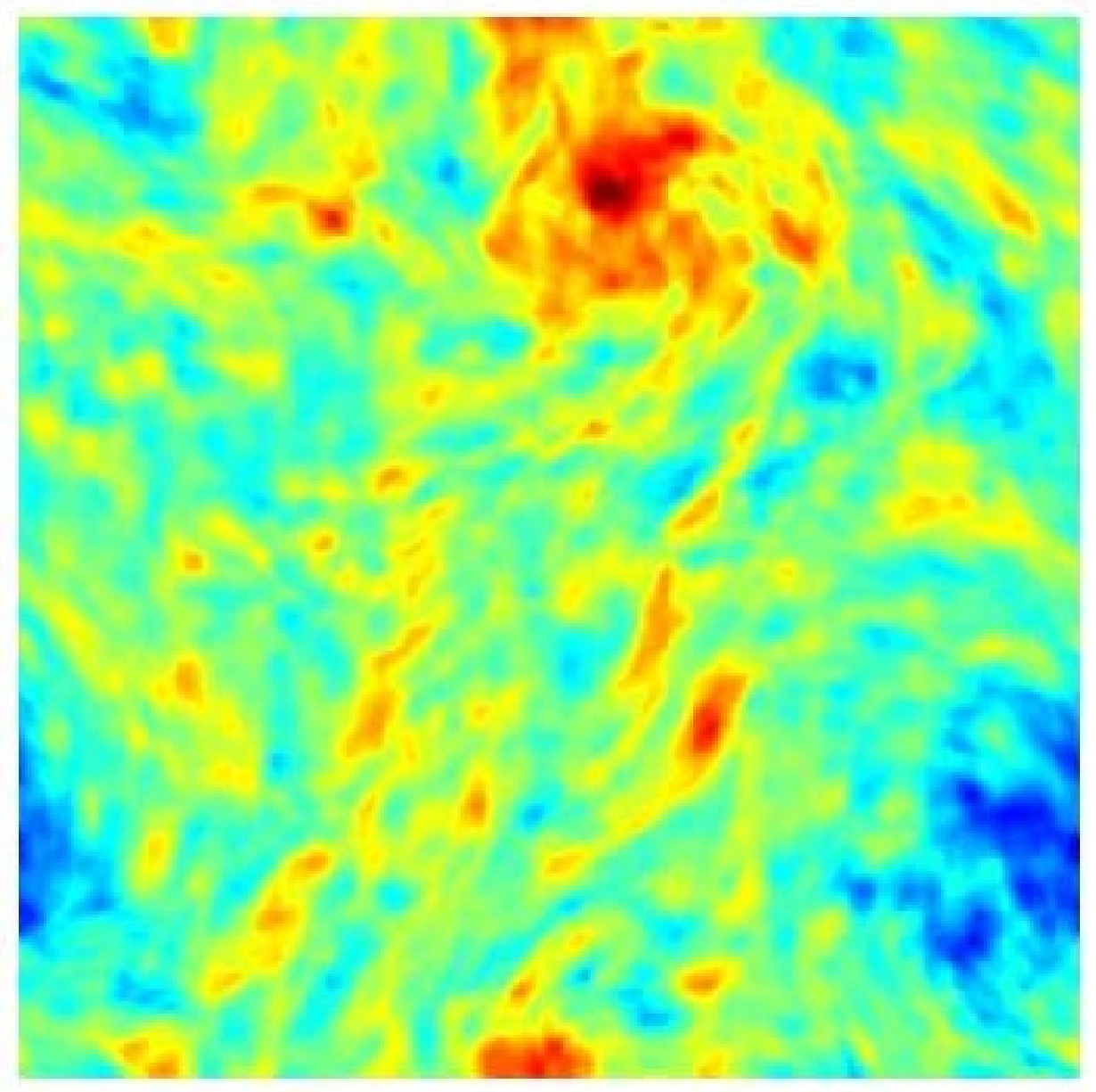}} \hspace{.5em} 
				\subfloat[t $\approx$ 54]{\includegraphics[width=0.23\textwidth]{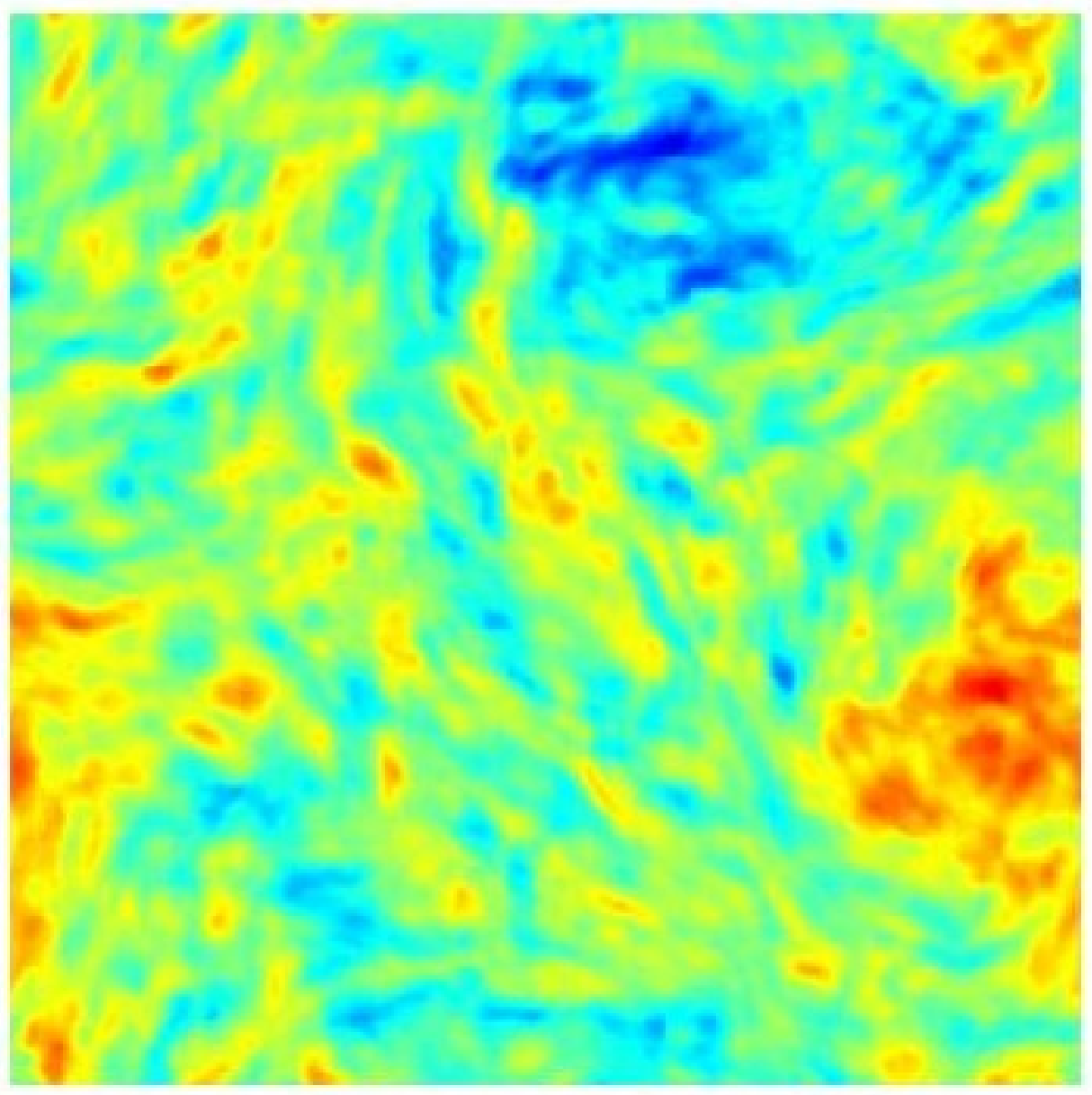}} \hspace{.75em} 
				\subfloat[t $\approx$ 58 $L^2/\nu$]{\includegraphics[width=0.23\textwidth]{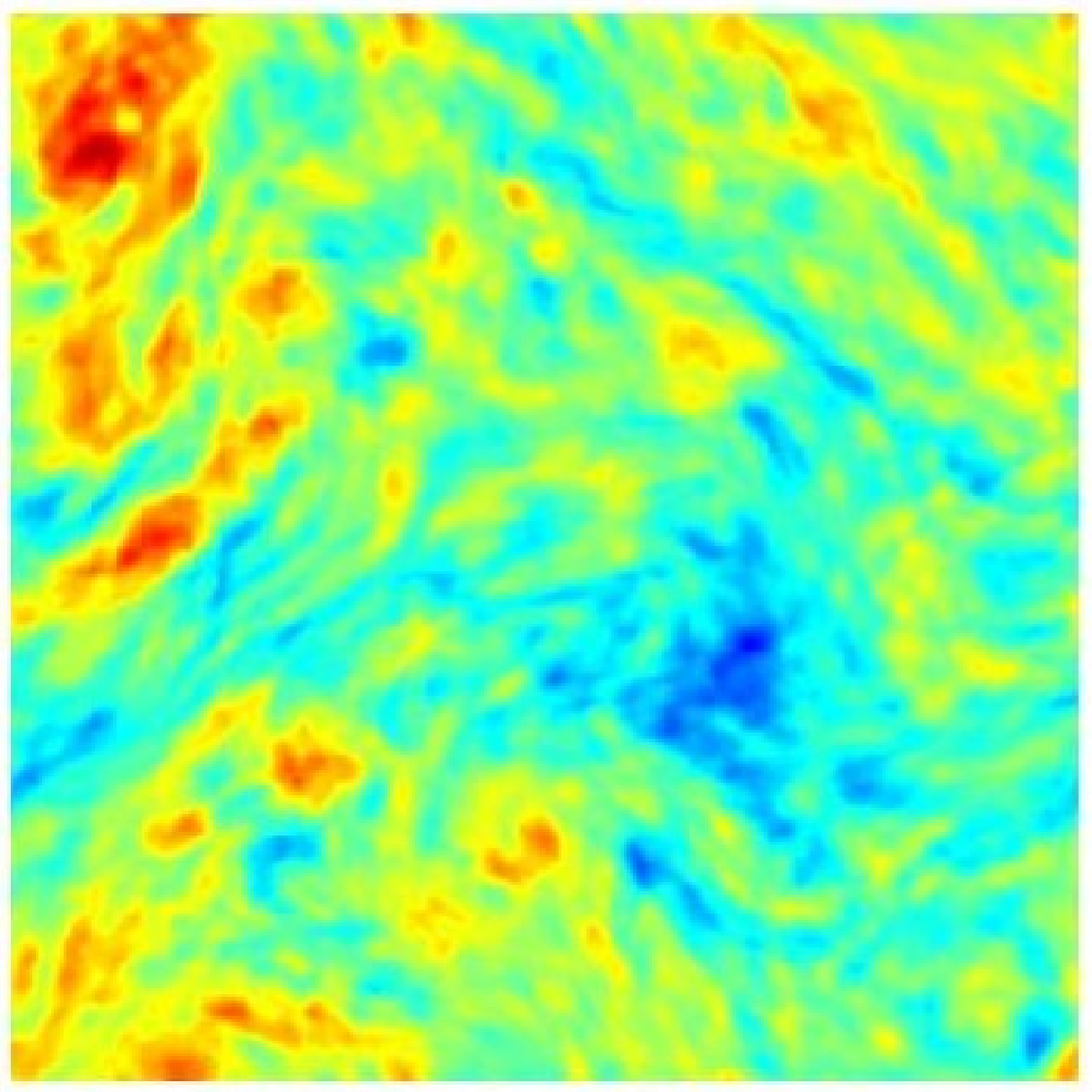}} 
	\caption{\small{Time series from the CNH-QGM of the total vertical vorticity $\zeta$ in the GT regime at $\widetilde{Ra}=90$, $\Pran=1$ with NS-NS boundaries at $E=10^{-7}$. The bottom row shows the barotropic (depth-averaged) vorticity $\langle \zeta \rangle$ scaled from -60 (blue) to 60 (red). The structures that form for two no-slip boundaries vary with time in both strength and coherence.}}
	\label{fig:time_series}
\end{figure}

%

\begin{figure}
	\centering
	\includegraphics[width=0.5\textwidth]{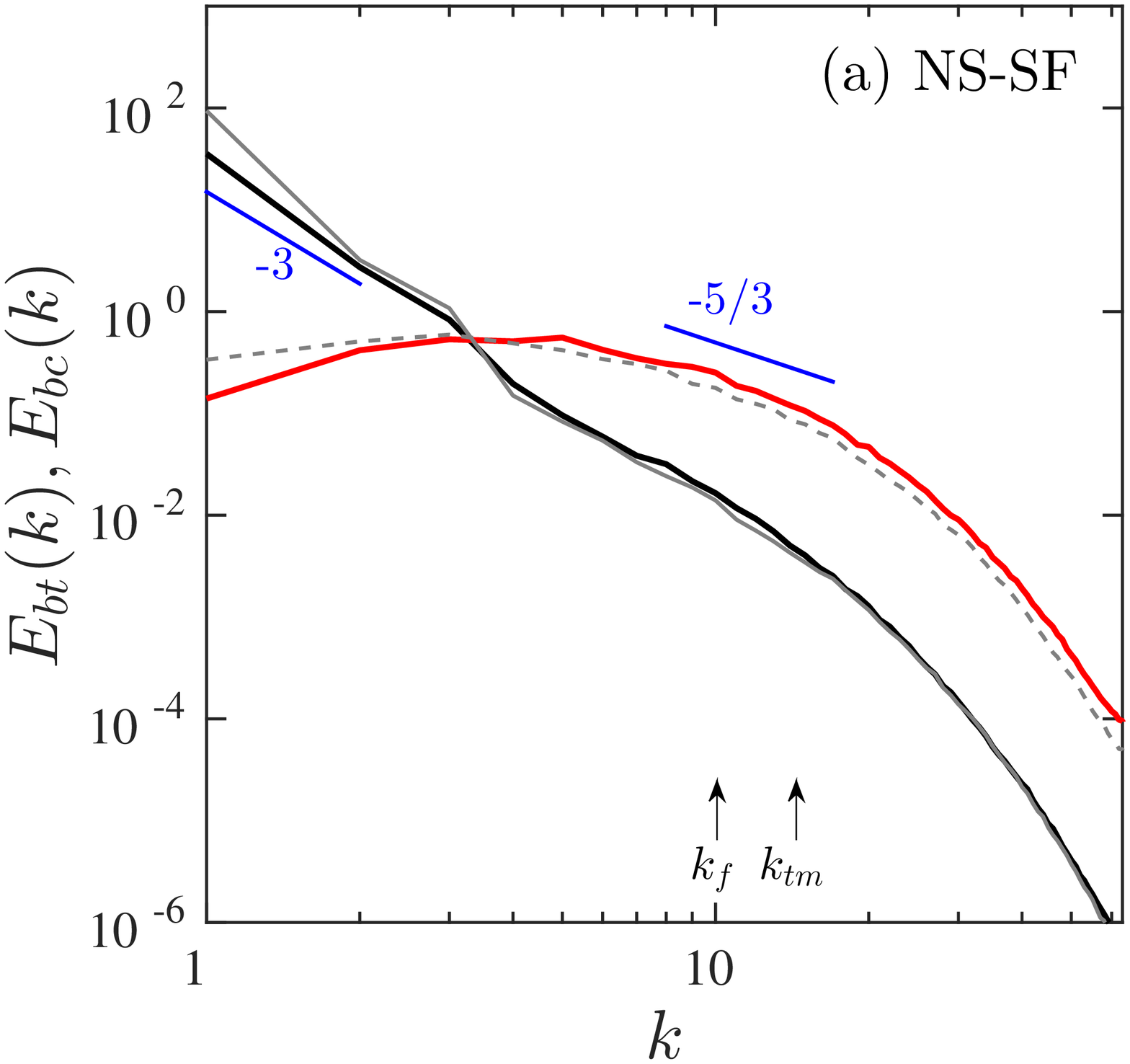}
		\hspace{-1em}
		\includegraphics[width=0.5\textwidth]{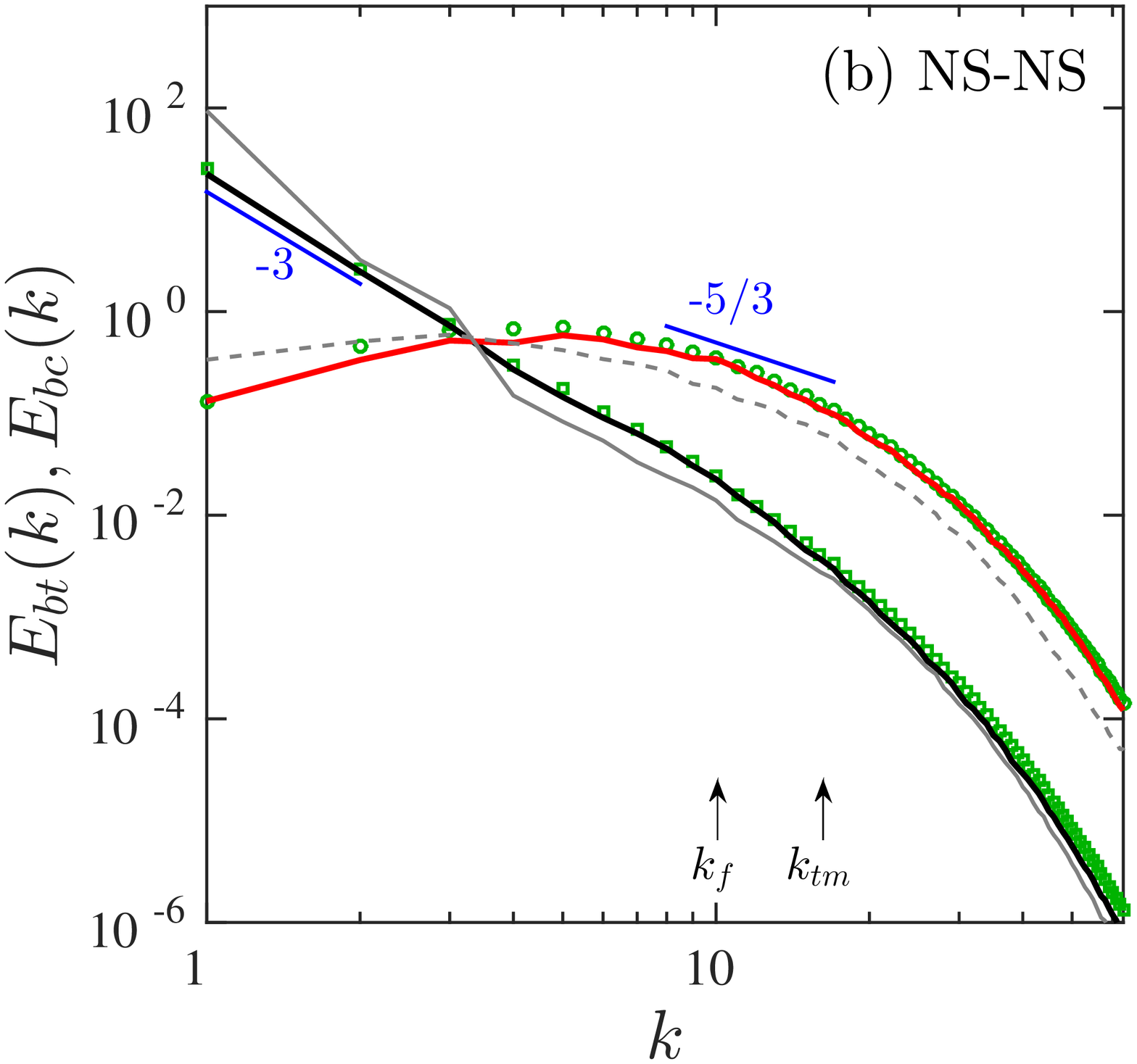}
	\caption{\small{The time-averaged KE spectra for (a) mixed NS-SF and (b) NS-NS simulations with $\Pran=1$, $\widetilde{Ra} = 90$ and $E=10^{-7}$.  Wavenumbers $k$  are normalized to the box scale $L=10L_c$, so $k=10 \tilde{k}/\tilde{k}_c$ where $\tilde{k}_c=1.3048$.	The black line shows the barotropic KE spectra $E_{bt}(k)$ with a reference slope of -3 for guidance. The red line shows the horizontal baroclinic KE spectra $E_{bc}(k)$ taken at $Z = 1/2$ with the slope of $-5/3$ plotted for reference. For comparison, the gray lines indicate the baroclinic (dashed) and barotropic (solid) components for the case of SF-SF boundaries. DNS  NS-NS  results (green symbols) are in agreement (plot b).
	The effects of Ekman friction on $E_{bt}(k)$ and Ekman pumping on $E_{bc}(k)$  are maximal for the NS-NS case. 
	The spectra exhibit scaling transitions reminiscent of \cite{nastrom}. The arrows mark the energy injection at box wave number $k_f$ and the Taylor microscale at $k_{tm}$.
	}}
	\label{fig:energy}
\vspace{-2ex}
\end{figure}

It has recently been demonstrated that geostrophic turbulence in RRBC with \textit{stress-free boundaries} yields a nonlocal inverse energy cascade that drives the formation of large scale dipole condensate consisting of a cyclonic and anticyclonic vortex pair (figure \ref{fig:barotropic}a). This was established using the asymptotic NH-QGM \citep{Julien12,Rubio} and  confirmed by DNS \citep{bF14,cG14,Stellmach}. We note that in the Rayleigh-B\'enard configuration the inverse cascade is unaffected by the box size, and it is found that energy always cascades to the largest scales permitted and the associated coherent structures continually grow to box-scale. However, it presently remains an open question whether (i) such a cascade persists in the presence of boundary layer friction, and (ii) if present, how its effects on the fluid dynamics may be quantified (cf. \cite{Kunnen16}).
Figure \ref{fig:bargrowth} illustrates the growth in barotropic kinetic energy occurs for all combinations of boundary conditions used in the NH-QGM. 
With only bulk dissipation present, slow saturation is observed for SF-SF simulations \citep{Rubio}. However, in the presence of NS-NS or NS-SF boundaries it can be seen that an inverse energy cascade \textit{still persists} but is both diminished and saturated by Ekman friction
(figure \ref{fig:bargrowth}).  For mixed NS-SF boundaries, the dipole condensate persists in a diminished stationary state  (figure \ref{fig:barotropic}b). In the presence of NS-NS boundaries, visualizations  (figure \ref{fig:barotropic}c) show that no persistent large-scale condensate is established. Only structures that are intermittent in time are observed  (figure~\ref{fig:time_series}). 
Here, despite the perpetual flow of energy to large scales, we find that Ekman friction continually causes the large vortical structures to break up. The time series in figure~\ref{fig:time_series} indicate that no large dipole forms and that the depth-independent structures vary with time in intensity and coherence.


Figure~\ref{fig:energy} illustrates the 1D spectra of horizontal kinetic energy  decomposed into the barotropic and baroclinic components, $E_{bt}(k)$ and $E_{bc}(k)$, as defined in (\ref{eq:1Denergy}).  
We note that the characteristics of the vertical and horizontal baroclinic spectra are similar.  We also note that the ability to build an extensive inertial range by a forward cascade in rotating convection is challenging.
 Specifically, the baroclinic inertial range is bounded between the injection scale comparable to the linear onset scale $L/H\sim 4.82 E^{1/3}$ and the Taylor micro scale. The Taylor micro scale estimates the viscous distortion of eddies and, by estimation from isotropic turbulence theory, is given as $\sqrt{10/\varepsilon}\sim 2.71E^{1/3} $, where $\varepsilon$ is defined as in (\ref{eq:kol}).
At the thermal forcing of $\widetilde{Ra}=90$, this implies a ratio of $1$:$1.77$ (see figure~\ref{fig:energy}), thus the convective forcing scale still operates very close to the diffusive scales. Furthermore, the inertial range for the inverse cascade is bounded from above by the box size $L_\mathrm{box}$.  Figure~\ref{fig:energy} therefore illustrates the direct and inverse cascades, each with a wavenumber range that is less than decadal.

Despite having inertial ranges that are less than decadal, we find a compelling picture emerges upon consideration of the properties of energy transfer and the \textit{instantaneous} power law exponents for the energy spectra. Hereinafter, wavenumbers $\tilde{k}$ are normalized to the box scale $L_{box}=\lambda L_c$, such that $k=\tilde{k}/\tilde{k}_{box}=\lambda\tilde{k}/\tilde{k}_{c}$ with $\lambda=10$ or $20$  (see Table~\ref{tab:TABLEOfValues})  and $\tilde{k}_{c}=1.3048$ \citep{Chandrar}.  The barotropic component dominates the energy spectra at low wavenumbers for all combinations of boundary conditions. Moreover, we find that the baroclinic kinetic energy $E_{bc}(k)$ saturates in all cases. 
For SF-SF simulations \citep{Julien12,Rubio, Stellmach}, the barotropic kinetic energy $E_{bt}(k)$  saturates on a longer time scale and exhibits the steepest power law.  In the presence of Ekman friction, the barotropic kinetic energy saturates and the power law exponent is reduced with $E_{bt}(k) \sim k^{-3}$.
 The horizontal baroclinic spectra  $E_{bc}(k)$ gains dominance at higher wavenumbers $(k\approx 3)$,
 it also exhibit shallower instantaneous power law  exponents $E_{bc}(k) \sim k^{-5/3}$.   
The total kinetic energy spectra $E_{bt}+E_{bc}$  exhibits  a steep to shallow transition in the power law exponents, a result reminiscent of the Nastrom-Gage spectra of atmospheric and oceanic measurements with a $-3$ to $-5/3$ power law transition \citep{nastrom}.
 For both plots in figure~\ref{fig:energy},  it can be seen that  the spectra is very similar in comparison with the SF-SF simulation results (displayed in gray) but with a few important distinctions due to the presence of Ekman pumping and Ekman friction. The effect of Ekman friction is evident in the decreased power in the barotropic kinetic energy at low wavenumbers, which is maximal for the NS-NS case.  The effect of Ekman pumping is also evident  by the increased power levels of the baroclinic kinetic energy at high wavenumbers. Excellent agreement is achieved in comparison with NS-NS DNS results in figure~\ref{fig:energy}b (open green squares and  circles).  The quantitative agreement in the kinetic energy spectra thus establishes the validity of the inverse cascade in the presence of frictional boundaries and is a third marker for the fidelity of the CNH-QGM.

\begin{figure}
	\centering 
%
	    \includegraphics[width=0.3\textwidth]{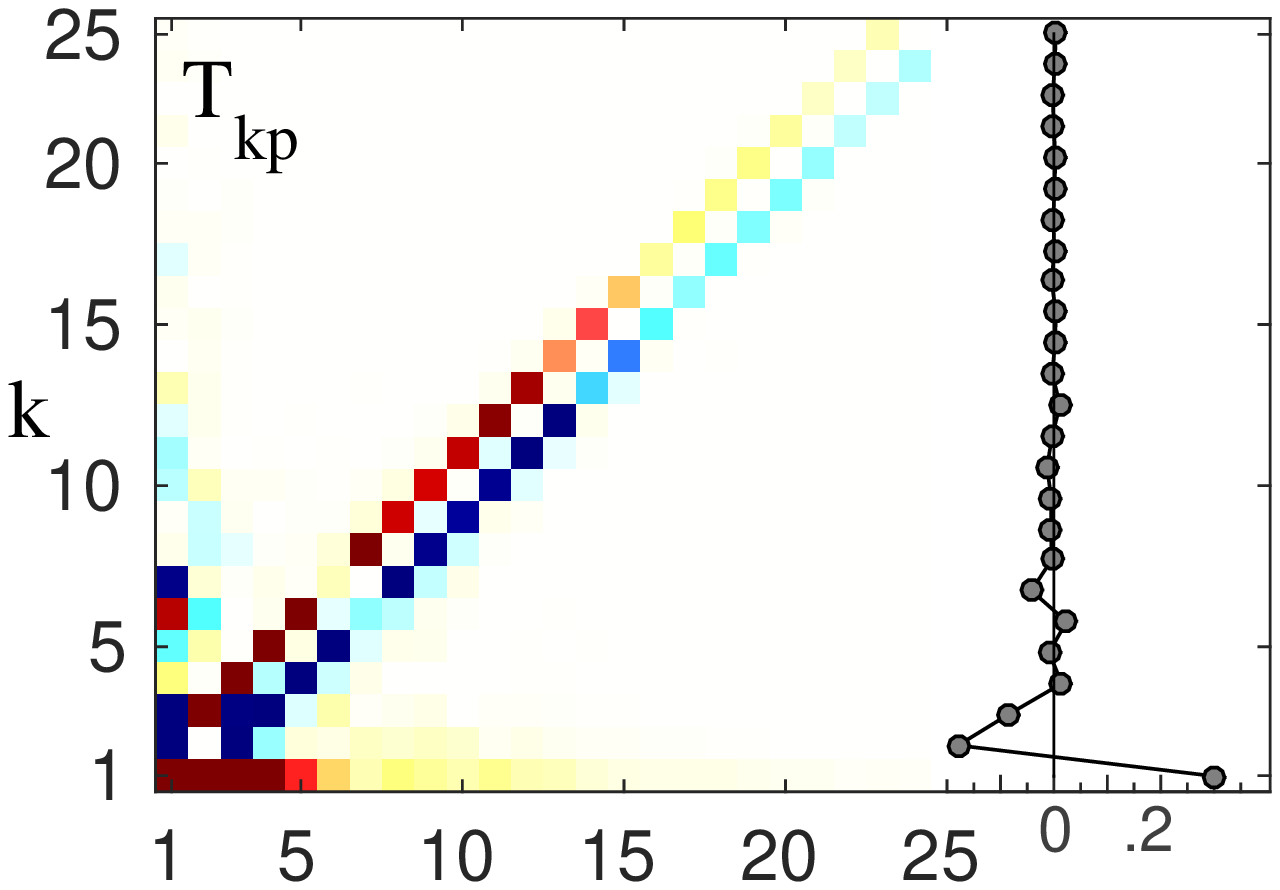} \hspace{-1em}
		\includegraphics[width=0.3\textwidth]{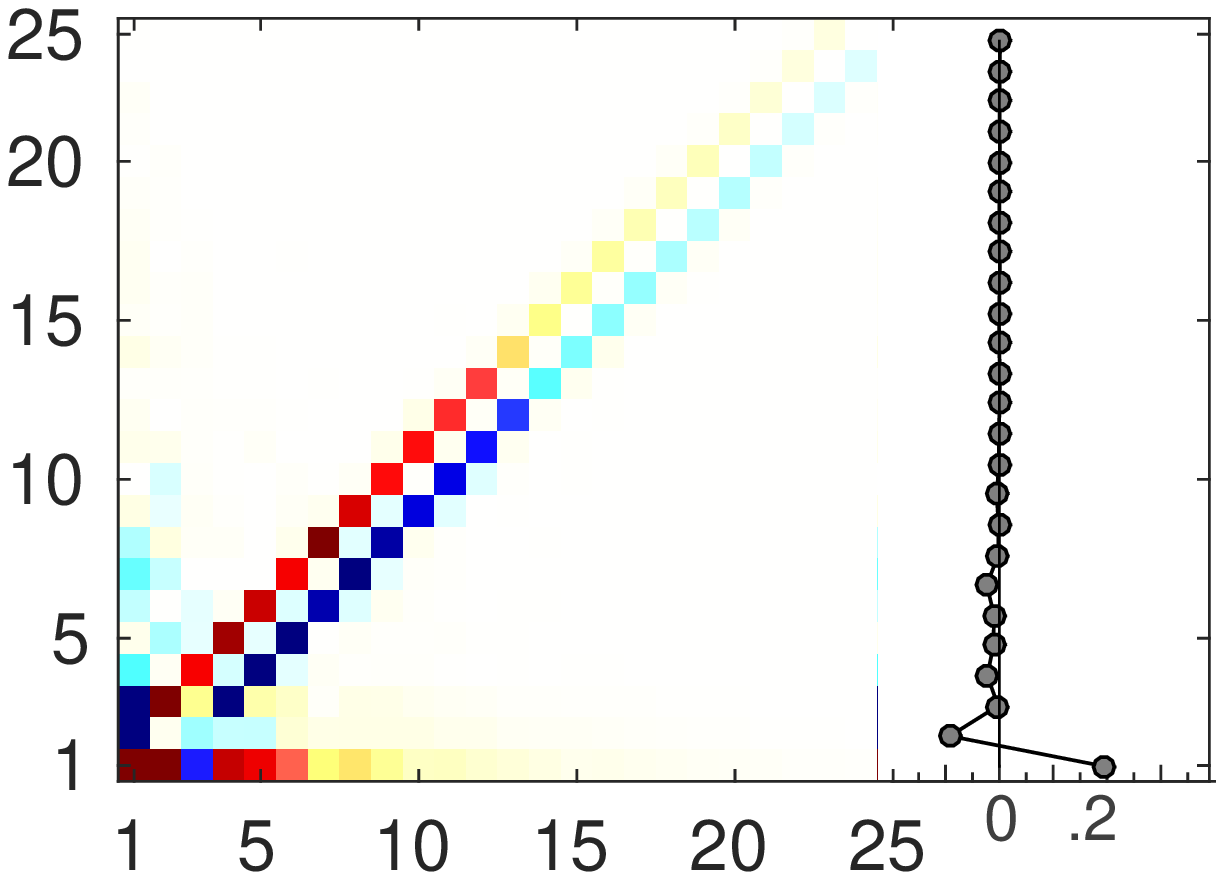}\hspace{-.5em}
		\includegraphics[width=0.3\textwidth]{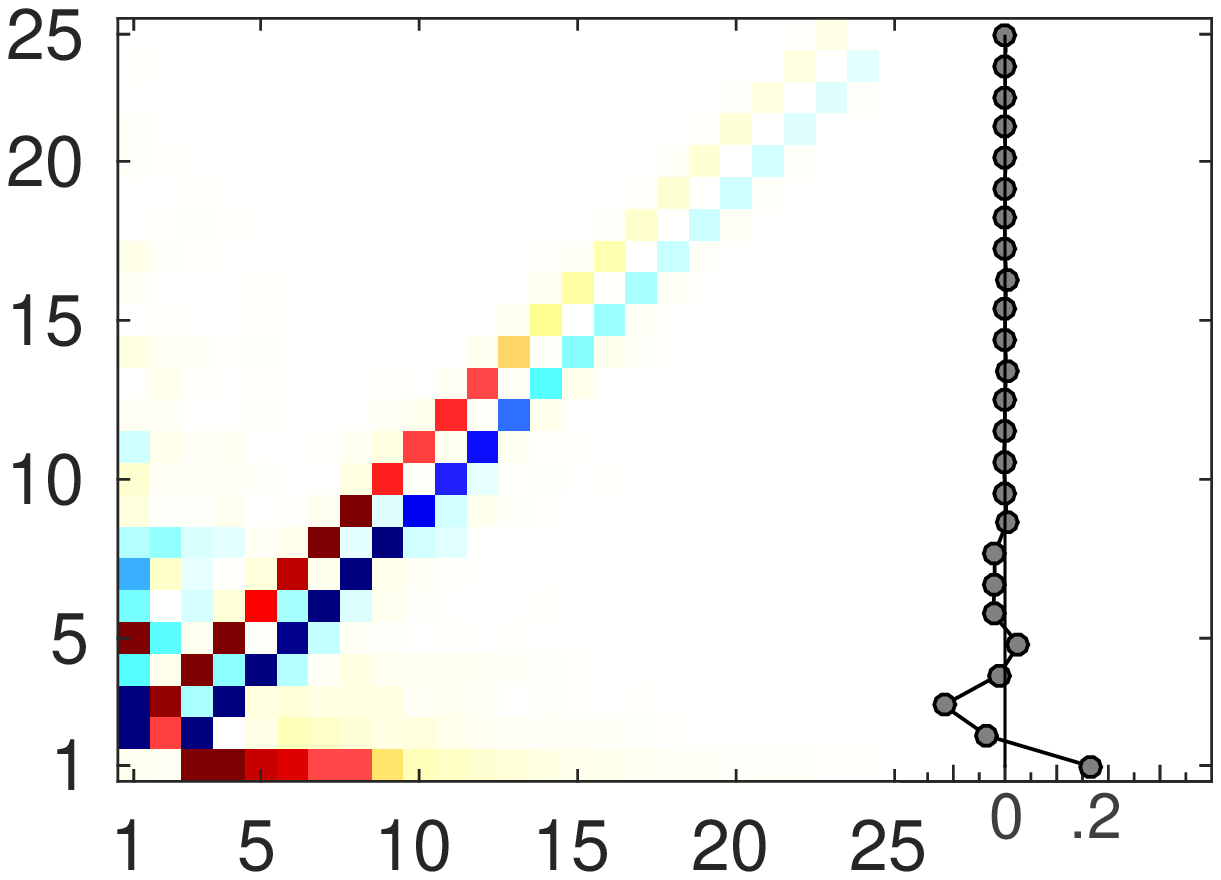} \hspace{-1em}
		\includegraphics[width=0.085\textwidth]{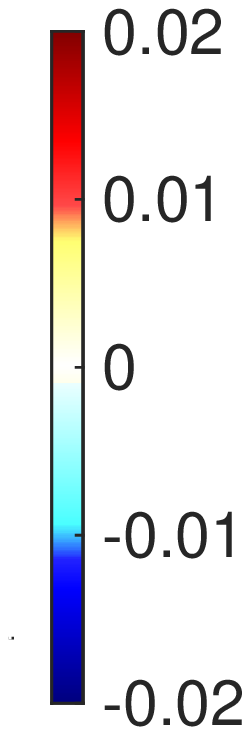}
		\\
		\vspace{-3ex}
		\subfloat[SF-SF]{\includegraphics[width=0.3\textwidth]{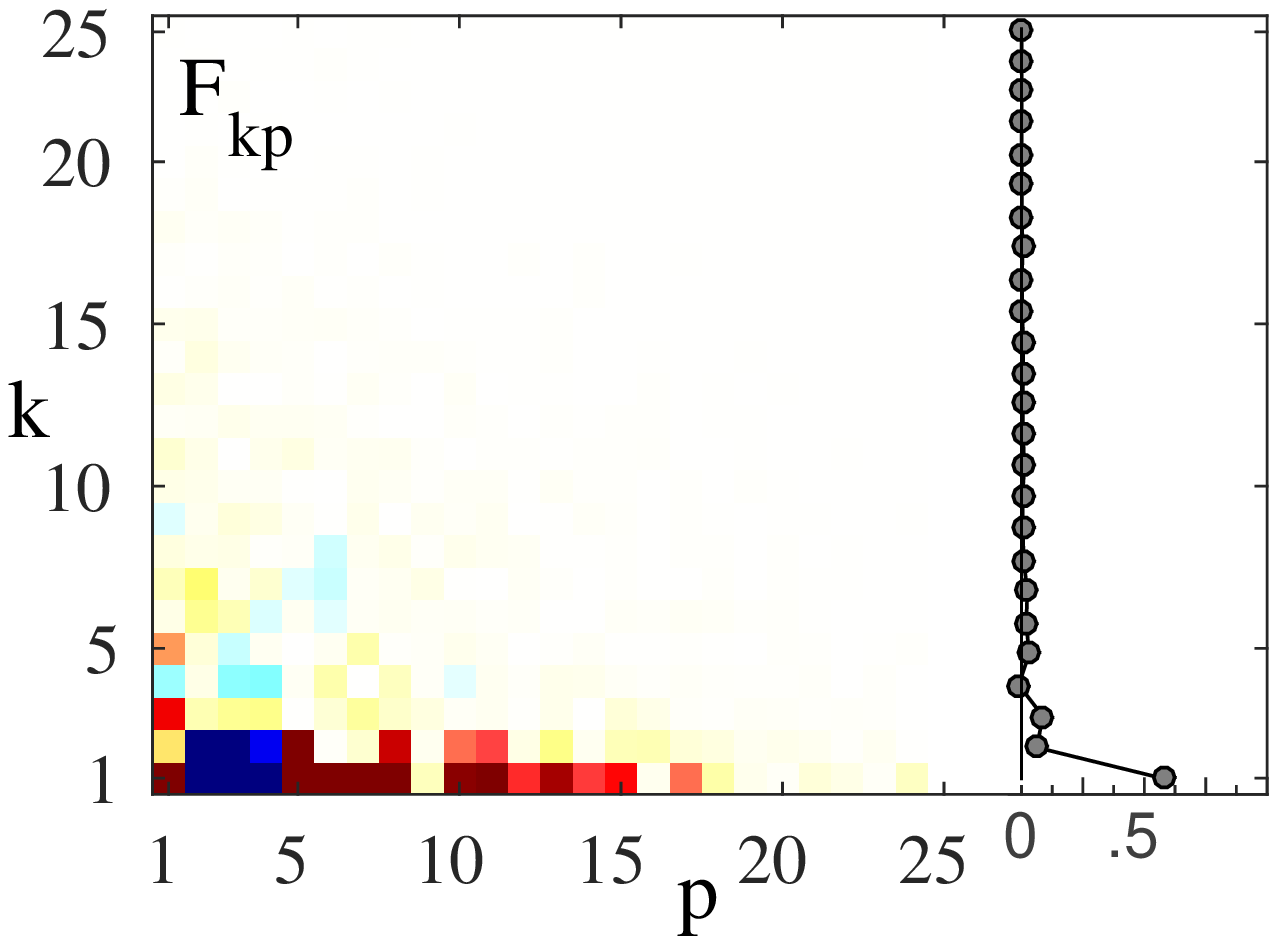}} \hspace{-.5em}
	         \subfloat[NS-SF]{\includegraphics[width=0.3\textwidth]{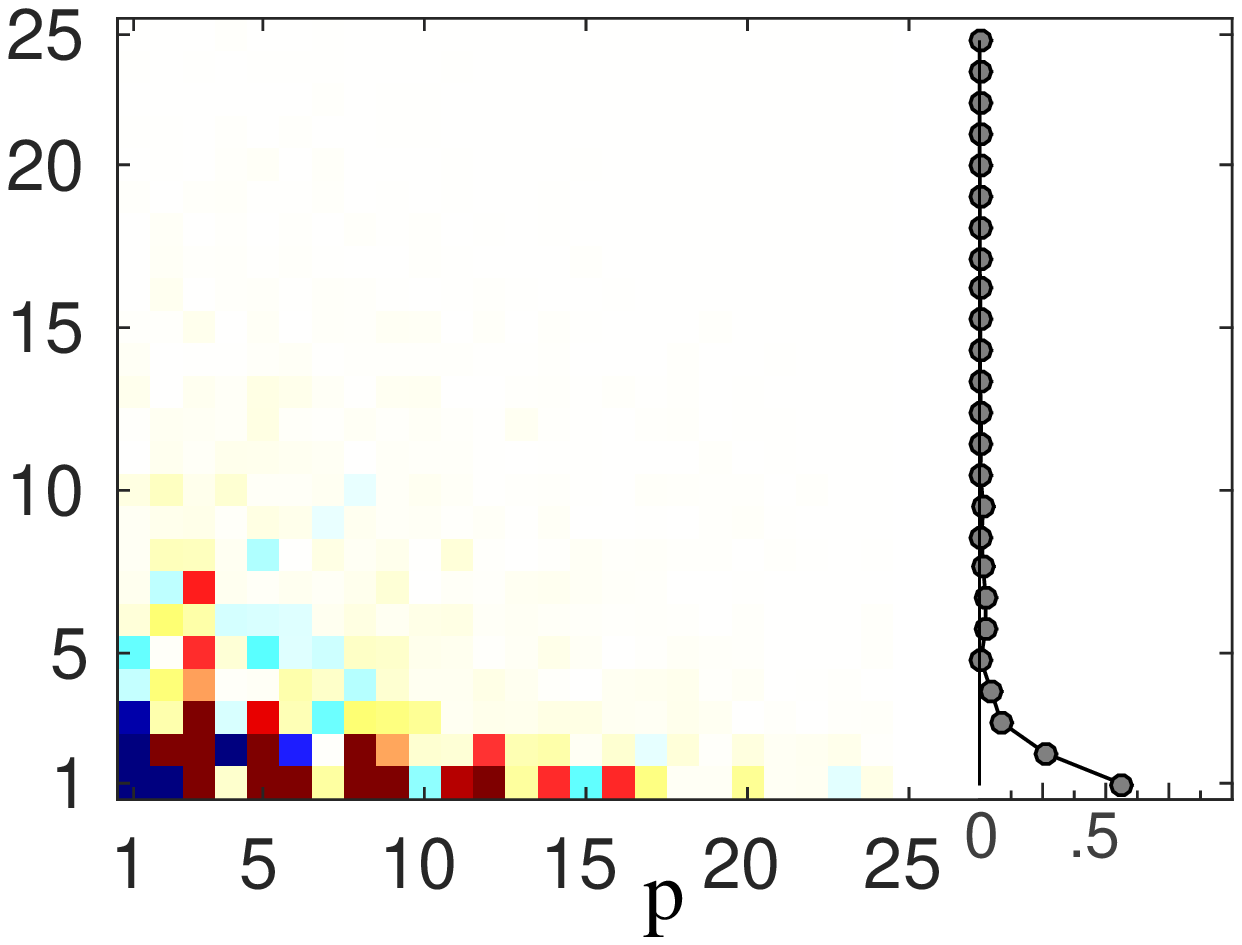}}   \hspace{-.5em}
	         \subfloat[NS-NS]{\includegraphics[width=0.3\textwidth]{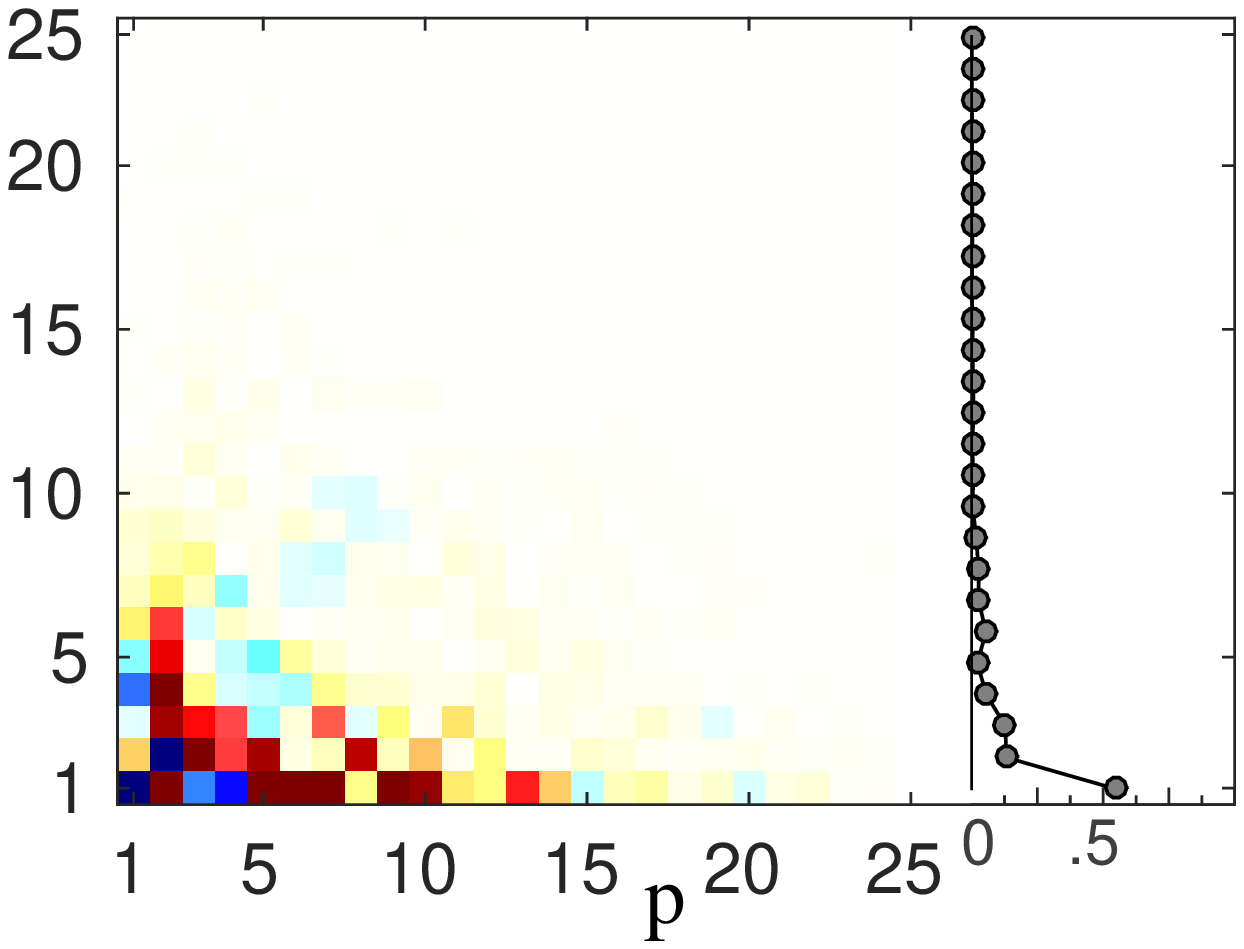}} \hspace{-1em}
	         \includegraphics[width=0.086\textwidth]{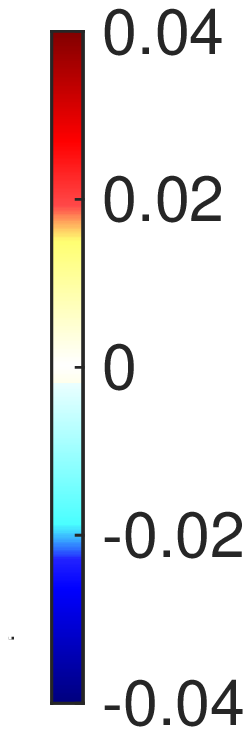}	         
	         \vspace{-1ex}
		\caption{\small{The transfer maps for barotropic self interaction $T_{kp}$ on the top row and  baroclinic-to-barotropic forcing $F_{kp}$ along the bottom row, each averaged over several times for a case with $\Pran=1$, $\widetilde{Ra} = 90$. Plots show (a) SF-SF $E=0$, (b) mixed NS-SF boundaries $E=10^{-7}$, and (c) NS-NS boundaries $E=10^{-7}$. Colorbars indicate how power moves from mode $p$ to mode $k$. The vertical lines show the integrated values (summed over $p$) for each $k$.}}
	\label{fig:energyMaps}
	\vspace{-3ex}
\end{figure}

The transfer maps for the nonlinear barotropic self interaction $T_{kp}$ and the baroclinic forcing of the barotropic mode $F_{kp}$  obtained from (\ref{eq:tkp}) are  displayed in figure~\ref{fig:energyMaps} for all combinations of boundary conditions. 
Each plot indicates how spectral power is transferred from wavenumber  mode ${p}$ to barotropic mode ${k}$ via triadic interaction with catalyst wavenumber mode ${q}$ such that  ${\boldsymbol{k}+\boldsymbol{p}+\boldsymbol{q}=\boldsymbol{0}}$.
Several common characteristics are evident in all plots. 

The spectral power in  $T_{kp}$ is negligible in all maps for all $(k,p) \ge 15$ (Figure~\ref{fig:energyMaps}, upper panels).
The  power signature in the super and sub off-diagonal lines in all barotropic self-interaction $T_{kp}$ maps 
indicate the existence of 
a forward or direct cascade, i.e., spectral power being directly transferred from low to high $k$-wavenumbers at constant wavenumber $p$. For $p\gg k$, we also find strong evidence for the nonlocal inverse cascade in that power is transferred  from the high $p$ wavenumbers directly to $k\approx1$. Similarly, for $p\ll k$, we find energy is being extracted from high $k$ wavenumbers and transferred to $p\approx 1$. 

The presence of large scale barotopic motions strongly influences the source of its origin, i.e. the baroclinic forcing $- \langle J[  \Psi^\prime ,  \zeta^\prime ]  \rangle$ in (\ref{eq:bxi}).   This term may be interpreted as barotropic collinearity between $\boldsymbol{u}_{\perp}$ and $\nabla_\perp\zeta'$. The transfer maps 
(Figure~\ref{fig:energyMaps}, lower panels) clearly indicate a direct forcing of the low $k$ wavenumber  baroclinic modes by high wavenumber $p$ modes.  Particularly,  one sees that the line plots for $F(k)=\sum_{p} F_{kp}$ are positive and non-negligible for $k\lesssim4$ when friction is present and non-negligible when $k=1$ for SF-SF boundaries. The increased wavenumber range in $F(k)$ for NS-NS and NS-SF  cases is consistent with the fact that it must be balanced by  the energy extracted by barotropic friction $Ef^{bt}_{k}$ which is directly proportional to $E_{bt}({\boldsymbol{k}})$ and therefore dominates at low wavenumbers (see (\ref{eq:barotropicE})).
One can also see that  the total nonlinear transfer $T(k)=\sum_{p} T_{kp}$  is zero for $k\gtrsim 8$ (line plots in Figure~\ref{fig:energyMaps})  indicating the beginning of  an inertial regime from $8 \lesssim k \lesssim 15$ where the observed forcing $F_{\boldsymbol{k}}$, dissipation $D_{\boldsymbol{k}}$, and $Ef_{\boldsymbol{k}}$  are negligible and the energy flux is constant.  For smaller wavenumber values $k\lesssim 8$ the barotropic $T(k)$ cascade first removes energy  before adding energy at the smallest $k$. Energy in the $k=1$ mode is maximal and minimal in the SF-SF and NS-NS cases respectively.

Here, we remark on the efficiency by which the barotropic mode extracts energy from the baroclininc dynamics. If the forcing term $- \langle J[  \Psi^\prime ,  \zeta^\prime ]  \rangle$ is removed from (\ref{eq:bxi}), thereby suppressing the barotropic cascade, the alterations to the baroclinic kinetic energy spectra are minimal. This result, first established by \cite{Rubio}, indicates a symbiotic relationship between baroclinic convective motions and the large scale  barotropic dynamics, i.e.,  convective fluctuations adjust in the presence of barotropic dynamics without amplitude suppression. This 
adjustment can be contrasted to the more extreme cases of predator-prey, competitive dynamics, in which relaxation oscillations result in large amplitude variations in both baroclinic and barotropic modes. 

\subsection{Low Ekman Simulations}
\label{sec:LowEk}
In this section, we extend our simulations to Ekman numbers below those presently attainable by DNS and experiments ($E \approx 10^{-7}$) to explore the heat transport as $E \rightarrow 0$. We recall that the asymptotic model CNH-QGM is valid within 
$\widetilde{Ra} = Ra E^{4/3} \lesssim E^{-1/3}$. Moreover in \citet{Julien15}, it was shown that the transition from a regime adequately described by stress-free boundaries (i.e., $E=0$) to a regime dominated by 
Ekman pumping occurred at $Ra_t E^{4/3} \sim E^{-1/9}$ for the special class of single-mode solutions.  These results may be of particular utility in physical systems such as Earth's core where $E \approx 10^{-15}$ \citep{Soderlund}. 
Specifically, we consider the cases $E=10^{-7}, 10^{-9}$, and $10^{-11}$ for Prandtl numbers $7$ and $1$. To account for the most restrictive resolution requirement in the vertical (i.e., the $\mathcal{O}(E^{1/3})$ thermal wind layer), a  resolution of $288$~x~$288$~x~$288$ with 6 points within the thermal wind layer was chosen for $E = 10^{-9}$ and $384$~x~$384$~x~$576$ with 5 points in the thermal wind layer was chosen for $E = 10^{-11}$.
\begin{figure}
	\centering 
	         \subfloat[$Pr = 7$]{\includegraphics[width=0.5\textwidth]{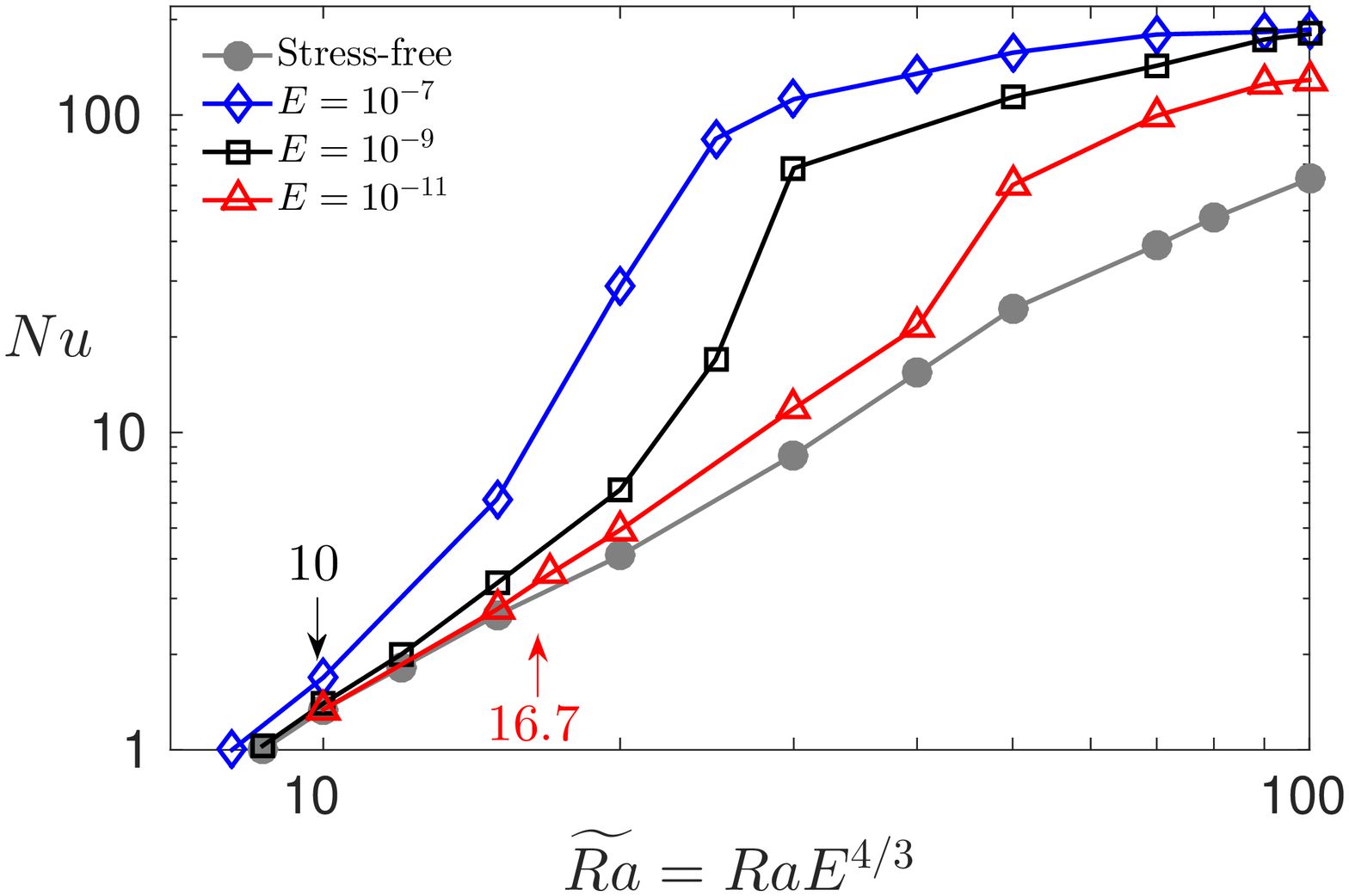}} \hspace{-1em}
	         \subfloat[$Pr = 1$]{\includegraphics[width=0.5\textwidth]{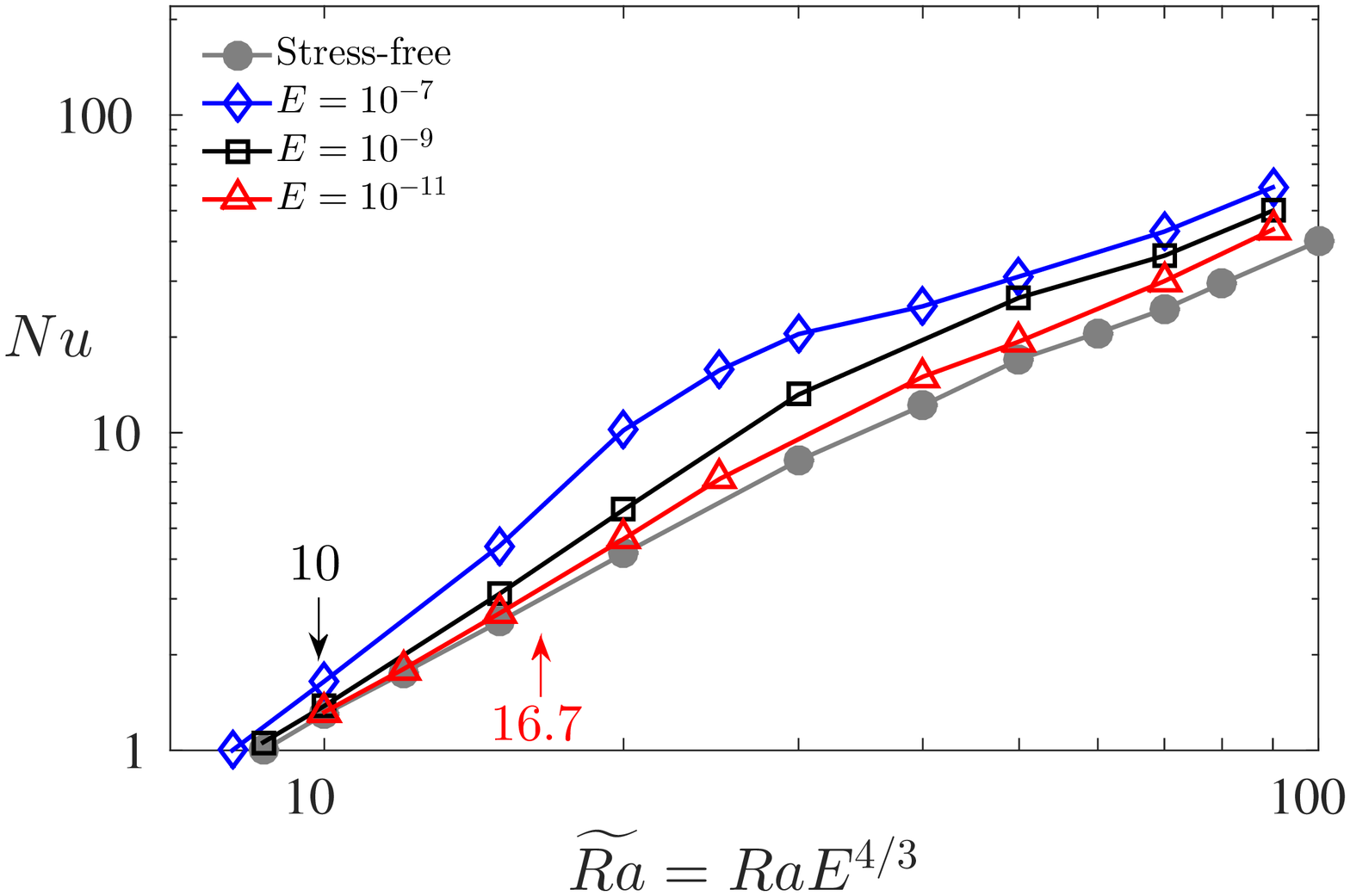}}
		\caption{\small{Heat transfer $Nu$ as a function of $\widetilde{Ra}$ for CNH-QGM no-slip simulations with (a) $Pr = 7$ and (b) $Pr =1$.
Comparison with the stress-free ($E = 0$) results (gray circles) are made for Ekman numbers $E= 10^{-7}$ (blue diamonds), $E = 10^{-9}$ (black squares), and $E = 10^{-11}$ (red triangles). The arrows indicate the locations of the predicted departure at $\widetilde{Ra_t} = E^{-1/9}$ \citep{Julien15} away from the stress-free curve for $E = 10^{-9}$ and $E = 10^{-11}$. The departure for $E = 10^{-7}$ is below the onset $\widetilde{Ra}$ and is not shown. }}
	\label{fig:lowE}
	\vspace{-3ex}
\end{figure}

 Figure~\ref{fig:lowE} shows the Nusselt number $Nu$ as a function of $\widetilde{Ra}$ for (a) $Pr = 7$ and (b) $Pr =1$. No-slip boundaries are considered with Ekman numbers $E=10^{-7}, 10^{-9},$ and $10^{-11}$ and compared to stress-free results with $E=0$ (gray curves).
 For increasing $\widetilde{Ra}$, it is observed that the curves associated with no-slip boundaries initially coincide with the stress-free result before departure to enhanced values of $Nu$. 
The departure locations are consistent  with the $\widetilde{Ra_t} = c E^{-1/9}$ prediction by \citet{Julien15}, which we approximate simply with $c = 1$.  Specifically, with $c =1$, their theory predicts a departure from the stress-free curves at $\widetilde{Ra_t} \approx 5.99$ for $E=10^{-7}$ (not shown), $\widetilde{Ra_t} \approx 10$ for $E=10^{-9}$ (black arrow), and $\widetilde{Ra_t} \approx 16.7$ for $E=10^{-11}$ (red arrow). The transitional value at $E =10^{-7}$ is below critical onset and thus departure occurs immediately. 
 
 In Figure~\ref{fig:lowE}a, the departures from the stress-free results are apparent for both $E = 10^{-9}$ and $E = 10^{-11}$. The shape of these curves is largely similar to those predicted by the single mode solutions, which provide an upper bound estimate of the heat transfer \citep[cf. Figure 10 in ][]{Julien15}. These results show a strong departure to higher heat transfer, and the curves for $E=10^{-7}$ and $E=10^{-9}$ seem to be crossing near $\widetilde{Ra} = 100$.
 The results with $Pr = 1$ in Figure~\ref{fig:lowE}b show a smaller departure away from the stress-free curve.  We note that the vertically coherent, single mode solutions are less relevant for low $Pr$. Here the flow is dominated by geostrophic turbulence and the small departures from the stress-free may indicate that the stress-free curve is a good leading-order approximation for the heat transfer of low $Pr$, geostrophically turbulent flows. 


 

\section{Conclusions}

 Understanding the details and implications of rotating thermal convection in the geophysically and astrophysically relevant, but extreme, high Rayleigh $Ra$ - low Rossby $Ro$ number regime remains one of the outstanding challenges for both laboratory experiments and direct numerical simulations. Attainable bounds in Ekman number for an adequate survey across a sufficient range of $Ra$ presently reside at  $E\gtrsim 10^{-7}$ \citep{Ecke14,Cheng} compared to planetary and stellar values of $\lesssim 10^{-15}$ \citep{Soderlund}. The utilization of asymptotically reduced models derived from the governing Navier-Stokes equations to gain access to such regimes has a long tradition in geophysical and astrophysical  fluid dynamics beginning  with the seminal introduction of the classical  hydrostatic quasigeostrophic equations of \cite{jC48,Charney}. His model, which captures the asymptotic behavior of geostrophically balanced large-scale eddies in a stably-stratified layer while filtering fast inertio-gravity waves, remains a vital theoretical tool in investigating geophysical turbulence. Asymptotically reduced models for unstably-stratified (i.e. nonhydrostatic) columnar turbulence  have also played a vital role when the directions of gravity and rotation are orthogonal \citep{pR68,fB70,fB86,mC13}. Orthogonality aids the Proudman-Taylor constraint in this configuration, which can co-exist with strong thermal forcing without conflict. Only recently has this constraint of orthogonality been relaxed to produce asymptotic (nonhydrostatic quasigeostrophic) models valid as $E\rightarrow0$ for stress-free bounding plates  \citep{kJ98a,kJ06,Sprague,Julien12,Rubio}. However, comparisons with laboratory investigations and geophysical settings, where no-slip bounding surfaces are inevitable, require an understanding of the roles of Ekman pumping and Ekman friction in the extrapolated limit $E\rightarrow0$.  

In this regard, the composite nonhydrostatic quasigeostrophic model was derived by \cite{Julien15} to include the effect of no-slip boundaries. Through an investigation into single mode solutions, they verified that the presence of Ekman layers formed along a no-slip boundary have a dominant positive effect on the convective  heat flux and transport in the RRBC system within the geostrophic regime. 
This occurs, despite the increasingly narrow widths  ($\propto E^{1/2} H$) and pumping magnitudes ($\propto E^{1/2}\nu/H$) as $E \rightarrow 0$, as a consequence of the increasing magnitude of thermal fluctuations in an $\mathcal{O}(E^{1/3} H)$  thermal wind boundary layer. The increased thermal fluctuations are produced by the vertical advection of the mean temperature gradient and effectively compensate the decreasing magnitude of Ekman pumping to yield O(1) convective fluxes. 

In this present study, the fidelity of the  reduced composite model is explored through full numerical simulations capturing all dynamically unstable modes, and results are compared against DNS. To surmise the findings, excellent qualitative and quantitative comparisons are found for three metrics: (i) flow visualizations and transitions between all geostrophic regimes, (ii) enhanced heat transport  due to Ekman pumping in the Nusselt-Rayleigh relation, and (iii)  the inverse barotropic cascade of kinetic energy. The latter in particular has only recently been established and quantified in the presence of stress-free boundaries \citep{Rubio,cG14,bF14,Stellmach}. The advantage of the model over DNS is the relaxation of the stiffness imposed by (i) the multi-time scale properties of the thermal field, and often, the explicit time representation of the Coriolis force, 
and (ii)  spatially resolving the Ekman boundary layers (see Table~\ref{tab:TABLEOfValues}). Parameterized Ekman pumping boundary layers may also be employed in the DNS, however, any gain is ultimately restricted by the explicit time stepper constraints on the non-linear advection terms due to the large velocity amplitudes at the computational boundary.

For $Pr> 1$,  good qualitative agreement is achieved in flow visualizations (Figure~\ref{fig:Flows}) and  good quantitative agreement  in heat transport is obtained in the cellular regime and as the geostrophic turbulence regime is approached  (Figure~\ref{fig:Comparison}).
Plausible explanations for discrepancies in the heat transport for the intermediate convective Taylor column and plume regimes at $E=10^{-7}$ are twofold.
First, the value $E=10^{-7}$ remains asymptotically too large, a fact that motivates future efforts in the Rayleigh-B\'enard community
at more extreme Ekman numbers (e.g., the Eindhoven TROCONVEX, the Gottingen U-boot, and the UCLA Spinlab rotating convection experiments). Secondly, discrepancies may occur as a result of the subdominant effects of nonlinear advection of momenta and temperature omitted in the asymptotical model but inherently retained in the DNS.
Resolution to each conjecture 
requires systematic studies of either lowering of the Ekman number in DNS or inclusion of subdominant terms in the asymptotic equations, both of which are beyond the scope and intent of the present manuscript.

For $Pr\le1$, where a direct transition from the cellular to the geostrophic turbulence regime occurs, excellent agreement with DNS (Figure~\ref{fig:Comparison}) is found.
 In the geostrophic turbulence regime, a nonlocal inverse barotropic cascade \textit{persists} for all types of boundaries but is both diminished and rapidly saturated by Ekman friction.
A Nastrom-Gage kinetic spectrum is found that exhibits a shallow to steep, i.e.  $\approx -3$ to $\approx-5/3$, transition in the power law exponent.  In the presence of NS-NS boundaries, large scale vortices are no longer statistically stationary but are found to vary both in time and coherence. We note that this finding, from both the CNH-QGM and the DNS, is contrary to \cite{Kunnen16}, whose simulations argue that the inverse cascade vanishes for no-slip plates.  

The investigation using low Ekman numbers ($E\leq 10^{-7}$) demonstrates the effect of decreasing $E$ on the heat transfer enhancements. The departure from the stress-free curve remains large  for simulations with $Pr>1$, even at $E = 10^{-11}$. However, for $Pr = 1$, the results suggest that the stress-free curve is a fair leading-order approximation at low $E$.  The capacity of this model to probe smaller $E$ provides a new avenue for explorations of RRBC, one that should be of utility to DNS and laboratory experiments, both presently limited to $E\geq 10^{-7}$, going forward. Moreover, future studies at lower $E$ and higher $Pr$ should provide insight into geophysical systems. 

Large scale structures are persistent features of rapidly rotating stars and planets  that remain largely  ill-understood. The non-local inverse cascade represents a new mechanism for generating persistent large scale structures now captured  in DNS both with \citep{cG15,rY15} and without magnetic fields  \citep{bF14, cG14, Stellmach}. We contend that a systematic exploration of such a phenomenon will require a  collective approach involving laboratory experiments, DNS, and asymptotic simulations. In this regard and in a similar spirit to investigations using the classical quasigeostrophic equations,  the CNH-QGE represents a promising fundamental model that will complement emerging laboratory experiments and DNS for future investigations at more extreme Ekman numbers.

\vspace{-2ex}
\section*{Acknowledgments} 
This work was supported by the National Science Foundation under EAR grants \#1320991 \& CSEDI \#1067944 (K.J., P.M.) and by NASA Headquarters under the NASA Earth and Space Science Fellowship Program - Grant 15-PLANET15F-0105 (M.P.).  We wish to thank the anonymous referees for their useful comments on the manuscript.  S.S. gratefully acknowledges the Gauss Centre for
Supercomputing (GCS) for providing computing time through the John von Neumann
Institute for Computing (NIC) on the GCS share of the supercomputer JUQUEEN at
Jülich Supercomputing Centre (JSC) in Germany. This work utilized the Janus supercomputer, which is supported by the National Science Foundation (award number CNS-0821794) and the University of Colorado Boulder. 
\vspace{-2ex}

\bibliographystyle{jfm}
\bibliography{mybib}

\begin{thebibliography}{38}
\expandafter\ifx\csname natexlab\endcsname\relax\def\natexlab#1{#1}\fi
\def\au#1{#1} \def\ed#1{#1} \def\yr#1{#1}\def\at#1{#1}\def\jt#1{\textit{#1}}
  \def\bt#1{#1}\def\bvol#1{\textbf{#1}} \def\vol#1{#1} \def\pg#1{#1}
  \def\publ#1{#1}\def\arxiv#1{#1}\def\org#1{#1}\def\st#1{\textit{#1}}

\bibitem[Aurnou {\em et~al.\/}(2015)Aurnou, Calkins, Cheng, Julien, King,
  Nieves, Soderlund \& Stellmach]{jmA15}
{\sc \au{Aurnou, J.~M.}, \au{Calkins, M.~A.}, \au{Cheng, J.~S.}, \au{Julien,
  K.}, \au{King, E.~M.}, \au{Nieves, D.}, \au{Soderlund, K.~M.} \&
  \au{Stellmach, S.}} \yr{2015}  \at{Rotating convective turbulence in earth
  and planetary cores}.  \jt{Physics of the Earth and Planetary Interiors}
  \bvol{246},  \pg{52--71}.

\bibitem[Busse(1970)]{fB70}
{\sc \au{Busse, F.~H.}} \yr{1970}  \at{Thermal instabilities in rapidly
  rotating systems}.  \jt{J. Fluid Mech.}  \bvol{44},  \pg{441--460}.

\bibitem[Busse(1986)]{fB86}
{\sc \au{Busse, F.~H.}} \yr{1986}  \at{Asymptotic theory of convection in a
  rotating, cylindrical annulus}.  \jt{J. Fluid Mech.}  \bvol{173},
  \pg{545--556}.

\bibitem[Calkins {\em et~al.\/}(2013)Calkins, Julien \& Marti]{mC13}
{\sc \au{Calkins, M.~A.}, \au{Julien, K.} \& \au{Marti, P.}} \yr{2013}
  \at{Three-dimensional quasi-geostrophic convection in the rotating
  cylindrical annulus with steeply sloping endwalls}.  \jt{Journal of Fluid
  Mechanics}  \bvol{732},  \pg{214--244}.

\bibitem[Chandrasekhar(1961)]{Chandrar}
{\sc \au{Chandrasekhar, S.}} \yr{1961} {\em Hydrodynamic and Hydromagnetic
  Stability\/}.  \publ{Oxford: Oxford University Press}.

\bibitem[Charney(1971)]{Charney}
{\sc \au{Charney, J.}} \yr{1971}  \at{Geostrophic turbulence}.  \jt{Journal of
  Atmoshperic Science}  \bvol{28},  \pg{1087--1095}.

\bibitem[Charney(1948)]{jC48}
{\sc \au{Charney, J.~G.}} \yr{1948}  \at{On the scale of atmospheric motions}.
  \jt{Geofys. Publ.}  \bvol{17},  \pg{3--17}.

\bibitem[Cheng {\em et~al.\/}(2015)Cheng, Stellmach, Ribeiro, Grannan, King. \&
  Aurnou]{Cheng}
{\sc \au{Cheng, J.~S.}, \au{Stellmach, S.}, \au{Ribeiro, A.}, \au{Grannan, A.},
  \au{King., E.~M.} \& \au{Aurnou, J.~M.}} \yr{2015}  \at{Laboratory-numerical
  models of rapidly rotating convection in planetary cores}.  \jt{Geophysics
  Journal International}  \bvol{201},  \pg{1--17}.

\bibitem[Ecke \& Niemela(2014)]{Ecke14}
{\sc \au{Ecke, R.~E.} \& \au{Niemela, J.~J.}} \yr{2014}  \at{{Heat transport in
  the geostrophic regime of rotating Rayleigh-B\'{e}nard convection}}.
  \jt{Phys. Rev. Lett.}  \bvol{113},  \pg{114301}.

\bibitem[Favier {\em et~al.\/}(2014)Favier, Silvers \& Proctor]{bF14}
{\sc \au{Favier, B.}, \au{Silvers, L.~J.} \& \au{Proctor, M. R.~E.}} \yr{2014}
  \at{{Inverse cascade and symmetry breaking in rapidly rotating Boussinesq
  convection}}.  \jt{Phys. Fluids}  \bvol{26}~(096605).

\bibitem[Gastine {\em et~al.\/}(2014)Gastine, Heimpel \&
  Wicht]{gastine2014zonal}
{\sc \au{Gastine, T.}, \au{Heimpel, M.} \& \au{Wicht, J.}} \yr{2014}  \at{Zonal
  flow scaling in rapidly-rotating compressible convection}.  \jt{Physics of
  the Earth and Planetary Interiors}  \bvol{232},  \pg{36--50}.

\bibitem[Greenspan(1968)]{Greenspan}
{\sc \au{Greenspan, H.~P.}} \yr{1968} {\em The Theory of Rotating Fluids\/}.
  \publ{Cambridge, United Kingdom: Cambridge University Press}.

\bibitem[Guervilly {\em et~al.\/}(2014)Guervilly, Hughes \& Jones]{cG14}
{\sc \au{Guervilly, C.}, \au{Hughes, D.~W.} \& \au{Jones, C.~A.}} \yr{2014}
  \at{{Large-scale vortices in rapidly rotating Rayleigh--B{\'e}nard
  convection}}.  \jt{Journal of Fluid Mechanics}  \bvol{758},  \pg{407--435}.

\bibitem[Guervilly {\em et~al.\/}(2015)Guervilly, Hughes \& Jones]{cG15}
{\sc \au{Guervilly, C.}, \au{Hughes, D.~W.} \& \au{Jones, C.~A.}} \yr{2015}
  \at{Generation of magnetic fields by large-scale vortices in rotating
  convection}.  \jt{Physical Review E}  \bvol{91}~(4),  \pg{041001}.

\bibitem[Heimpel {\em et~al.\/}(2016)Heimpel, Gastine \& Wicht]{mH2015}
{\sc \au{Heimpel, M.}, \au{Gastine, T.} \& \au{Wicht, J.}} \yr{2016}
  \at{Simulation of deep-seated zonal jets and shallow vortices in gas giant
  atmospheres}.  \jt{Nature Geoscience}  \bvol{9},  \pg{19–--23}.

\bibitem[Horn \& Shishkina(2015)]{Horn}
{\sc \au{Horn, S.} \& \au{Shishkina, O.}} \yr{2015}  \at{{Toroidal and poloidal
  energy in rotating Rayleigh-B\'enard convection}}.  \jt{J. Fluid Mech.}
  \bvol{762},  \pg{232--255}.

\bibitem[Julien {\em et~al.\/}(2016)Julien, Aurnou, Calkins, Knobloch, Marti,
  Stellmach \& Vasil]{Julien15}
{\sc \au{Julien, K.}, \au{Aurnou, J.}, \au{Calkins, M.}, \au{Knobloch, E.},
  \au{Marti, P.}, \au{Stellmach, S.} \& \au{Vasil, G.}} \yr{2016}  \at{A
  nonlinear model for rotationally constrained convection with ekman pumping}.
  \jt{Journal of Fluid Mechanics}  \bvol{798},  \pg{50--87}.

\bibitem[Julien \& Knobloch(2007)]{Julien07}
{\sc \au{Julien, K.} \& \au{Knobloch, E.}} \yr{2007}  \at{Reduced models for
  fluid flows with strong constraints}.  \jt{Journal of Mathematical Physics}
  \bvol{48}~(6).

\bibitem[Julien {\em et~al.\/}(2006)Julien, Knobloch, Milliff \& Werne]{kJ06}
{\sc \au{Julien, K.}, \au{Knobloch, E.}, \au{Milliff, R.} \& \au{Werne, J.}}
  \yr{2006}  \at{Generalized quasi-geostrophy for spatially anistropic
  rotationally constrained flows}.  \jt{J. Fluid Mech.}  \bvol{555},
  \pg{233--274}.

\bibitem[Julien {\em et~al.\/}(2012{\natexlab{{\em a\/}}})Julien, Knobloch,
  Rubio \& Vasil]{kJ12b}
{\sc \au{Julien, K.}, \au{Knobloch, E.}, \au{Rubio, A.~M.} \& \au{Vasil,
  G.~M.}} \yr{2012{\natexlab{{\em a\/}}}}  \at{{Heat transport in
  Low-Rossby-number Rayleigh-B\'enard Convection}}.  \jt{Phys. Rev. Lett.}
  \bvol{109}~(254503).

\bibitem[Julien {\em et~al.\/}(1998)Julien, Knobloch \& Werne]{kJ98a}
{\sc \au{Julien, K.}, \au{Knobloch, E.} \& \au{Werne, J.}} \yr{1998}  \at{A new
  class of equations for rotationally constrained flows}.  \jt{Theoret. Comput.
  Fluid Dyn.}  \bvol{11},  \pg{251--261}.

\bibitem[Julien {\em et~al.\/}(2012{\natexlab{{\em b\/}}})Julien, Rubio, Grooms
  \& Knobloch]{Julien12}
{\sc \au{Julien, K.}, \au{Rubio, A.~M.}, \au{Grooms, I.} \& \au{Knobloch, E.}}
  \yr{2012{\natexlab{{\em b\/}}}}  \at{{Statistical and physical balances in
  low Rossby number Rayleigh-B\'{e}nard convection}}.  \jt{Geophysical \&
  Astrophysical Fluid Dynamics}  \bvol{106}~(4-5),  \pg{392--428}.

\bibitem[Julien \& Watson(2009)]{Watson}
{\sc \au{Julien, K.} \& \au{Watson, M.}} \yr{2009}  \at{{Efficient
  multi-dimensional solution of PDEs using Chebyshev spectral methods}}.
  \jt{Journal of Computational Physics}  \bvol{228},  \pg{1480--1503}.

\bibitem[King \& Aurnou(2013)]{King13}
{\sc \au{King, E.~M.} \& \au{Aurnou, J.~M.}} \yr{2013}  \at{Turbulent
  convection in liquid metal with and without rotation}.  \jt{Proceedings of
  the National Academy of Sciences}  \bvol{110}~(17),  \pg{6688--6693},
  \arxiv{arXiv: http://www.pnas.org/content/110/17/6688.full.pdf}.

\bibitem[Kunnen {\em et~al.\/}(2016)Kunnen, Ostilla-Mónico, van~der Poel,
  Verzicco \& Lohse]{Kunnen16}
{\sc \au{Kunnen, R. P.~J.}, \au{Ostilla-Mónico, R.}, \au{van~der Poel, E.~P.},
  \au{Verzicco, R.} \& \au{Lohse, D.}} \yr{2016}  \at{Transition to geostrophic
  convection: the role of the boundary conditions}.  \jt{Journal of Fluid
  Mechanics}  \bvol{799},  \pg{413--432}.

\bibitem[Marshall \& Schott(1999)]{jM99}
{\sc \au{Marshall, J.} \& \au{Schott, F.}} \yr{1999}  \at{Open-ocean
  convection: Observations, theory, and models}.  \jt{Reviews of Geophysics}
  \bvol{37}~(1),  \pg{1--64}.

\bibitem[Miesch(2005)]{mM05}
{\sc \au{Miesch, M.~S.}} \yr{2005}  \at{Large-scale dynamics of the convection
  zone and tachocline}.  \jt{Living Reviews in Solar Physics}  \bvol{2}~(1).

\bibitem[Nastrom \& Gage(1985)]{nastrom}
{\sc \au{Nastrom, G.~D.} \& \au{Gage, K.~S.}} \yr{1985}  \at{A climatology of
  atmospheric wavenumber spectra of wind and temperature observed by commercial
  aircraft}.  \jt{Journal of the atmospheric sciences}  \bvol{42}~(9),
  \pg{950--960}.

\bibitem[Nieves {\em et~al.\/}(2014)Nieves, Rubio \& Julien]{Nieves}
{\sc \au{Nieves, D.}, \au{Rubio, A.~M.} \& \au{Julien, K.}} \yr{2014}
  \at{{Statistical classification of flow morphology in rapidly rotating
  Rayleigh-B\'{e}nard convection}}.  \jt{Physics of Fluids}  \bvol{26}~(8).

\bibitem[Pope(2000)]{sP00}
{\sc \au{Pope, S.~B.}} \yr{2000} {\em Turbulent Flows\/}.  \publ{Cambridge:
  Cambridge University Press}.

\bibitem[Roberts(1968)]{pR68}
{\sc \au{Roberts, P.~H.}} \yr{1968}  \at{On the thermal instability of a
  rotating-fluid sphere containing heat sources}.  \jt{Phil. Trans. R. Soc. A}
  \bvol{263},  \pg{93--117}.

\bibitem[Rubio {\em et~al.\/}(2014)Rubio, Julien, Knobloch \& Weiss]{Rubio}
{\sc \au{Rubio, A.~M.}, \au{Julien, K.}, \au{Knobloch, E.} \& \au{Weiss,
  J.~B.}} \yr{2014}  \at{Upscale energy transfer in three-dimensional rapidly
  rotating turbulent convection}.  \jt{Phys. Rev. Lett.}  \bvol{112},
  \pg{144501}.

\bibitem[Schubert \& Soderlund(2011)]{Soderlund}
{\sc \au{Schubert, G.} \& \au{Soderlund, K.~M.}} \yr{2011}  \at{Planetary
  magnetic fields: Observations and models}.  \jt{Physics of the Earth and
  Planetary Interiors}  \bvol{187}~(3),  \pg{92--108}.

\bibitem[Spalart {\em et~al.\/}(1991)Spalart, Moser \& Rogers]{SMR91}
{\sc \au{Spalart, P.~R.}, \au{Moser, R.~D.} \& \au{Rogers, M.~M.}} \yr{1991}
  \at{Spectral methods for the {N}avier-{S}tokes equations with one infinite
  and two periodic directions}.  \jt{J. Comp. Phys.}  \bvol{96},
  \pg{297--324}.

\bibitem[Sprague {\em et~al.\/}(2006)Sprague, Julien, Knobloch \&
  Werne]{Sprague}
{\sc \au{Sprague, M.}, \au{Julien, K.}, \au{Knobloch, E.} \& \au{Werne, J.}}
  \yr{2006}  \at{Numerical simulation of an asymptotically reduced system for
  rotationally constrained convection}.  \jt{Journal of Fluid Mechanics}
  \bvol{551},  \pg{141--174}.

\bibitem[Stellmach \& Hansen(2008)]{sS08}
{\sc \au{Stellmach, S.} \& \au{Hansen, U.}} \yr{2008}  \at{{An efficient
  spectral method for the simulation of dynamos in Cartesian geometry and its
  implementation on massively parallel computers}}.  \jt{Geochem. Geophys.
  Geosys.}  \bvol{9}~(5).

\bibitem[Stellmach {\em et~al.\/}(2014)Stellmach, Lischper, Julien, Vasil,
  Cheng, Ribeiro, King \& Aurnou]{Stellmach}
{\sc \au{Stellmach, S.}, \au{Lischper, M.}, \au{Julien, K.}, \au{Vasil, G.},
  \au{Cheng, J.~S.}, \au{Ribeiro, A.}, \au{King, E.~M.} \& \au{Aurnou, J.~M.}}
  \yr{2014}  \at{Approaching the asymptotic regime of rapidly rotating
  convection: Boundary layers versus interior dynamics}.  \jt{Phys. Rev. Lett.}
   \bvol{113},  \pg{254501}.

\bibitem[Yadav {\em et~al.\/}(2015)Yadav, Gastine, Christensen \&
  Reiners]{rY15}
{\sc \au{Yadav, R.~K.}, \au{Gastine, T.}, \au{Christensen, U.~R.} \&
  \au{Reiners, A.}} \yr{2015}  \at{Formation of starspots in self-consistent
  global dynamo models: Polar spots on cool stars}.  \jt{Astronomy \&
  Astrophysics}  \bvol{573},  \pg{A68}.

\end{thebibliography}
\end{document}